\documentclass{layout}

\begin{document}

\thispagestyle{empty}
\begin{center}
  \renewcommand{\baselinestretch}{2.00}
  \Large\sffamily
  Department of Physics and Astronomy\\
  \large Heidelberg University
  \par\vfill\normalfont
  Master thesis\\
  in Physics\\
  submitted by\\
  Sarah Schuhegger\\
  born in Traunstein\\
  2021
\end{center}

\newpage\null\thispagestyle{empty}\newpage

\thispagestyle{empty}
\begin{center}
  \renewcommand{\baselinestretch}{2.00}
  \Large\bfseries\sffamily
    Body Part Regression for CT Images\\

  \par
  \vfill
  \large\normalfont
  This Master thesis has been carried out by Sarah Schuhegger\\
  at the\\
  German Cancer Research Center\\
  under the supervision of\\
  Prof. Dr. Klaus H. Maier-Hein, \\
  Dr. Fabian Isensee, \\
  Lisa Kausch, \\
  and \\
  Prof. Dr. Carsten Rother \\
  Visual Learning Lab  
\end{center}\par
\vspace{1\baselineskip} 

 
\newpage\null\thispagestyle{empty}\newpage

\thispagestyle{empty}
\begin{center}
  \begin{minipage}[c][0.38\textheight][b]{0.9\textwidth}
    \small
    \textbf{
      Body Part Regression for CT Images:
    }\par
    \vspace{\baselineskip}
One of the greatest challenges in the medical imaging domain is to successfully transfer deep learning models into clinical practice. Since models are often trained on a specific body region, a robust transfer into the clinic necessitates the selection of images with body regions that fit the algorithm to avoid false-positive predictions in unknown regions. Due to the insufficient and inaccurate nature of manually-defined imaging meta-data, automated body part recognition is a key ingredient towards the broad and reliable adoption of medical deep learning models.
While some approaches to this task have been presented in the past, building and evaluating robust algorithms for fine-grained body part recognition remains challenging. So far, no easy-to-use method exists to determine the scanned body range of medical Computed Tomography (CT) volumes. In this thesis, a self-supervised body part regression model for CT volumes is developed and trained on a heterogeneous collection of CT studies. Furthermore, it is demonstrated how the algorithm can contribute to the robust and reliable transfer of medical models into the clinic. Finally, easy application of the developed method is ensured by integrating it into the medical platform toolkit Kaapana and providing it as a python package on GitHub.

  \end{minipage}\par
   \par\bigskip 
   \par\bigskip 
   \begin{minipage}[c][0.45\textheight][b]{0.9\textwidth}
    \small
    \textbf{
      Körperbereichs-Regression für CT Bilder: 
    }\par
    \vspace{\baselineskip}
Eine der größten Herausforderungen im Bereich der medizinischen Bildgebung besteht darin, Deep-Learning-Modelle erfolgreich in die klinische Praxis zu übertragen. Da Modelle häufig auf einer bestimmten Körperregion trainiert werden, erfordert eine robuste Übertragung in die Klinik die Auswahl von Bildern mit Körperregionen, die zum Algorithmus passen, um falsch-positive Vorhersagen in unbekannten Regionen zu vermeiden. Aufgrund der typischerweise unzureichenden Informationen des aufgenommenen Körperbereichs in den Bildmetadaten, stellt die automatische Erkennung von Körperbereichen ein Schlüsselelement für die breite und zuverlässige Einführung von medizinischen Deep-Learning-Modellen dar. Obwohl in den letzten Jahren einige Ansätze für diese Aufgabe vorgestellt wurden, bleibt das Entwickeln und die Evaluation robuster Algorithmen für die genaue Körperbereichserkennung eine Herausforderung. Bisher gibt es keine einfach zu handhabende Methode zur Bestimmung des gescannten Körperbereichs von medizinischen Computertomografie (CT)-Volumina. In dieser Arbeit wird ein selbstüberwachtes Körperbereichs-Regressionsmodell für CT-Volumen entwickelt und auf einer heterogenen Sammlung von CT-Studien trainiert. Weiterhin wird gezeigt, wie der Algorithmus zur robusten und zuverlässigen Übertragung von medizinischen Modellen in die klinische Praxis beitragen kann. 
Schließlich wird die Methode auf dem medizinischen Plattform-Toolkit Kaapana und als einfach zu bedienendes Python-Paket auf GitHub bereitgestellt.

  \end{minipage}
\end{center}

\newpage\null\thispagestyle{empty}\newpage

\pagenumbering{arabic}
\setcounter{page}{1}

\tableofcontents

\chapter{Introduction} 
\section{Motivation}
One of the greatest challenges in the medical imaging domain is to successfully transfer deep learning models from homogeneous development environments into clinical practice. Most algorithms are trained on a specific body region, disease, or modality. If the models are then required to process images that are out of their scope, one can no longer rely on their results to be accurate. This can lead to potentially harmful and unintended consequences for the clinical environment. 

One important strategy to avoid these issues is to only present images to the models close to their respective training distribution and reject images or image regions that could be considered out-of-distribution. For example, in the medical domain, the modality and scanned body part in the image meta-data could be leveraged to decide whether an image fits a given algorithm. 

The standard format for storing radiologic images and their meta-data is the \ac{DICOM} format. The meta-data is stored in the DICOM header and includes, e.g., the patient's name, examination date, examined device settings, or modality. Furthermore, it includes fields with the information of the scanned body part called \textit{Body Part Examined} and \textit{Anatomic Structure}. While image meta-data can, to some extent, address the above-outlined problem, it is inherently unreliable in practice \cite{gueld2002quality} and cannot solve cases where the desired anatomical region is located somewhere within a larger image, such as a whole-body CT scan. 

Automatic body part recognition can pave the way towards a broader application of models in hospitals and can become an essential part of medical deep learning software. Moreover, a better estimate of the examined body part than provided by the meta-data can be obtained through automatic body part recognition. This can lead to a significant time reduction in preprocessing, sorting, and filtering large medical datasets. 

So far, no easy-to-use method exists for automatic body part recognition in \ac{CT} images. In addition, no easy applicable and fully automated technique exists to crop  \ac{CT} images to the scope of medical deep learning algorithms to provide robust end-to-end computer vision tools for the clinic. 

Previous publications delivered the proof of concept of self-supervised body part regression for automatic body part recognition \cite{yan2018deep, yan2018unsupervised, tang2021body}. This thesis will investigate the generalizability of the approach considering a heterogeneous collection of CT studies and deploy the method into an easily applicable framework. Finally, it will be shown how the method can be used as an essential pillar for \mbox{robust clinical end-to-end computer vision pipelines}.

\section{Contributions}\label{sec:aims}
Specifically, the work done in this thesis will focus on three main contributions: 

\begin{enumerate}
    \item[] \textbf{C1: } Obtaining insights into the underlying optimization problem of robust body part regression to train a better-performing and generalizable body part regression model for CT images.

    \item[] \textbf{C2: }  Proposing a more fine-grained, thorough and robust evaluation strategy for cross-model performance comparison for body part regression models. 

    \item[] \textbf{C3: } Proposing three use cases of body part regression and simplify their application for the research community and the clinical environment. The presented use cases tackle the challenges of estimating the examined body part, cropping the known region from a CT volume for an algorithm, and carrying out basic data sanity checks. 
\end{enumerate}

\chapter{Background}
In this thesis, a deep learning based body part regression model for \ac{CT} volumes is trained and deployed on GitHub and Kaapana. 
The foundations of deep learning will be explained in section \ref{sec:deep-learning}, and the medical background regarding the human anatomy and Computed Tomography will be explained in section \ref{sec:medical-background}. The software toolkit Kaapana which tries to build the bridge between the deep learning research community and the clinical environment, is explained in section 
\ref{sec:background-deployment}.

\section{Deep Learning}\label{sec:deep-learning}
\begin{figure}[t]
    \centering
    \includegraphics[width=0.5\textwidth]{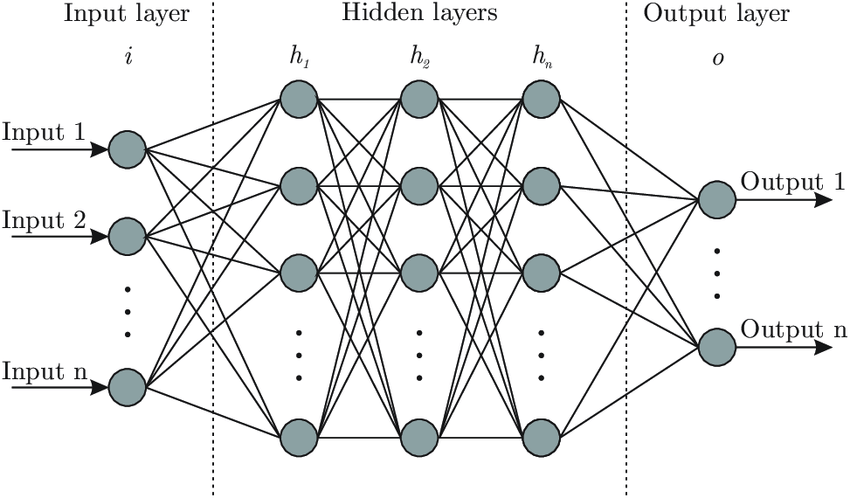}
    \caption{Simplified version of a fully connected feedforward neural network. The Image was taken from \cite{neuralnetwork}.}
    \label{fig:feedforward}
\end{figure}

Deep learning is a subfield of machine learning that uses multi-layer \textbf{neural networks} to solve optimization tasks numerically. Neural networks are concatenations of linear and non-linear functions. They are a cascading of many simple functions, which in the end learns complex representations of the input \cite{DeepVision,bishop2006pattern}. For example, in figure \ref{fig:feedforward} a simplified version of a standard fully connected feedforward neural network with one input layer, three hidden layers, and one output layer can be seen. Each hidden layer consists of several neurons, and each neuron consists of a series connection of a linear function of the input values and a subsequent non-linear function. 

\textbf{Universal Function Approximation Theorem: } In general, what makes neural networks so powerful, is that they are trainable parametric universal function approximators. It means that they can learn any function with arbitrary precision as long as the complexity of the neural network is high enough. This holds because of the Universal Function Approximation Theorem, which states that every feedforward neural network with at least one single hidden layer is sufficient to represent any continuous function with arbitrary precision \cite{DeepVision}. Feedforward neural networks are networks where connections between nodes do not form a cycle like, e.g., the network in figure \ref{fig:feedforward}. 
If the network has just one hidden layer, it may be extremely big and fail to learn. Therefore, from a practical point of view, it is better to use deeper models. This leads to more complex representations of the input, which helps the model to generalize better \cite{krizhevsky2012imagenet}.

\begin{figure}
    \centering
    \includegraphics[width=0.6\textwidth]{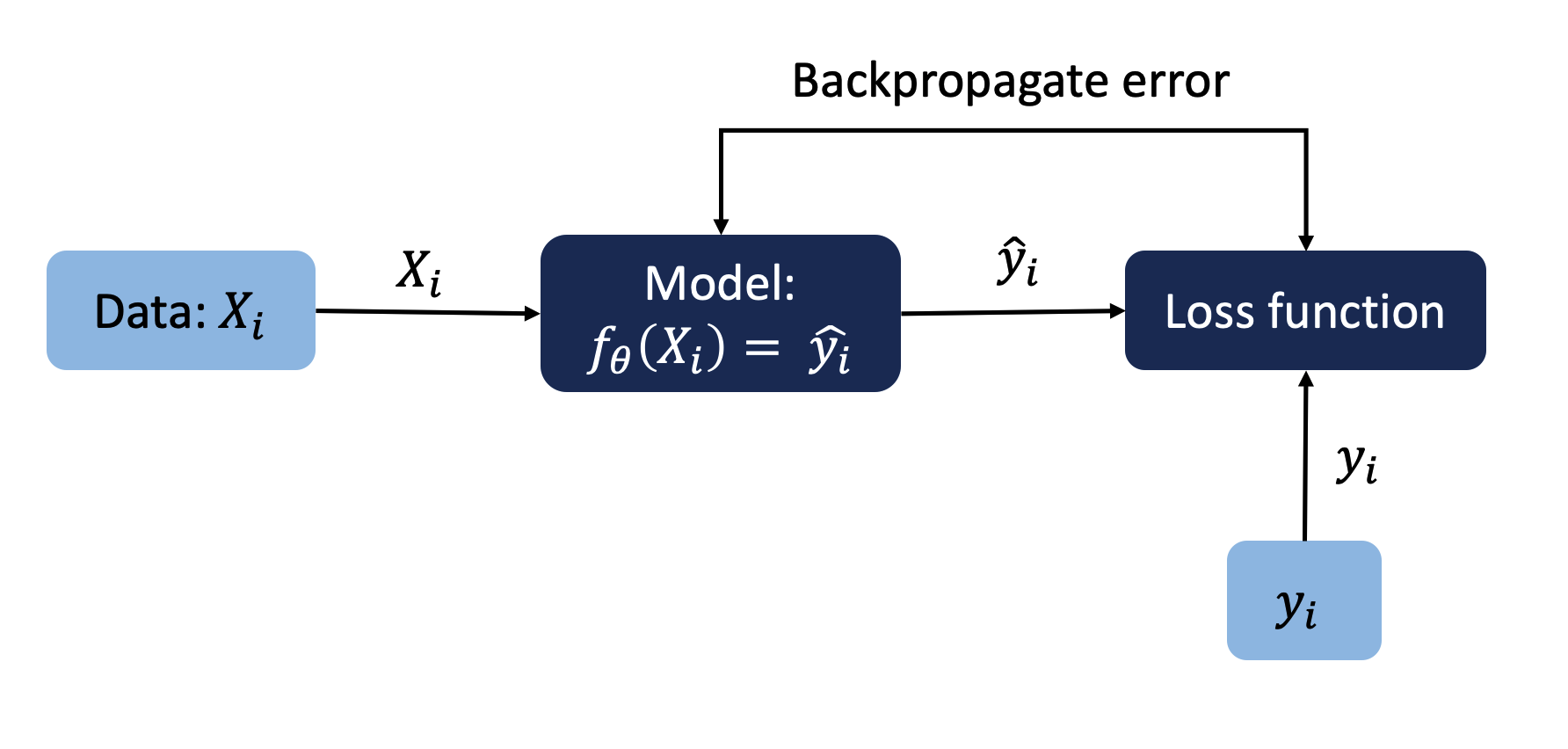}
    \caption{Neural network learning framework for mini-batch size of $B=1$ and neural network $f_{\boldsymbol{\boldsymbol{\theta}}}$.}
    \label{fig:nn-framework}
\end{figure}

\textbf{Learning Framework: } 
From the Universal Function Approximation Theorem, we know that a deep neural network $f_{\boldsymbol{\theta}}$ is, in theory, able to represent any function with arbitrary precision, given a sufficient number of linear and non-linear components. The main task that remains is to find parameters $\boldsymbol{\theta}$ for the network to solve an optimization problem 
reasonably good. 
The optimization problem is defined by an \textbf{objective function} $\Phi$ 
\begin{equation*}
    \Phi(\boldsymbol{\theta}) = \frac{1}{N} \sum_{i=1}^{N} l(f_{\boldsymbol{\theta}}(\mathbf{X}_{i}), y_{i}), 
    \qquad f_{\boldsymbol{\theta}}(\mathbf{X}_{i})=\hat{y}_{i}, 
\end{equation*}
where $N$ is the amount of available data pairs $(\mathbf{X}_{i}, y_{i})$ and $l$ the \textbf{loss function} per sample. 
For supervised tasks, the objective function normally compares the predicted output from the neural network $\hat{y}_{i}$ and the ground truth label $y_{i}$. 
The loss is high if the prediction from the neural network and the ground truth label differ and low if similar. 
The objective function is numerically minimized by \textbf{Gradient Descent} and \textbf{Backpropagation}. Gradient Descent is a numerical optimization method to find extreme values of a function by following the negative gradient. The numerical update step of classic gradient descent is given by
\begin{equation*}
    \boldsymbol{\theta} = \boldsymbol{\theta} - \epsilon \nabla_{\boldsymbol{\theta}}\Phi(\boldsymbol{\theta}), 
\end{equation*}
where $\epsilon$ is the learning rate. The learning rate is a hyperparameter that controls the step size of the parameter updates along the negative gradient. If the learning rate is too low, the convergence may be slow. If the learning rate is too high, the gradients might diverge. 
Backpropagation describes how the gradient $\nabla_{\theta_{i}}L(\theta)$ for each parameter $\theta_{i}$ can be found in a neural network through the chain rule \cite{DeepVision}. 
For training neural networks, a robust approximation of gradient descents is used to save computation time. One common optimization strategy to gain robust gradient estimates is called \textbf{Adam} \cite{kingma2014adam}. The Adam optimizer approximates the gradients by averaging over a mini-batch $B$ and computing adaptive learning rates $\epsilon$ for stabilization \cite{kingma2014adam}. 
The adaptive learning rates are estimated from the first and second moment of the gradients (for further details, refer to \cite{DeepVision}).
A simplified version of the learning framework for a neural network $f_{\boldsymbol{\theta}}(\mathbf{X}_{i}) = \hat{y}_{i}$  with the mini-batch size of $B=1$ can be seen in figure \ref{fig:nn-framework}. 
For the optimization problem, it is important to have many independent example data pairs $(\mathbf{X}_{i}, y_{i})$ for training so that the neural network can find suitable parameters and generalize on unseen new data. In general, before training neural networks the total dataset of input-output pairs  $(\mathbf{X}_{i}, y_{i})$ is split into three independent datasets: the \textbf{training dataset}, \textbf{validation dataset} and \textbf{test dataset}. The training dataset is used to train the neural network. The validation dataset is used to choose the best hyperparameters for the neural network based on the network's performance on the validation set. Finally, the test set is the dataset with which the neural network's generalization to new data is estimated. The dataset is separated at the beginning to avoid overestimating the generalizability of the model. 

\begin{figure}
    \centering
    \includegraphics[width=0.6\textwidth]{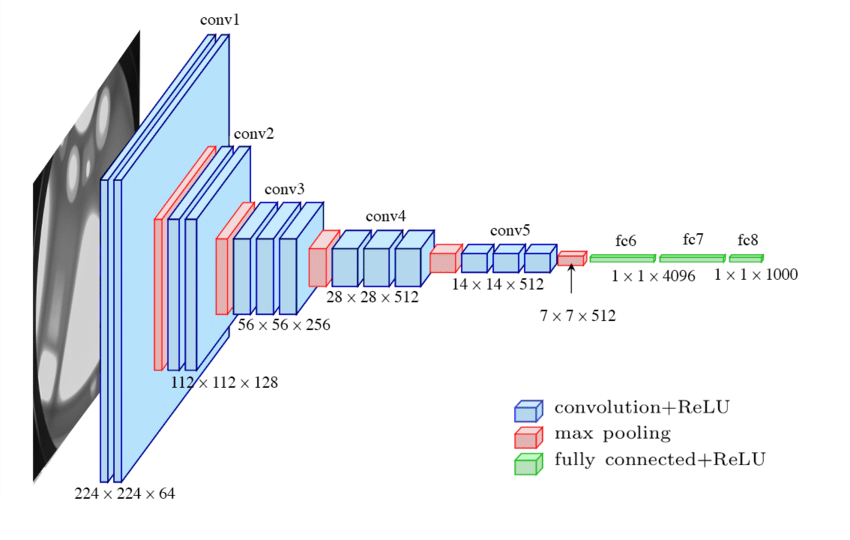}
    \caption{Visualization of the VGG16 architecture. The image was taken from \cite{ferguson2017automatic}.}
    \label{fig:vgg16}
\end{figure}

\begin{figure}
    \centering
    \includegraphics[width=\textwidth]{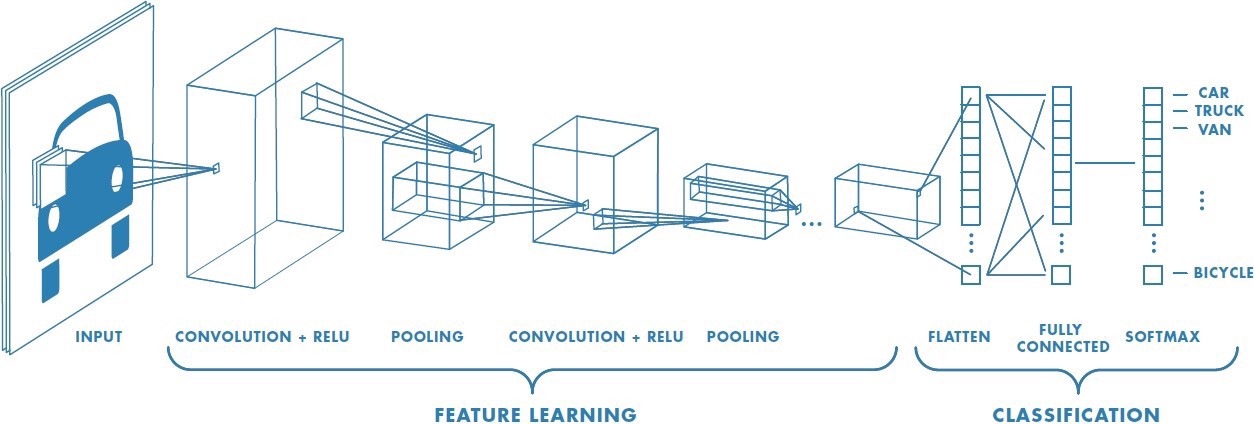}
    \caption{Example of a CNN for predicting the vehicle visible in an image. At the beginning, convolutional layers learn a good feature representation for separating the classes. At the end fully connected layers take over the classification similar to logistic regression. The image was taken from \cite{cnn}.}
    \label{fig:CNN}
\end{figure}

\textbf{Convolutional Neural Networks: }Different types of feedforward neural networks $f_{\boldsymbol{\theta}}$ are distinguished based on their architecture and input-output types. The state-of-the-art neural network architectures for analyzing images are called Convolutional Neural Networks (CNNs). This thesis concentrates on analyzing images. Therefore, we will focus on discussing the architecture of CNNs in more detail.
In the later part of the work, we will come back to the VGG16 network, which is one of the mose common CNN architectures introduced by Simonyan and Zisserman \cite{simonyan2014very}. The network architecture can be found in figure \ref{fig:vgg16}. Besides this, a standard architecture of a CNN is shown in figure \ref{fig:CNN}. 
In general, an image is feed into multiple CNN building blocks, which are stacked together. One building block consists of a convolutional layer, an activation layer, which is normally the \ac{ReLU} activation function, and a pooling layer \cite{DeepVision,bishop2006pattern}. The convolutional layer performs a discrete convolution on the input image. The ReLU function is a simple non-linear function given by 
\begin{equation}
    \text{ReLU}(x) = \text{max(}0, x). 
\end{equation}
Additionally, the pooling layer is a downsampling layer that reduces the size of the input matrix. 
The output is flattened and used as input for a fully connected neural network at the end of these building blocks. 
In summary, the convolutional body transforms the input image into a dense feature map encoding spatially separated semantic information. Finally, the fully connected model head combines the semantic information from various regions to produce a final prediction.
For further details regarding the CNN architecture and functionality, 
\mbox{refer to \cite{DeepVision}}.


\subsection{Learning Tasks}
We can divide deep learning tasks by the format of their target. This thesis will concentrate on three main deep learning tasks: classification, regression, and semantic segmentation. 

\textbf{Classification: }
In classification tasks, categorical information is predicted. For example, a neural network that predicts if a cat or a dog is seen in an image would be a neural network for classification. 
Neural networks for classification have a softmax function following the last layer, which maps 
continuous values to values between zero and one. For a multi-class classification problem, the softmax function $f$ is given by 
\begin{equation*}
    \begin{split}
    & f: \quad \mathbb{R}^{m} \rightarrow \mathbb{R}^{m}: \quad
    \mathbf{z} \mapsto \frac{\text{exp}(\mathbf{z})}
    {\sum_{j} \text{exp}(z_{j})} = \mathbf{\hat{p}}, \\
    &\mathbf{z}= \mathbf{W}^{T}\mathbf{x} + \mathbf{b}; \quad 
    z_{i} = (\mathbf{W}^{T} \mathbf{x} + \mathbf{b})_{i}, 
    \end{split}
\end{equation*} 
with
\begin{equation*}
    \sum_{i=1}^{m} \hat{p}_{i} = 1, 
\end{equation*}
where $m$ is the number of distinct classes and $\mathbf{W}$ the weights of the last fully connected layer and $\mathbf{b}$ the bias. 
With the outputs $\hat{p}_{i} \in (0, 1)$ of the last fully connected layer, the loss is computed. For classification tasks, the standard loss function is the \textbf{Cross-Entropy} loss (see sec. \ref{sec:loss-functions}).

\textbf{Regression: }
If the target variable is continuous, we have a regression problem to solve.
The network architecture of classification networks and regression networks look quite similar. The regression network has a linear layer at the end of the network instead of a softmax function. Moreover, the standard loss function for regression models is the \textbf{\ac{MSE}} (see sec. \ref{sec:loss-functions}). 

\begin{figure}
    \centering
    \begin{subfigure}[b]{0.45\textwidth}
    \centering
        \includegraphics[width=0.6\textwidth]{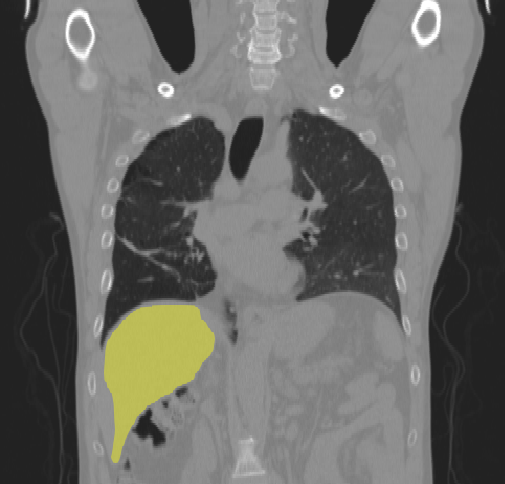}
        \caption{Liver segmentation}
    \end{subfigure}
    \begin{subfigure}[b]{0.45\textwidth}
        \centering
        \includegraphics[width=0.6\textwidth]{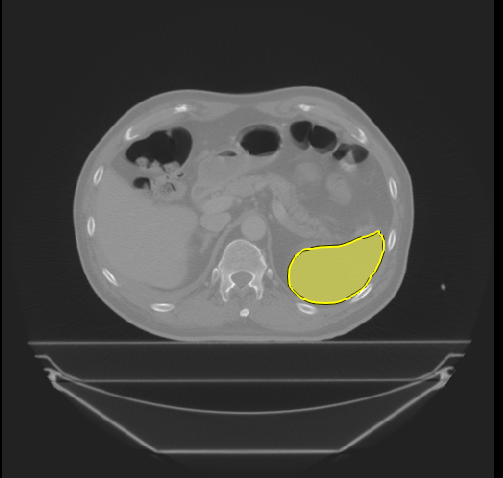}
        \caption{Spleen segmentation}
    \end{subfigure}
    \caption{Example segmentations of liver and spleen on the SegTHOR dataset  \cite{lambert2020segthor}. The visualization was done with \ac{MITK} \cite{wolf2005medical}.}
    \label{fig:semantic-segmentation}
\end{figure}

\textbf{Semantic Segmentation: }
Classification on pixel level is called semantic segmentation. Semantic segmentation is mostly related to organ and disease (e.g., tumor) segmentation in the medical domain. Example images of organ segmentation can be found in figure \ref{fig:semantic-segmentation}. The standard baseline model for medical image segmentation is the \textbf{\ac{nnU-Net}}. The nnU-Net was developed by Isensee et al. \cite{isensee2021nnu}. 
It is an out-of-the-box segmentation framework for organ and tumor segmentation for radiological images. High performing semantic segmentation algorithms can be easily trained without manual interventions for a variety of different dataset settings \cite{isensee2021nnu}. 

\subsection{Learning Methodologies}\label{sec:learning-methodologies}
There are many deep learning methodologies, such as supervised learning, unsupervised learning, semi-supervised learning, reinforcement learning, and self-supervised learning. In this thesis, the two most relevant learning methodologies are supervised learning and self-supervised learning. Therefore, these two methodologies will be explained in the following.

\textbf{Supervised Learning: } 
In supervised learning, the neural network maps an input to an output based on many annotated input-output data points. Examples of supervised learning are: 
\begin{itemize}
    \item Classifying color images with the ImageNet database. The ImageNet database has many different targets as dog, sailboat, human, and jet \cite{deng2009imagenet}.
    \item Organ segmentation, where the network is trained on lots of pairs of three-dimensional volumes and manually annotated segmentation masks.
    \item Predicting the brain age through a regression network and using a database with pairs of MRI images and the brain age \cite{jonsson2019brain}. 
\end{itemize}
\textbf{Self-Supervised Learning: } In self-supervised learning, the basic idea is that the neural network architecture generates itself a kind of supervisory signal through a proxy objective function to solve a task. Self-supervised learning is often used to find a good initialization for supervised learning tasks, where only a small annotated dataset is available. Through self-supervised learning, meaningful representations of the data should be learned without the need for annotated data. The self-supervised pretraining task is referred to as "\textbf{pretext task}", and the actual task to be solved is referred to as "\textbf{downstream task}". 
Examples for self-supervised learning tasks are: 
 \begin{itemize}
     \item Transform gray scale images to color images \cite{zhang2016colorful}. 
     \item Ordering image patches correctly. Cut puzzle pieces out of the image, shuffle them and predict the correct permutation \cite{noroozi2016unsupervised}. 
     \item Ordering movie frames correctly. Shuffle movie frames and predict the chronological order \cite{lee2017unsupervised}. 
 \end{itemize}

\subsection{Loss Functions}\label{sec:loss-functions}
For training a neural network, the loss function is crucial. The main loss functions for classification (Cross-Entropy loss) and for regression (Mean Square Error) will be explained in this section. 

\textbf{Cross-Entropy Loss: } The Cross-Entropy loss is the main loss function for classification tasks. For $M$ classes and data point $i$ it is given by 
\begin{equation*}
    l_{i} =  \sum_{c=1}^{M} y_{ic} \cdot \text{ln}(\hat{p}_{ic}), 
\end{equation*}
where $y_{ic}\in [0, 1]$ is the ground truth of data point $i$ and 
$\hat{p}_{ic}$ is the predicted probability for $p(y_{ic}=1)$.
For $N$ data points, the corresponding Cross-Entropy loss function is respectively given by 
\begin{equation*}
    L = \sum_{i=1}^{N} l_{i}(\mathbf{y_{i}}, \mathbf{\hat{p}_{i}}).
\end{equation*}
To understand the idea and the derivation behind the Cross-Entropy loss better, we will derive the Cross-Entropy loss for the example of a binary image classification problem with the ground truth label $y_{i} \in [0, 1]$ (e.g. cat or dog). We assume that the class $y_{i}$ is Bernoulli distributed 
\begin{equation*}
\begin{split}
    & y_{i} \sim \mathcal{B}(p_{i}),  \\ 
    & \Rightarrow P(y_{i} = 1| \mathbf{X}_{i}) = p_{i}, \quad P(y_{i} = 0| \mathbf{X}_{i}) = 1 - p_{i}, \\
    & \Rightarrow P(y_{i}|\mathbf{X}_{i}) = p_{i}^{y_{i}}\cdot (1 - p_{i})^{1 - y_{i}}.
\end{split}
\end{equation*}
The probability for the positive class $p$ will be estimated by a neural network $f_{\boldsymbol{\theta}}$ 
\begin{equation*}
    f_{\boldsymbol{\theta}}(\mathbf{X}_{i}) = \hat{p}_{i},\quad  \hat{p}_{i} \in (0, 1),
\end{equation*}
where $\mathbf{X}_{i}$ is the image with the ground truth label $y_{i}$. 
For this problem the likelihood is given by 
\begin{equation*}
\begin{split}
    \mathcal{L}(\{\mathbf{X}, y\}_{i \in [0, ..., N]} | \ \{\hat{p}_{i}\}_{i \in [0, ..., N]}) 
    &= \prod_{i=1}^{N} P(y=y_{i}|\hat{p}_{i}), \\ 
    &= \prod_{i=1}^{N} \hat{p}_{i}^{y_{i}}\cdot (1 - \hat{p}_{i})^{1 - y_{i}}.
\end{split}
\end{equation*}
The corresponding log-likelihood is given by the natural logarithm of the likelihood function 
\begin{equation}\label{eq:Log-Likelihood-ce}
    \text{ln }\mathcal{L} = \sum_{i=1}^{N} {y_{i}} \cdot \text{ln(}\hat{p}_{i}) 
    + (1 - y_{i})\cdot \text{ln(}1 - \hat{p}_{i}).
\end{equation}
Maximizing the log-likelihood of equation \ref{eq:Log-Likelihood-ce} is equivalent of minimizing the Cross-Entropy given by
\begin{equation}\label{eq:ce-loss}
    L = - \sum_{i=1}^{N} y_{i}\cdot \text{ln(} \hat{p}_{i}) - (1 - y_{i}) \cdot \text{ln(}1 - \hat{p}_{i}), 
\end{equation}
where $N$ is the number of observed data points $(\mathbf{X}, \ y)_{i}$. 

\textbf{Mean Square Error: } The Mean Square Error (MSE) is the main loss function for regression tasks. For $N$ samples and the ground truth $y_{i} \in \mathbb{R}$ the MSE is given by 
\begin{equation}
     L = \frac{1}{N}\sum_{i=1}^{N}||y_{i} - f_{\boldsymbol{\theta}}(\mathbf{X}_{i})||_{2}^{2}. 
\end{equation}
Minimizing the Mean Square Error is equivalent to maximizing the likelihood of the regression problem, if a Gaussian error $\boldsymbol{\epsilon}$ is assumed \cite{DeepVision}: 
\begin{equation*}
    \mathbf{y} = f_{\boldsymbol{\theta}}(\mathbf{X})+ \pmb{\epsilon},\quad \pmb{\epsilon} \sim N(0, \ \sigma^{2}\mathds{1}). 
\end{equation*} 

\subsection{Regularization}
\begin{figure}
    \centering
    \includegraphics[width=0.8\textwidth]{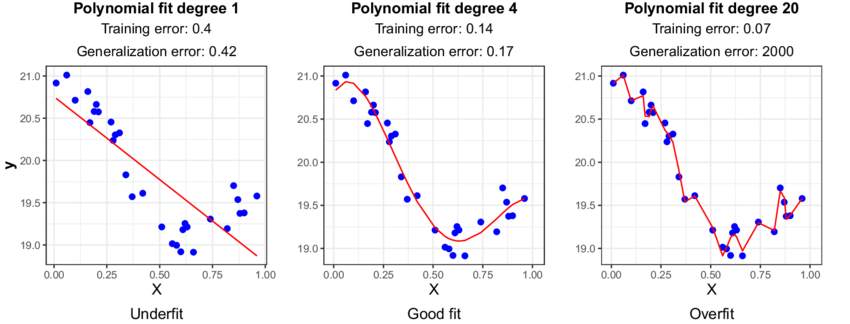}
    \caption{Visualization of the overfitting vs. underfitting problem on a simple regression model. The problems are based on an inappropriate model complexity. The image was taken from \cite{badillo2020introduction}.}
    \label{fig:overfitting-underfitting}
\end{figure}

One of the main pitfalls in training a neural network is to overfit or to underfit the problem. A model that overfits the training data performs excellent on the training data itself but fails if it sees new data. A model which underfits on the training data is performing poorly on the training data and unseen data. If a model underfits a problem, the model complexity is chosen too low to capture the problem's complexity (see fig. \ref{fig:overfitting-underfitting}). A higher model complexity will resolve the underfitting problem. If a model overfits on the training data, the model complexity is too high (see fig. \ref{fig:overfitting-underfitting}). If the complexity is too high, the model learns the noise in the training data as well. This leads to a poor generalization on new data. 
The key method to combat overfitting is regularization. With regularization techniques, the complexity of a model is reduced, and the model performance on unseen data can be improved.
Regularization sets additional constraints to the optimization problem or adds additional noise to the training process to reduce the complexity of the learned model. There are different ways of regularization as for example, using data augmentations, a regularization loss, or dropout. For further details regarding regularization techniques, refer to \cite{DeepVision}. 

Besides preventing overfitting, regularization is also used for ill-posed problems, which are problems with no unique solution. Through regularization, the solution space shrinks, and the numerical optimization is more stable \cite{buhlmann2011statistics}. 

\section{Medical Background}\label{sec:medical-background}
For this thesis, \ac{CT} scans are used as the data basis. Moreover, the annotation of robust anatomical landmarks is an important pillar of this work. Therefore, the anatomical foundations will be discussed in section \ref{sec:anatomical-foundations} and CT will be explained in section \ref{sec:Computed-Tomography}. 

\subsection{Anatomical Foundations}\label{sec:anatomical-foundations}
\begin{figure}
    \centering
    \includegraphics[width=0.5\textwidth]{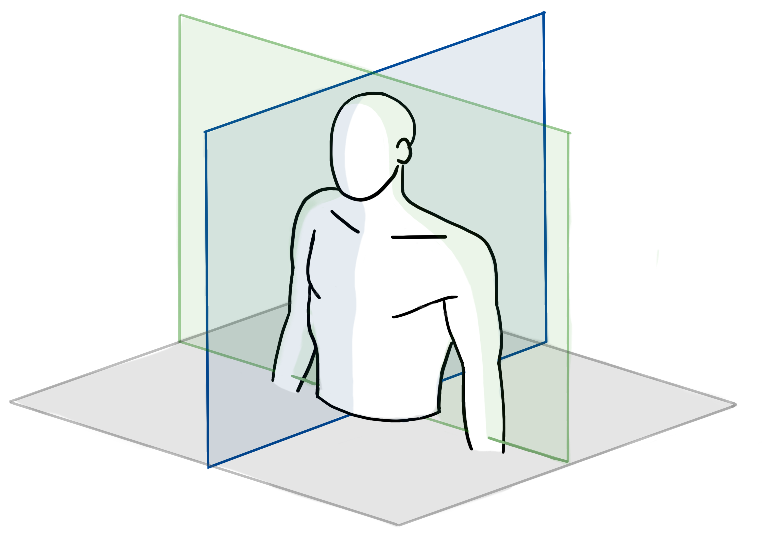}
    \caption{Visualization of axial (gray), sagittal (blue) and coronal (green) plane.}
    \label{fig:axial-sagittal-coronal}
\end{figure}

For a better orientation in the human body, \textbf{anatomical planes} are used. They are hypothetical planes to better describe locations and areas in the human body. A distinction is made between the following planes (see fig. \ref{fig:axial-sagittal-coronal}): 
\begin{itemize}
    \item The \textbf{axial} plane is parallel to the ground and separates the head from the feet. 
    \item The \textbf{sagittal} plane is perpendicular to the ground and separates the left arm from the right one. 
    \item The \textbf{coronal} plane is perpendicular to the ground and separates the chest from the back. 
\end{itemize}
In figure \ref{fig:anatomy} the appearance of different body parts in a CT scan is visualized. For each anatomical region, three different axial CT slices from different patients can be seen. This demonstration emphasizes the inter-patient variability in each anatomical region. The inter-patient variability of organs is much higher than the inter-patient variability of bones. One reason for this is that organs can move, and the position can depend, e.g., on the breathing cycle \cite{schiebler2013anatomie}. Therefore, for robust inter-patient anatomical landmarks, bone landmarks are beneficial compared to organ landmarks. The central support system of the human body which is suitable to define bone landmarks is the spine. 

\begin{figure}
    \centering
    \includegraphics[width=0.9\textwidth]{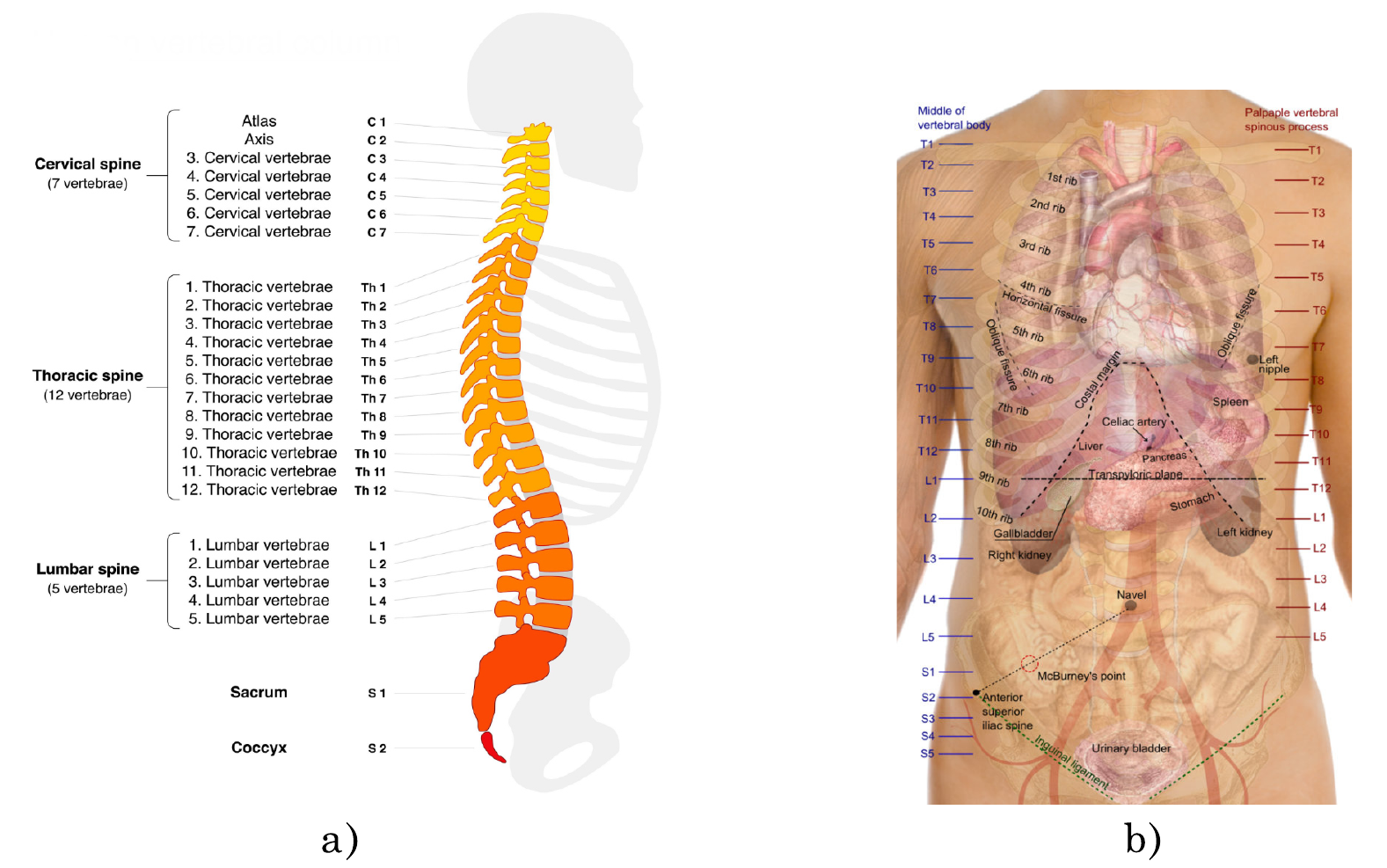}
    \caption{The anatomy of the spine is visible in figure a) and the surface projection of the organs of the trunk is visible in figure b). The images were taken from \cite{spine, anatomy-spine-projection}. }
    \label{fig:spine}
\end{figure}

\textbf{Spine: } The spine consists of five regions: Cervical, Thoracic, Lumbar, Sacral, and Coccygeal \cite{schiebler2013anatomie}. The Cervical spine consists of seven vertebrae, the Thoracic spine of twelve vertebrae, and the Lumbar spine of five vertebrae (see fig. \ref{fig:spine}). Each Thoracic vertebra carries one rib. The Lumbar vertebrae need to carry the most weight. Therefore, the volume of the Lumbar vertebrae is bigger than the vertebrae of the Thoracic spines or the Cervical spine \cite{schiebler2013anatomie}. 
In return, the vertebrae of the Cervical spine are more flexible \cite{schiebler2013anatomie}. 
 One of the primary diseases of the spine is scoliosis, where the spine has a lateral curvature \cite{kuznia2020adolescent}. This can lead to a bad posture and can affect breathing, and movement \cite{weinstein2008adolescent,yang2016early}. 
It occurs in about 3 \% of \mbox{humans \cite{shakil2014scoliosis}}. 

In figure \ref{fig:spine} b) the projection of the organs to the trunk can be seen. The Thoracic vertebrae protect most of the upper body and upper abdomen organs, including the lungs, the spleen, and the liver. Therefore, the vertebrae positions can be used as a reference to describe organ positions in the upper body. 

\subsection{Computed Tomography}\label{sec:Computed-Tomography}
\begin{figure}
    \centering
    \includegraphics[width=0.5\textwidth]{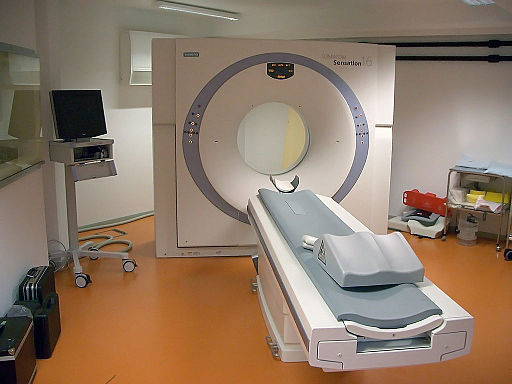}
    \caption{Clinical CT scanners consists usually of an adjustable patient table and a gantry in which detectors and X-ray tube rotate around the patient. The image was taken from
    \cite{ctscanner}.}
    \label{fig:ct-scanner}
\end{figure}

Computed Tomography (CT) is a quantitative medical imaging method, which produces images in \ac{HU}. It determines the X-ray absorption coefficient in a matrix of equal size volume elements named voxels. Figure \ref{fig:ct-scanner} shows a clinical CT scanner. It consists of a patient table and a gantry, including the detectors and at least one X-ray tube. The detectors and the X-ray tubes rotate around the patient during measurement time, and the patient table moves horizontally. The X-ray tubes emit X-rays, and the detectors measure the attenuation of the X-rays after passing through the patient. 
The Beer-Lambert law can describe the attenuation of the radiation 
\begin{equation*}
    I = I_{0} \cdot e^{-\int_{0}^{d}\mu(\mathbf{r}) dr}, 
\end{equation*}
where $I_{0}$ is the original beam intensity, $I$ the final beam intensity, $d$ the distance between sender and receiver and $\mu(\boldsymbol{r})$ the absorption coefficient at the position $\mathbf{r}$. 
The CT scanner measures the radiation attenuation of a variety of different directions. From the ensemble of all data points, the distribution of $\mu(\mathbf{r})$ can be reconstructed. Finally, the distribution of $\mu(\mathbf{r})$ is linear transformed into Hounsfield units and visualized in gray values. A typical CT-volume consists of multiple slices with a layer thickness of 0.5 - 10 mm \cite{alkadhi2011funktioniert}. Each slice typically has the size of  512 x 512 pixels, with a corresponding pixel spacing of $0.2 \text{mm/pixel}$ to $1 \text{mm/pixel}$ \cite{alkadhi2011funktioniert}. These settings lead to a measured cross-section of 10 cm to 50 cm. 

\textbf{Hounsfield Unit: } From the X-ray absorption coefficients $\mu$, the Hounsfield unit can be calculated by
\begin{equation*}
    \text{HU-value} = 1000 \cdot \frac{\mu - \mu_{w}}{\mu_{w}}, 
\end{equation*}
where $\mu_{w}$ is the X-ray absorption coefficient of water. This is given by  $\mu_{w} \approx \text{0.192/cm}$ \cite{alkadhi2011funktioniert}. It follows that water always has the CT-value of 0 HU. Air has a CT-value of about - 1000 HU, because its absorption coefficient lies around zero. In table \ref{tab:hu} typical HU values for different tissue types and organs are shown. In a typical CT-volume, the data is saved in an unsigned 12 bit format with an offset of 1024 HU \cite{schlegel2018medizinische}. Therefore, the HU values range between - 1024 HU till 3071 HU. 
In the first approximation, the attenuation coefficients are proportional to the density of the tissue. Therefore, the CT values can be interpreted in good approximation as density values. The quantitative physical interpretation of the intensity units of a CT image opens up the opportunity of setting intensity ranges for visualization, called \textit{windowing}. Windowing can be beneficial to enhance the visual contrast for particular organs or pathology and filtering useless information.
\begin{table}
    \centering
    \caption{Organs and tissue types and their typical HU-values \cite{schlegel2018medizinische}.}.
    \begin{tabular}{ll|ll}
        \hline
        Tissue $\qquad \qquad$ & HU-value & Tissue $\qquad \qquad$ & HU-value \\
        \hline
        Air  & - 1000 HU & Pancreas & 20 HU to 50 HU \\
        Lung & -900 HU to -500 HU & Kidney & 20 HU to 40 HU  \\
        Fat & -100 HU to -70 HU & Liver & 40 HU to 70 HU \\
        Water & 0 HU & Bone (cancellous) & 70 HU to 350 HU \\
        Blood & 30 HU to 60 HU & Bone (cortical) & 350 HU to 2000 HU \\
        \hline
    \end{tabular}
    \label{tab:hu}
\end{table} 

\textbf{Variability: } Although the intensity values of a CT scan have a physical interpretation, variability across different CT scans from the same subject exists. One reason for the variability in CT scans are artifacts. Artifacts can be distinguished into physics-based, hardware-based, patient-based and setup-related artifacts
\cite{boas2012ct,barrett2004artifacts}. They can cause rings, noise, stripes and blurring in the CT scan \cite{boas2012ct,barrett2004artifacts}.
Variability in contrast and intensity exists as well across different scans of the same scanner based on different acquisition parameters \cite{berenguer2018radiomics}. Moreover, there is variability across different CT scanners. Inter-scanner variability can be traced back to different image reconstruction techniques, different scanner designs, and initial scanner settings, but are not fully understood yet \cite{mackin2015measuring,berenguer2018radiomics,choe2019deep}. 

\textbf{DICOM: } 
The standard format for storing and exchanging CT scans and other radiological images is the Digital Imaging and Communication in Medicine (DICOM) format.  DICOM files contain additional meta information, such as information about the scanner, the patient, and the imaged body part. In hospitals, the DICOM files are saved in the \ac{PACS}, the hospital's internal medical database for patient-related data. 

\section{Deployment}\label{sec:background-deployment} 
One main challenge for the medical computer vision research community is to provide the developed algorithms in the clinical environment. The software toolkit \textbf{Kaapana} aims to build the bridge between the deep learning research community and the clinical environment. Kaapana is an open-source toolkit for platform provisioning in the field of radiological data analysis \cite{kaapana}. It is not a ready-to-use software but rather a toolkit that can be used to build customized medical platforms for data analysis and AI applications. Because Kaapana builds upon Docker containers, users are enabled to add and remove tools from the platform. 
Docker containers are isolated from one another, including their own libraries, configuration files, and software. For further details to Docker container refer to \cite{docker}.  
Processing pipelines are build in Airflow, where a workflow is called \textbf{\ac{DAG}} \cite{airflow}. The bricks of the pipeline are called operators. Each operator triggers a particular Docker container that runs a specific task.
Kubernetes is used to manage the Docker containers and Kibana to trigger the DAG workflows (for further details, refer to \cite{kubernetes,kibana}). 
To implement new image analysis pipelines in Kaapana, a DAG needs to be built. 

\begin{figure}
    \centering
    \includegraphics[width=0.8\textwidth]{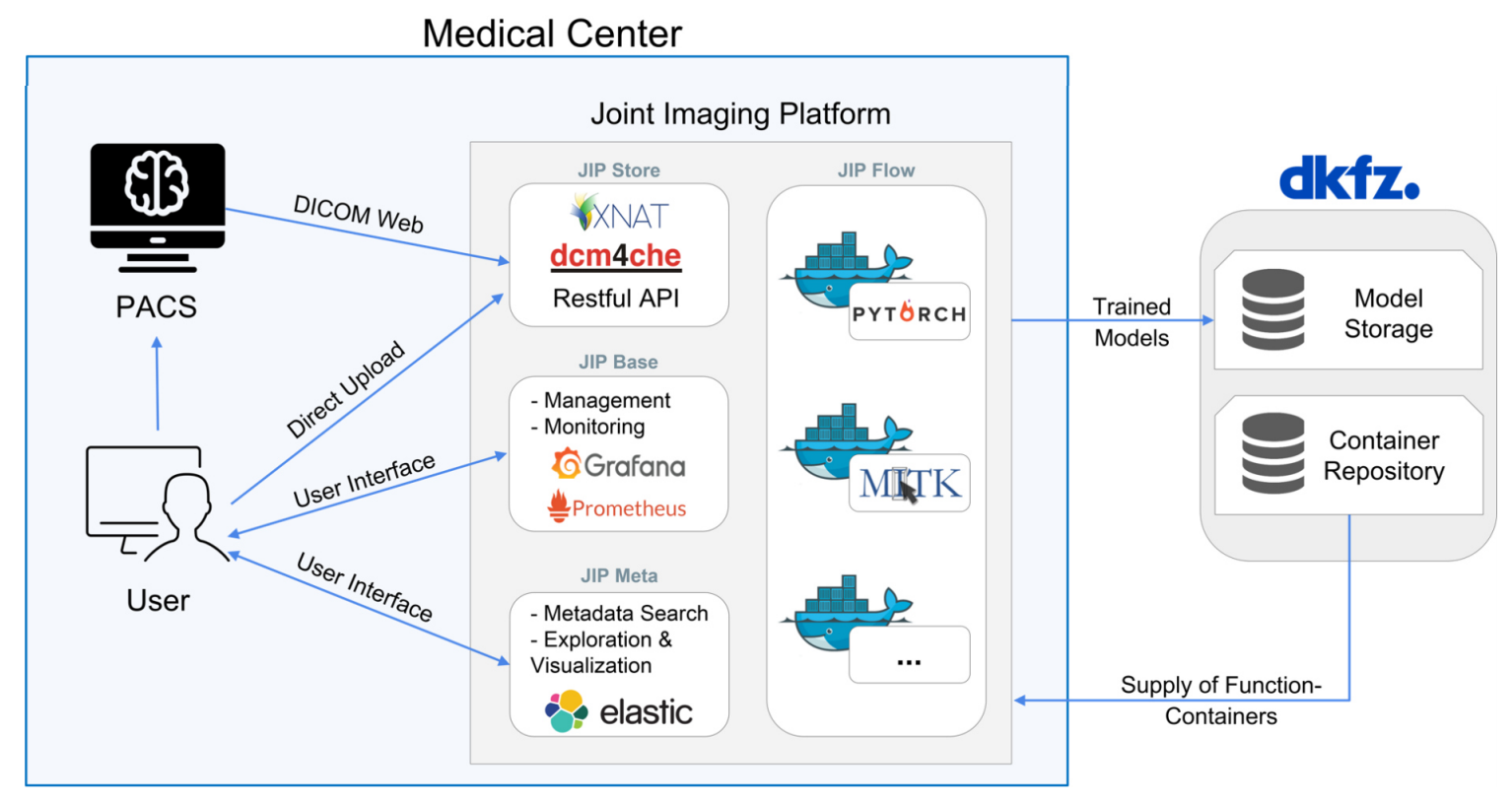}
    \caption{Infrastructure architecture of the Joint Imaging Platform (JIP), which is build on Kaapana \cite{scherer2020joint}. The image was taken from \cite{JIP}. }
    \label{fig:jip}
\end{figure}

Figure \ref{fig:jip} depicts the infrastructure of the \ac{JIP} developed by the German Cancer Research Center. The JIP is a medical platform for data analysis that was built on Kaapana and is used in several German university hospitals \cite{scherer2020joint, JIP}. 

\chapter{Related Work}\label{sec:related-work}
This work will refer to previously trained body part recognition models explained in section \ref{sec:related-work-bpr} and their evaluation methods explained in section \ref{sec:related-work-evaluation-methods}.
The previously stated use cases for body part regression are presented in section \ref{sec:related-work-deployment}.

\section{Body Part Regression}\label{sec:related-work-bpr}
For automatic body part recognition, several approaches exist. For example, through slice-wise body part classification, each slice in a body part gets predicted to a particular region. 
The classification approach was studied already in several studies \cite{roth2015anatomy,yan2016multi,de20162d,de2017convnet}. 
For example, Roth et al.  \cite{roth2015anatomy} implemented a CNN-based body part classifier, which takes two-dimensional axial CT slices as input and distinguishes between 5 classes (legs, pelvis, liver, lungs \& neck). They have shown that data augmentation is essential for training a robust body part classifier and achieved an accuracy of 94.1 \%. Most misclassifications appeared between two neighboring classes because of ambiguous class labels. Yan et al. \cite{yan2016multi} extended the work of Roth et al. \cite{roth2015anatomy} and trained a CNN-based classifier with more classes to gain a fine-grained prediction of the imaged body parts. In total, they defined 12 different classes (nose, teeth, neck, shoulder, lung apex, sternal, aorta arch, cardiac, liver upper, liver middle, abdomen/kidney, and ilium). 
A multi-stage deep learning framework was proposed and applied for body part recognition. In the first stage, the network discovers local discriminative regions, and in the second stage, it learns a slice-level classification on the local regions. With this approach, an $F_{1}-$score of about 92.23 \% was reached. 
For the slice-wise body part classification approach, an extensive dataset with annotated images is needed. Furthermore, ambiguities will exist for slices that lie in between two anatomic regions. This deliberation leads to the assumption that anatomy-specific classification may not be the best choice to convert human anatomy into a machine-readable form. The human body is a whole continuous object instead of multiple disjoint classes. This consideration strengthens the approach of a regression model to represent the natural order in the human body. 

There exist multiple approaches to create a body part regression model with deep learning methods \cite{yan2018unsupervised,yan2018deep,zhang2017self,criminisi2010regression}. Most of these approaches use the intrinsic order information in three-dimensional radiological volumes to train the models in a self-supervised manner.
Self-supervised methods have a crucial advantage in their independence to manually annotated data. As the acquisition of annotated data is usually time-consuming and expensive, self-supervised approaches shine when applied to large datasets for which no manual annotations are available. 
Therefore, in the following sections, the \ac{SSBR} model proposed by Yan et al. \cite{yan2018unsupervised,yan2018deep} and the \ac{BUSN} model proposed by Tang et al. \cite{tang2021body} which are the most state-of-the-art self-supervised body part regression models are explained. This thesis builds upon the \ac{SSBR} model. 

\subsection{Self-Supervised Body Part Regression}\label{sec:ssbr}
The  \ac{SSBR} model was proposed by Yan et al.\cite{yan2018unsupervised, yan2018deep}. 
The main goal of the \ac{SSBR} model is to map every axial two-dimensional slice $\mathbf{X}$ of a CT-image to a continuous score $s$
\begin{equation}\label{eq:regression-task}
   f: \mathbb{R}^{n \times n} \rightarrow \mathbb{R}, \quad \mathbf{X} \mapsto s, 
\end{equation}
where $n$ denotes the width of the axial slices. The slice score $s_{i}$ should increase monotonously with  slice index. Moreover, similar anatomical regions should map to similar scores. The slice index itself can not be used to gain body part related information, because CT images often have different scan ranges and therefore different anatomical start and endpoints or inter-slice intervals. Additionally, the size of people varies naturally. 

\textbf{Linear Relation: }By the intrinsic information of the two-dimensional slice ordering in a three-dimensional volume, the network should learn the mapping from  equation
\ref{eq:regression-task} by itself without any guidance through manual labels.
Every predicted slice score difference $\Delta s$ should be approximately proportional to the spatial distance between the slices \cite{yan2018unsupervised}. 
Formally, the desired target function corresponds to a linear mapping $g$ between slice index $j$ and \mbox{slice score $s_{ij}$}
\begin{equation*}
   g_{i}: \mathbb{R} \rightarrow \mathbb{R}, \quad
   j \mapsto s_{ij}
   \qquad
   \Rightarrow g_{i}(j) = s_{ij}, 
\end{equation*}
with: 
\begin{equation}\label{eq:linear}
    g_{i}(j) \approx m_{i}\cdot j + s_{i0},
\end{equation}
where the $s_{i0}$ indicates the start region of the CT-scan and the slope $m_i$ depends on the height of the patient and the inter-slice distance of the volume $\mathbf{X}_{i}$. 

\begin{figure}
    \centering
    \includegraphics[width=0.5\textwidth]{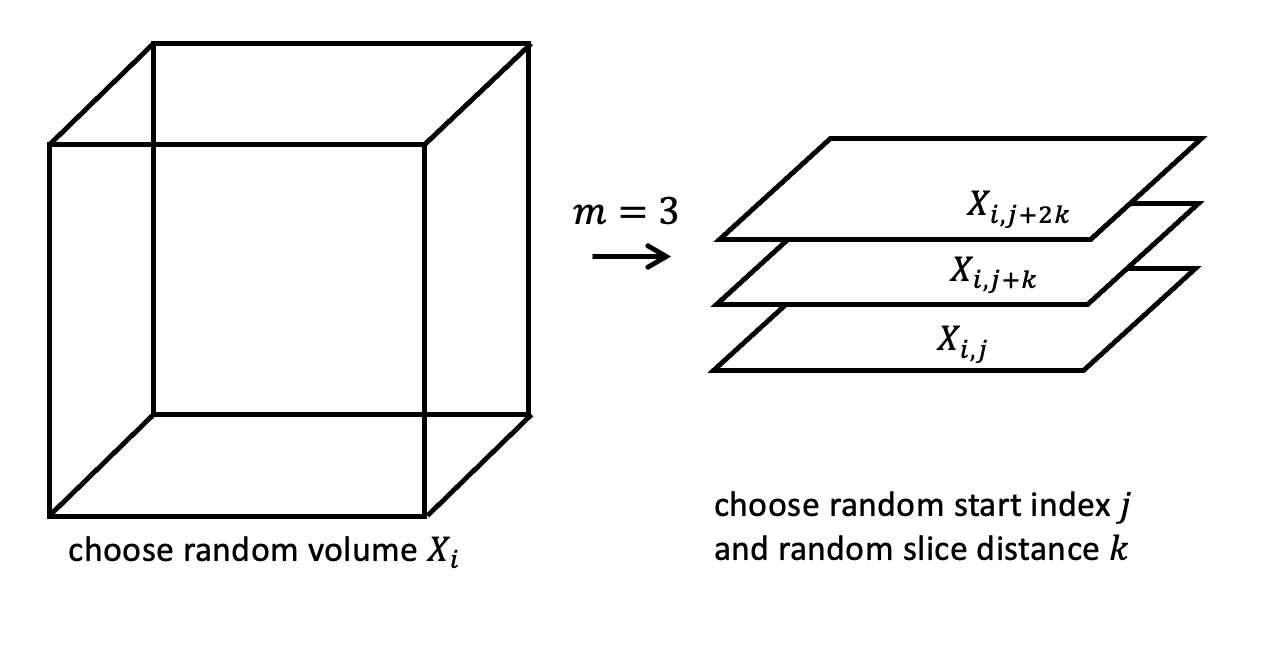}
    \caption{One data item for training the \ac{SSBR} model with the number of slices $m=3$, the start index $j$ and the slice index difference $k$. }
    \label{fig:dataitem}
\end{figure}

\textbf{Loss Functions: }One data item for training the \ac{SSBR} model consists of $m$ equidistant two-dimensional axial CT slices, as it can be seen in figure \ref{fig:dataitem}. 
Random cropping is applied to every slice. After that, the score $s$ is calculated by the network. With the help of a defined loss function, the network obtains feedback and can learn through backpropagation. The loss function consists of an order loss and a distance loss
\begin{equation*}
    L = L_{\text{dist}} + L_{\text{order}},
\end{equation*}
where the order loss $L_{\text{order}}$  is defined as 
\begin{equation}
    L_{\text{order}} = -\sum_{i=1}^{B}\sum_{j=1}^{m-1} \text{ln} \ \sigma(\Delta s_{ij}),  \quad
    \Delta s_{ij} = s_{i,j+1} - s_{ij},
\end{equation} 
where $\sigma$ is the sigmoid activation function, $B$ the batch size, $m$ the sampled slices per volume and $s_{ij}$ dedicates the predicted slice score of volume $i$ and slice index $j$ of the sampled data item. 
The sigmoid function is given by 
\begin{equation}\label{eq:sigmoid}
    \sigma(x) = \frac{1}{1+e^{-x}}. 
\end{equation}
The order loss is big if the slice scores do not monotonously increase with slice index. The distance loss is defined as 
\begin{equation} \label{eq:distance-loss}
    L_{\text{dist}} =\sum_{i=1}^{B}\sum_{j=1}^{m-2}
    l_{1}(\Delta s_{i, j+1} - \Delta s_{ij}),
\end{equation}
where $l_{1}$ is the smooth L1-loss \cite{ren2016faster}. 
The smooth L1-loss is given by 
\begin{equation}\label{eq:smooth-l1-loss}
    l_{1}(\Delta s_{i, j+1} - \Delta s_{ij}) = 
    \begin{cases}
    0.5(\Delta s_{i, j+1} - \Delta s_{ij})^{2} & 
    |\Delta s_{i, j+1} - \Delta s_{ij}|  < 1\\
    |\Delta s_{i, j+1} - \Delta s_{ij}| - 0.5 & 
    \text{otherwise}.
    \end{cases}
\end{equation}
The distance loss is big if consecutive slice differences $\Delta s_{ij}$ are particular different. The distance loss enforces the network to maintain a linear relation between slice index and slice score, as seen in equation \ref{eq:linear}. 
\begin{figure}
    \centering
    \includegraphics[width=0.8\textwidth]{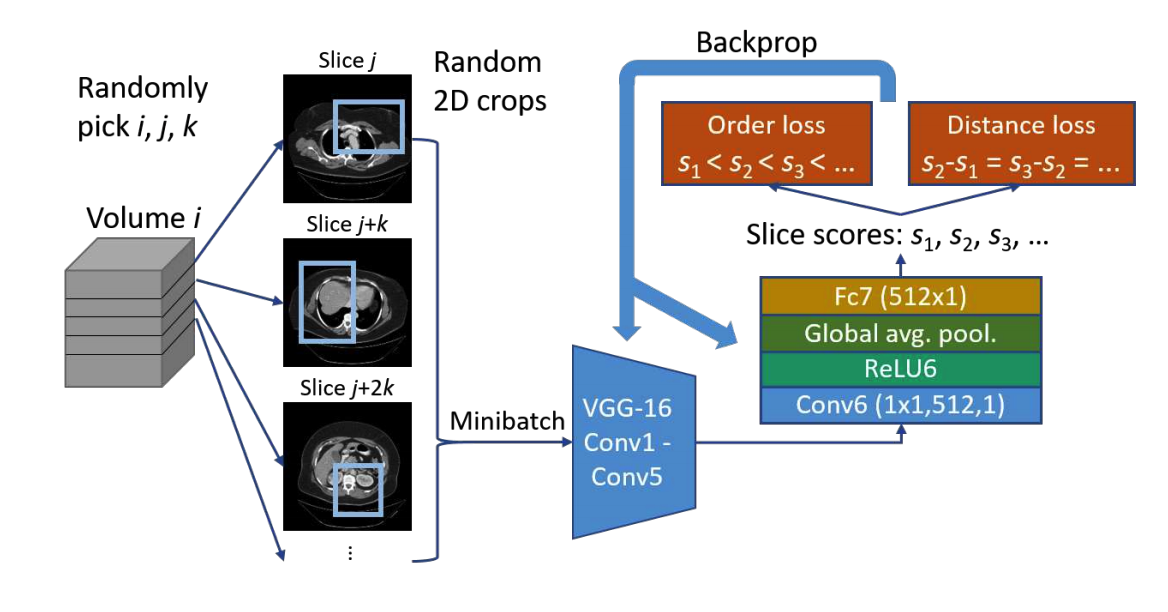}
    \caption{Proposed framework of the \ac{SSBR} model with random two-dimensional cropping as data augmentation step. The image was taken from \cite{yan2018deep}.}
    \label{fig:ssbr}
\end{figure} 

\textbf{Model Architecture: } Figure \ref{fig:ssbr} shows the structure of the \ac{SSBR} model. 
For the first convolutional layers, the Imagenet pre-trained VGG16 \cite{simonyan2014very} model was adopted. Then, an additional convolutional layer with 512 x 1 x 1 filters and a stride of one was added. After that, a ReLU layer, a global average pooling layer, and a fully connected layer follow to predict the slice score. 
The \ac{SSBR} model was trained on 800 CT volumes for body ranges between pelvis to chest \cite{yan2018deep}. The model was evaluated on a 3-class classification problem with the classes: pelvis, abdomen, and chest. 
The \ac{SSBR} model was able to achieve an accuracy of $95.99 \% $ on the test set. 

\textbf{Research Gap: } The theory behind the loss functions of the \ac{SSBR} model is not fully understood until now. Moreover, finding good hyperparameters and comparing \ac{SSBR} models with different hyperparameter settings is challenging. The accuracy metric is not  sensitive to minor deviations in prediction and makes it difficult to decide which model performs best. The scope of the \ac{SSBR} model is the pelvis to the chest body region. One of the remaining tasks is to expand the algorithm's scope to further body regions and to use more CT studies for training to improve generalization towards different institutions and scanner protocols. 

\subsection{Blind Unsupervised-Supervision Network}\label{sec:busn}
\begin{figure}
    \centering
    \includegraphics[width=\textwidth]{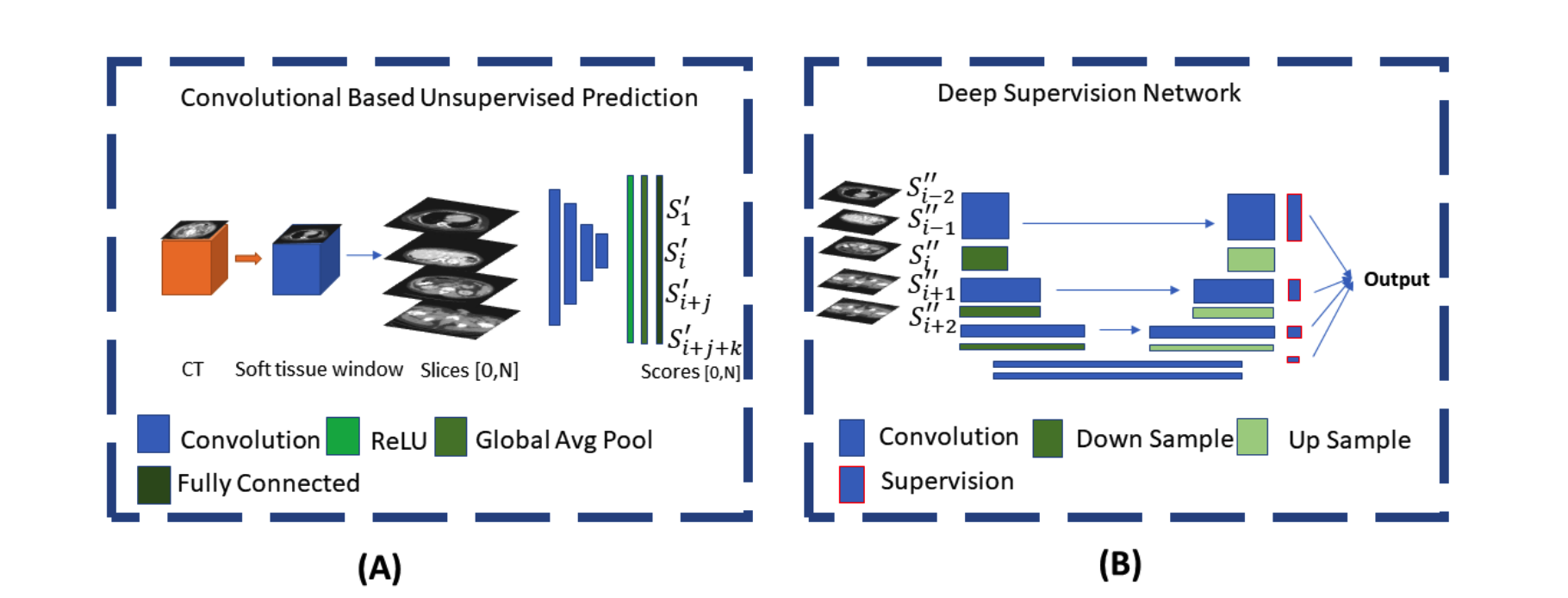}
    \caption{(A) The \ac{SSBR} model proposed by Yan et al. \cite{yan2018deep}. (B) Deep supervised network using the refined prediction scores from the robust regression. During test-time, only the second model part is required to perform body part regression. The image was taken from \cite{tang2021body}.}
    \label{fig:busn}
\end{figure}

Recently, Tang et al. \cite{tang2021body} proposed an additional modification to the \ac{SSBR} model. It is based on a two-stage approach, where in the first stage, the \ac{SSBR} model is trained. The predicted slice scores are corrected by a robust linear regression algorithm (RANSAC). In the second stage, a 2D U-Net model uses these corrected scores to train in a supervised fashion. An additional Neighbor Message Passing (NMP) method was introduced to improve the scores of a single slice by taking the information of the scores from the neighbor slices into account \cite{tang2021body}. The proposed modified model is referred to as \ac{BUSN} model. In figure \ref{fig:busn} an overview of the BUSN model can be found. The BUSN model was trained with 1030 CT volumes and was evaluated on about 100 CT scans of the BTCV dataset \cite{dataSynapse}. As evaluation metric, the $R^{2}$-score was used (see sec. \ref{sec:related-work-evaluation-methods}). 

\textbf{Research Gap: }Due to the novelty of the BUSN model, it was not yet been used and compared in other papers. 
The authors compared the proposed BUSN model with the original \ac{SSBR} model and reported relatively poor performance of the baseline approach. From the paper \cite{tang2021body}, it is not clear where the deviation in performance compared to the original SSBR paper \cite{yan2018unsupervised} comes from. 

\section{Evaluation Metrics}
\label{sec:related-work-evaluation-methods}
Evaluating a body part regression model is non-trivial because no ground truth labels are available. In the literature, different approaches for evaluating a body part regression model exist. 

\textbf{Accuracy: }
In the paper from Yan et al. \cite{yan2018unsupervised} the primary evaluation measure was the accuracy. Three classes were defined: pelvis, abdomen, and chest. For testing and validation, each slice was labeled manually. The abdomen class starts from the upper border of the ilium bone and ends at the upper border of the liver. With the help of two extra validation CT volumes, two thresholds to separate the three-body zones were identified. 
Based on the predicted slice score, each slice is assigned to a class $\hat{c}$. In the end, the accuracy $\psi$ compares the predicted classes $\hat{c}$ with the annotated ground truth classes $c$ by calculating
\begin{equation}\label{eq:accuracy}
    \begin{split}
    \psi = \frac{1}{N}\sum_{i=1}^{N}\frac{1}{M_{i}}\sum_{j=1}^{M_{i}}
    \chi(\hat{c}_{ij}, c_{ij}), \\
    \end{split}
\end{equation}
with
\begin{equation*}
        \chi(\hat{c}_{ij}, c_{ij}) = 
    \begin{cases} 1& \text{if } \hat{c}_{ij} = c_{ij}, \\
    0& \text{otherwise}, 
    \end{cases}
\end{equation*}
where $N$ is the number of CT volumes in the validation set and $M_{i}$ the number of slices in volume $i$.
The continuous output slice score is converted to discrete classes, implying that deviations of the scores within one class are neglected. 
Thus, one limitation of the accuracy metric is that it is not particularly sensitive. 

\textbf{$R^{2}$-Metric: }
The $R^{2}$-metric is used by Tang et al. \cite{tang2021body} to validate their two-stage body part regression approach. 
For each slice score curve, a linear curve $\hat{s}$ was fitted to the slice scores $s$. For a volume $X$ with $M$ slices, the $R^{2}$-metric is given by 
\begin{equation}
    R^{2}_{X} = \frac{\sum_{i=1}^{M} (s_{i} - \bar{s})^{2}}{\sum_{i=1}^{M}
    (\hat{s}_{i} - \bar{s})^{2}}
    = 1 -  \frac{\sum_{i=1}^{M} (\hat{s}_{i} - s_{i})^{2}}{\sum_{i=1}^{M} (\hat{s}_{i} - \bar{s})^{2}},
\end{equation}
where $s_{i}$ is the predicted slice score on index $i$, $\hat{s}_{i}$ is the linear fit at index $i$ and $\bar{s}$ the mean  slice score from the linear fit. On the one hand, an advantage of the $R^{2}$-metric is that no manually annotated data is needed. 
On the other hand, the assumption of linear correlation between body height and slice score does not consider inter-patient anatomical variation. Therefore, one limitation of the $R^{2}$-metric is that it does not measure if the same anatomical regions are mapped to the identical scores across patients. 

\textbf{Research Gap: } Up to now, no fine-grained metric was proposed for regression models that precisely measure if the same anatomical regions are predicted to similar scores. Moreover, there exists no evaluation metric for body part regression models on which the research community agreed on. Instead, each research group uses a different evaluation metric, and no comparison of the proposed evaluation metrics exists to agree on a universal metric for cross-model comparison.

\section{Application and Deployment} \label{sec:related-work-deployment}
Due to the fine-grained recognition of body regions, body part regression models have a broad range of possible use cases for medical image analysis tasks. Some of them were already investigated in previous literature and will be introduced in this section.  

\textbf{Annoation: } 
One evident use case is to obtain body part region annotations through a body part regression model. For example, Yan et al. annotated the position of lesions in the human body with a body part regression model to structure a large lesion dataset \cite{yan2018deep}. Further, diverse papers showed that a body part regression model can predict rough estimates of organ positions \cite{tang2021body, proskurov2021fast}, and that it is capable of cropping certain body regions as the abdomen \cite{tang2021high, tang2020learning, xu2020outlier} or the pelvis \cite{ivanovska2021efficient} from a CT scan. 

\textbf{Anomaly Detection: } 
A further use case is anomaly detection in radiologic volumes by visualizing the predicted slice scores against the slice index \cite{yan2018unsupervised}. Yan et al. showed that for some volumes, the slice score curves exhibit discontinuities where the appearance of the CT slices is abnormal due to scanning artifacts or large lesions \cite{yan2018unsupervised}. 

\textbf{Preprocessing for Registration: } 
The body part regression model could also be used as pre-alignment in z-direction for registration purposes. Heine and Hering pointed out that preliminary results confirm that a fast and robust pre-alignment of CT volumes through body part regression is possible \cite{meine2019efficient}. 

\textbf{Preprocessing for Segmentation: } 
Tang et al. pointed out that it is possible to enhance the quality of nnU-Net organ segmentation models in terms of dice score by using body part regression  \cite{tang2021body}. 
Through introducing an additional preprocessing step, the volumes from the training, validation, and test set were spatially normalized with the body part regression model by automatically cropping the abdomen region \cite{tang2021body}. This preprocessing step led to a significant increase in segmentation performances \cite{tang2021body}.

\textbf{Research Gap: } 
Although several use cases were already proposed, a broad field of use cases remains not yet analyzed. For example, up to now, no one has analyzed how the body part regression model can be used for false-positive detection in medical algorithms. Medical algorithms are typically trained on images from a specific body part and lead to false-positive predictions in unknown regions. The body part regression model opens up new opportunities to detect false-positive predictions in unknown regions, which can lead to a significant increase in model performance of medical algorithms, such as, e.g., tumor segmentation. 

Yan et al. pointed out that finding outliers in CT volumes with the body part regression model through visual inspection is possible \cite{yan2018unsupervised}. Nevertheless, up to our knowledge, no automatic approach of finding outliers within CT volumes or CT datasets due to, e.g., improper data loading or incorrectly assigned modality exists. 

Moreover, up to now, the possibility of gaining a better estimate for the imaged body region than the \textit{BodyPartExamined} tag in the DICOM meta-data through the body part regression model was not analyzed in the previous literature. Gueld et al. emphasized that the DICOM body part meta-data usually is not sufficient to describe the examined body part correctly \cite{gueld2002quality}. Therefore, the body part regression model can bring significant benefits in predicting robust and precise body part examined tags for radiological images.

Regarding the deployment of a body part regression model, no user-friendly model exists with which the body part regression model can be easily used in the research community or the clinical environment.

\chapter{Materials and Methods}
In this chapter, the theory behind the trained body part regression models is explained (sec. \ref{sec:bpr-methods}) and the evaluation methods for cross-model comparison are introduced (sec. \ref{sec:methods-evaluation-methods}). 
Building on this, in section \ref{sec:methods-application-deployment} several use cases for the model, are presented. Further, the model deployment is explained to make the model publicly available for the clinical and research communities. 

\section{Body Part Regression}\label{sec:bpr-methods}
A self-supervised body part regression model is presented that automatically maps similar anatomical regions to the same score across different patients. For training a model with good generalization capabilities, a diverse dataset was selected (sec. \ref{sec:dataset}). For evaluating the model performance, robust anatomical landmarks were defined and annotated in selected CT volumes (sec. \ref{sec:annotation}). Additionally, the network architecture of the body part regression model is described (sec. \ref{sec:architecture}), different order loss terms for the model are introduced (sec. \ref{sec:order-loss}) and the used data augmentation techniques are explained (sec. \ref{sec:data-augmentation}).

\subsection{Dataset Selection}\label{sec:dataset}
In order to train a body part regression model with good generalization properties, it is essential to use a heterogeneous dataset with CT volumes from different hospitals, CT devices, and various diseases. 
The dataset used for this thesis was compiled from studies of \ac{TCIA} \cite{clark2013cancer} and different computer vision challenges (see tab. \ref{tab:train_data} and tab. \ref{tab:test_data}). Table \ref{tab:train_data} summarizes the 12 studies used for training and validation, resulting in a dataset size of 2192 CT scans for training and 100 CT scans for validation. In table \ref{tab:test_data} an overview of the compiled test dataset is shown, which consists of 10 different studies and in total 100 CT scans. The pixel spacing in the z-direction varies between 0.45 mm and 6 mm. Only images with a $z$-range greater than 12 cm and 42 or more slices were used for training. The original pixel spacing in the axial plane in x- and y-direction vary between 0.3 mm and 7.5 mm.   

To avoid overfitting on studies and certain body regions, CT volumes were limited to 150 CT images for the studies Task007 and LIDC-IDRI. For the datasets, HNSCC, QIN-HEADNECK, ACRIN-NSCLC-FDG-PET, the limit was increased to 400 CT scans since those studies cover almost the whole body. By selecting the 400 volumes from the greater population, care was taken to favor volumes showing rare body regions like the head or the thigh. Nearly all studies are publicly available. Only the "whole body CT" study is private. This study includes the knees and the feet as well. In nearly no other study, these body areas are visible.  
Adding the "whole body CT" study to the dataset has two advantages: on the one hand, the model sees more volumes in rare regions, and on the other hand, it could be interesting to investigate the model behavior on whole body CT volumes during validation. Nevertheless, the valid scope of the trained body part regression model will be the lower pelvis until the end of the head because there are too few volumes from the legs available to extend the valid scope of the model. Moreover, due to a lack of data, body regions of different modalities, children, or pregnant women are not in the scope of the trained model. 

\begin{table}
    \centering
    \caption{Overview of studies used for training and validation.
    For a subset of the training data and the whole validation set landmarks were annotated (see sec. \ref{sec:annotation}).  
    Some studies are quite heterogeneous in terms of the scanned body areas. For these datasets, the minimum and maximum body part of the scanned volumes is documented.}
    \begin{tabular}{llllll}
        \hline
         Study name & Body target & \makecell[l]{\# Train. \\data} & \makecell[l]{\# Annotated\\ train. data}&  \makecell[l]{\# Val. \\ data} & Source \\
         \hline
         Task003 & liver &177 & 4 & 9  & \cite{simpson2019large,bilic2019liver}\\
         Task007 & pancreas &150 & 4 &  9 & \cite{simpson2019large,attiyeh2018survival,attiyeh2019preoperative,chakraborty2018ct}\\
         Task051 & upper body & 42 & 4 & 8 & \cite{dataGrandChallange}\\
         Task062 & pancreas &74 & 4 &  8 & \cite{clark2013cancer,dataPancreas,roth2015deeporgan}\\
         whole body CT & foot - head & 45 & 4 & 6 & private \\
         \makecell[l]{CT\\COLONOGRAPHY} & colon & 74 & 4 & 8 & \cite{clark2013cancer,johnson2008accuracy, dataCOLONOGRAPHY}\\
         COVID-19-AR & lung & 113 & 5 & 9 & \cite{clark2013cancer,jenjaroenpun2021two,desai2020chest,dataCOVID}\\
         \makecell[l]{CT LYMPH\\ NODES} & pelvis - neck&167& 4 & 9 & \makecell[l]{\cite{clark2013cancer,dataLYMPH,roth2014new}\\\cite{seff20142d,seff2015leveraging}}\\
         LIDC-IDRI & lung & 150& 4 & 9 & \cite{clark2013cancer,dataLIDC,armato2011lung}\\
         HNSCC & pelvis - head & 400& 5 & 9 & \cite{clark2013cancer,dataHNSCC,grossberg2018imaging,elhalawani2017matched}\\
         QIN-HEADNECK & pelvis - head & 400 & 4 & 8 & \cite{clark2013cancer,dataQIN,fedorov2016dicom}\\
         \makecell[l]{ACRIN-NSCLC-\\FDG-PET}& pelvis - head & 400 & 4 &  8 & \cite{clark2013cancer,dataACRIN,machtay2013prediction}\\
         \hline
         \hline
         Sum & & 2192 & 50& 100& \\
         \hline
    \end{tabular} 
    \label{tab:train_data}
\end{table} 

\begin{table}
    \centering
    \caption{Overview of test dataset. For each study, 10 CT volumes were randomly selected and the positions of the defined landmarks  were annotated (see sec. \ref{sec:annotation}).}
    \begin{tabular}{llll}
        \hline
         Study name & Body target &\# Annotated data & Source \\
         \hline
         Task006 & lung & 10& \cite{simpson2019large,bakr2018radiogenomic,napel2014nsclc,gevaert2012non}\\
         Task008 & abdomen & 10& \cite{simpson2019large,pak2018quantitative,simpson2015texture,simpson2017computed,zheng2017preoperative}\\
         Task009 & spleen& 10& \cite{simpson2019large,simpson2015chemotherapy}\\
         Task010 & colon& 10& \cite{simpson2019large}\\
         Task017 & abdomen & 10&\cite{dataSynapse}\\
         Task018 & pelvis & 10 &\cite{dataSynapse}\\
         Task046 & abdomen & 10& \cite{gibson2018multi}\\
         Task049 & head-neck & 10 & \cite{dataGrandChallange}\\
         Task055 & thorax & 10& \cite{lambert2020segthor}\\
         Task064 & kidneys & 10& \cite{heller2019kits19}\\
         \hline 
         \hline 
         Sum & & 100 & \\
         \hline
    \end{tabular}
    \label{tab:test_data}
\end{table}
\textbf{Data Preprocessing: } The intensity values in Hounsfield unit were clipped between -1000 HU and 1500 HU and rescaled to -1 to 1. Axial slices were resampled to a pixel spacing of 3.5 mm. All images were resized to \mbox{128 px $\times$ 128 px} by zero-padding or center cropping. The pixel spacing was chosen to assure that the resulting field-of-view is  44.8 cm $\times$ 44.8 cm and covers the entire cross-section of an average person. 

Before down-sampling, a Gaussian filter was used to avoid aliasing artifacts \cite{rao2018digital,engelberg2008digital}. For a downsampling factor of $a=0.25$, the smoothing standard deviation $\sigma=0.8$ for the Gaussian kernel was experimentally determined by visual inspection of resized images. This Gaussian filter configuration reduces aliasing effects with small edge information loss. Based on the original pixel spacing of the axial slices, we gain different downsampling factors $a^{\prime}$. 
The most suitable smoothing standard deviation $\sigma^{\prime}$ for the Gaussian kernel is dependent on the downsampling factor. For bigger downsampling factors, we would assume stronger smoothing and vise versa. Therefore we used the following relation to gain the custom smoothing standard deviation $\sigma^{\prime}$ for a downsampling factor of $a^{\prime}$:
\begin{equation*}
    \sigma^{\prime} = \frac{a}{a^{\prime}}\sigma.
\end{equation*}

\subsection{Annotations}\label{sec:annotation}
For evaluating the model performance, anatomical landmarks were defined. Table \ref{tab:landmark-definitions} summarizes all defined 35 landmarks and also contains a short description of each landmark. All landmarks except the vertebrae landmarks were annotated manually. 
For annotating the centroid of the vertebrae, a nnU-Net \cite{isensee2021nnu} vertebrae segmentation algorithm was used, which was kindly provided by Fabian Isensee.
The segmentation algorithm was trained on the data from the Verse2020 segmentation challenge data \cite{sekuboyina2020verse,loffler2020vertebral}. With this segmentation algorithm, all vertebrae of the Cervical spine, Thoracic spine, and Lumbar spine are segmented. In addition, automatic vertebra landmark annotations were derived by computing the centroid of the predicted vertebra masks. Segmenting the vertebrae of one volume took about 20 minutes on an Nvidia V100 SXM2 GPU with 32 GB GPU memory. 

Figure \ref{fig:example-slices} shows an example slice for each landmark. 
The positions of the landmarks for 100 volumes of the validation set, 50 volumes of the training set, and 100 volumes of the test set were annotated.
It was taken care to annotate the volumes uniformly across different studies. Therefore, in the test dataset for every study 10 volumes were annotated, in the validation set 6 to 9 volumes and in the training set 4 to 5 volumes per study (see tab. \ref{tab:train_data} and tab. \ref{tab:test_data}). For validation, only six volumes were annotated from the "whole body CT" study to use more whole body CT volumes for training. 
A few misplaced landmarks were identified in a second pass through the annotated slices. The corresponding annotations were removed from the database. A summary of the number of annotated landmarks in the training, validation, and test dataset can be found in table \ref{tab:count-annotated-landmarks}.

\begin{figure}
    \centering
     \begin{subfigure}[b]{0.45\textwidth}
         \centering
         \includegraphics[width=\textwidth]{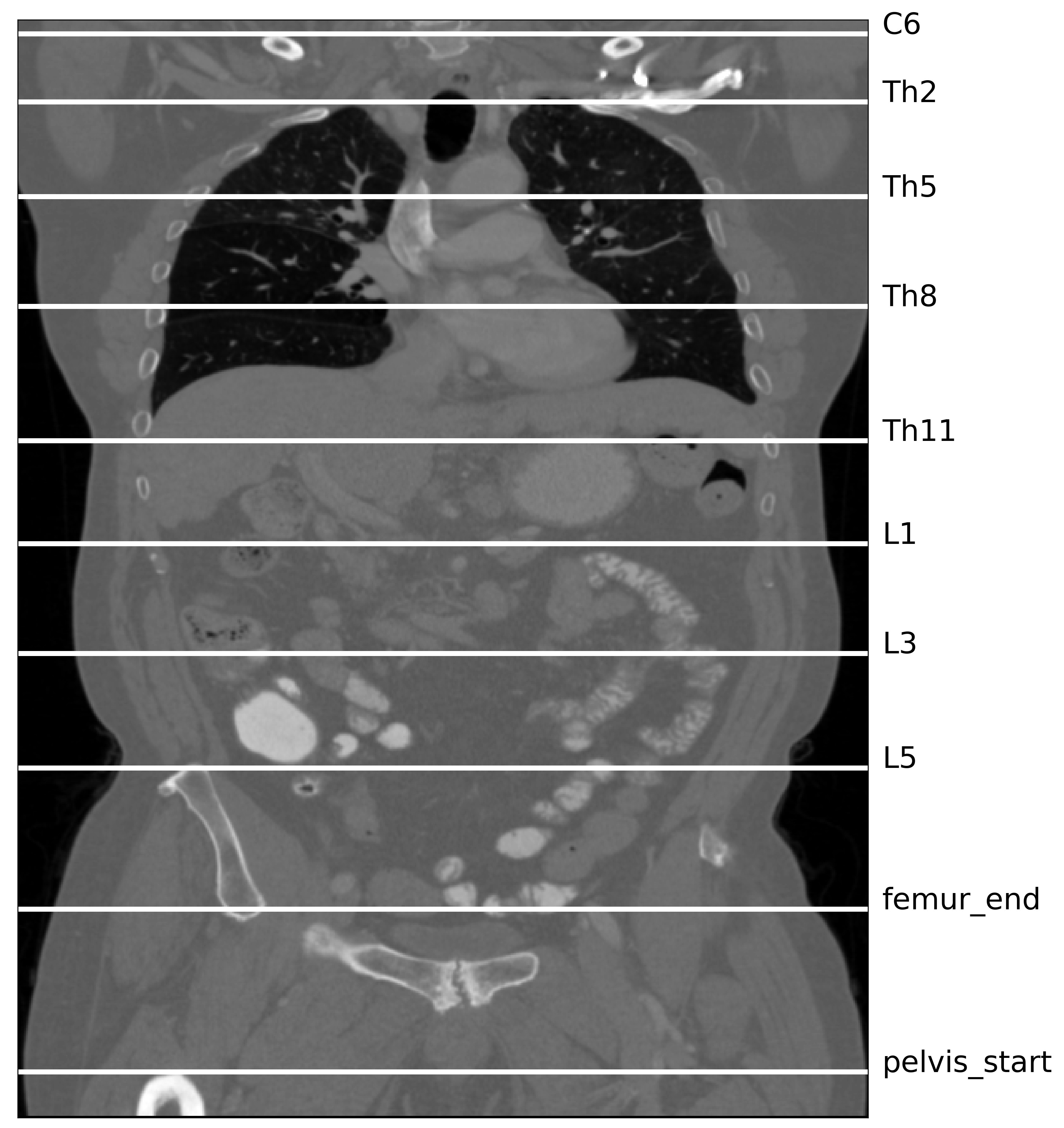}
     \end{subfigure}
     \hfill
     \begin{subfigure}[b]{0.45\textwidth}
         \centering
         \includegraphics[width=\textwidth]{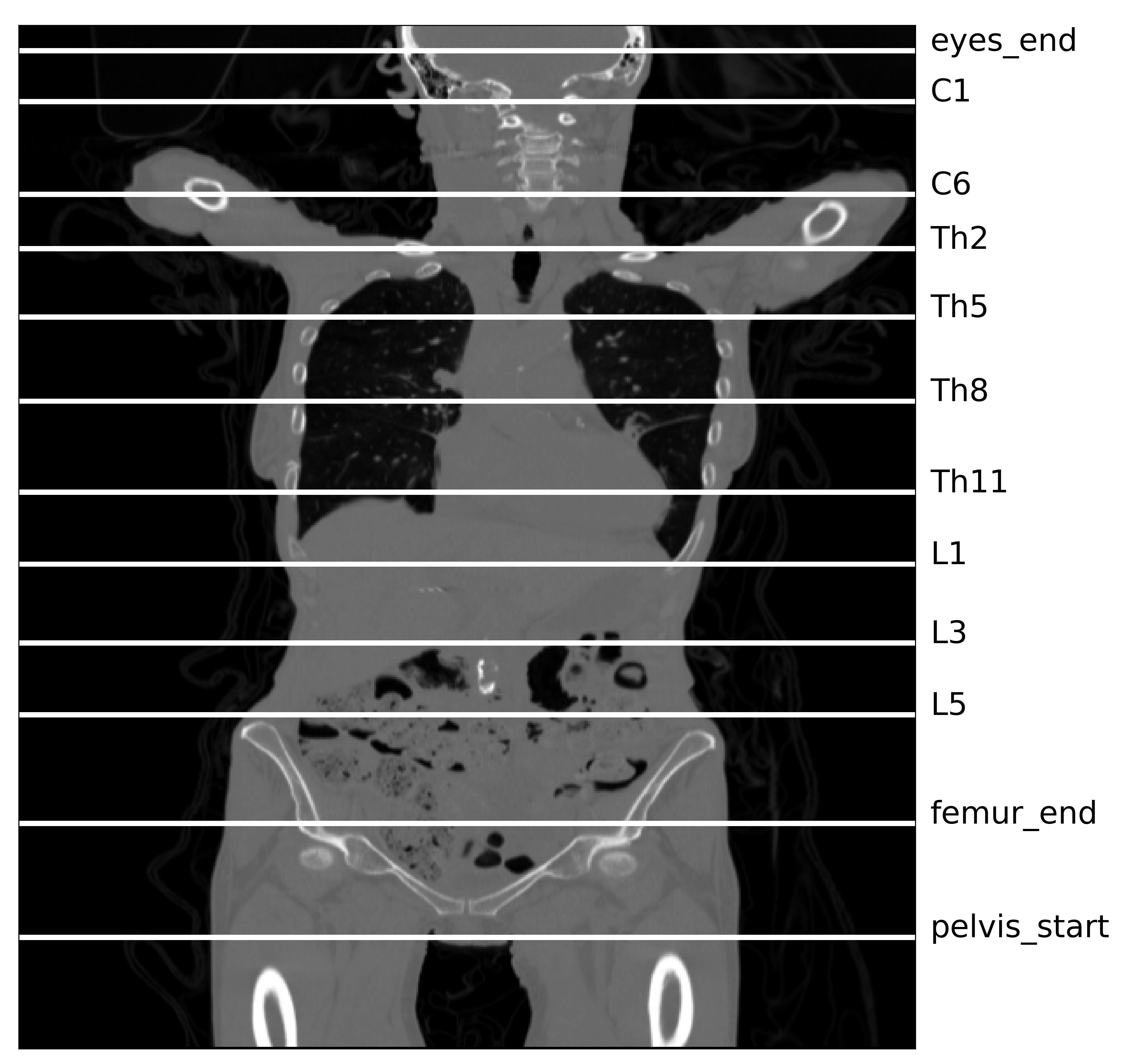}
     \end{subfigure}
     
    \caption{Two example coronal CT slices from the validation set with highlighted evaluation landmark positions. }
    \label{fig:evaluation-landmarks-examples-2}
\end{figure}

The anatomical variation of the different landmarks is highly diverse. In general, bone landmarks are anatomically more robust than organ landmarks. For the later evaluation of the models, only a subset of the defined bone landmarks was used. The evaluation landmarks are pelvis-start, femur-end, L5, L3, L1, Th11, Th8, Th5, Th2, C6, C1, and eyes-end. The evaluation landmarks will be referred to as $l_{k}$ with $k \in \{1, ..., 12\}$. 
It was taken care of the distance between evaluation landmarks to ensure that they have approximately the same distance to each other. A detailed definition for each landmark can be found in table \ref{tab:landmark-definitions}. Figure \ref{fig:evaluation-landmarks-examples-2} visualizes in two example images where the evaluation landmarks are located in the human body. 

\begin{table}
    \centering
    \caption{Used anatomical landmarks and defined  bone landmarks for evaluation.}
    \begin{tabular}{llcl}
        \hline
         \makecell[l]{Landmark \\ name} & Description & \makecell[l]{Eval. \\ landmark} & Notation \\
        \hline
        pelvis-start & Slice before ischium bone starts & \checkmark & $l_{1}$ \\
        femur-end & Slice after the femur bone ends & \checkmark & $l_{2}$ \\
        pelvis-end & Slice after the end of the illium bone &  & \\
        kidneys & First slice where both kidneys can be seen well& & \\
        lung-start &Slice before lung starts& & \\
        liver-end &Slice after end of liver& & \\
        lung-end&Slice after end of lung & & \\
        teeth &Middle slice of lower jaw& & \\ 
        nose & Slice where tip of nose can be seen the first time & & \\
        eyes-end &Slice after the end of the eye socket& \checkmark & $l_{3}$ \\
        head-end &Last slice where bone and brain can be seen & & \\
        L5 & Centroid of 5. Lumbar vertebrae &\checkmark & $l_{4}$ \\
        L4 & Centroid of 4. Lumbar vertebrae & & \\
        L3 & Centroid of 3. Lumbar vertebrae & \checkmark & $l_{5}$ \\
        L2 & Centroid of 2. Lumbar vertebrae & & \\
        L1 & Centroid of 1. Lumbar vertebrae & \checkmark & $l_{6}$ \\
        Th12 & Centroid of 12. Thoracic vertebrae & & \\
        Th11 & Centroid of 11. Thoracic vertebrae & \checkmark & $l_{7}$ \\
        Th10 & Centroid of 10. Thoracic vertebrae & & \\
        Th9 & Centroid of 9. Thoracic vertebrae &  & \\
        Th8 & Centroid of 8. Thoracic vertebrae & \checkmark & $l_{8}$ \\
        Th7 & Centroid of 7. Thoracic vertebrae &  & \\
        Th6 & Centroid of 6. Thoracic vertebrae &  & \\
        Th5 & Centroid of 5. Thoracic vertebrae & \checkmark & $l_{9}$ \\
        Th4 & Centroid of 4. Thoracic vertebrae &  & \\
        Th3 & Centroid of 3. Thoracic vertebrae &  & \\
        Th2 & Centroid of 2. Thoracic vertebrae & \checkmark & $l_{10}$ \\
        Th1 & Centroid of 1. Thoracic vertebrae &  & \\
        C7 & Centroid of 7. Cervical vertebrae &  & \\
        C6 & Centroid of 6. Cervical vertebrae & \checkmark & $l_{11}$ \\
        C5 & Centroid of 5. Cervical vertebrae &  & \\
        C4 & Centroid of 4. Cervical vertebrae &  & \\
        C3 & Centroid of 3. Cervical vertebrae &  & \\
        C2 & Centroid of Axis vertebrae &  & \\
        C1 & Centroid of Atlas vertebrae & \checkmark & $l_{12}$ \\
        \hline
    \end{tabular}
    \label{tab:landmark-definitions}
\end{table}
\subsection{Learning Procedure}\label{sec:architecture}
In this thesis, a body part regression model is trained. The learning procedure is based on the \ac{SSBR} model proposed by Yan et al. \cite{yan2018unsupervised, yan2018deep}. At the end of this section, a detailed overview of the differences to the original method will be given.

The final body part regression model $f_{\boldsymbol{\theta}}$ with parameters $\boldsymbol{\theta}$ should map CT slices $\mathbf{X}_{ij}$ of volume $i$ and slice index $j$ to anatomical slice scores $s_{ij}$: 
\begin{equation*}
    f_{\boldsymbol{\theta}}: \mathbb{R}^{n \times n} \rightarrow  \mathbb{R}: \quad f_{\boldsymbol{\theta}}(\mathbf{X}_{ij}) = s_{ij}.
\end{equation*}
The same anatomical region should map to the same slice score, independently of patient and study. This means that our function $f_{\boldsymbol{\theta}}$ should be invariant under certain transformations, which do not change the visible anatomy but may change study- or patient-related features. 
Moreover, each body part should be assigned to a unique slice score. Mathematically speaking, we want to learn an injective function which maps anatomical regions to slice scores. Therefore, we need a function that produces slice scores that increase strictly monotonic with slice index. 
For example, the patient's pelvis should map to a lower slice score than the head of a patient. Finally, we have two main conditions for our mapping function $f_{\boldsymbol{\theta}}$: 
\begin{enumerate}
    \item[] \textbf{1. Independence condition: }The same anatomical region should map to the same score, independently of patient and study.
    \item[] \textbf{2. Monotony condition: }The anatomical slice scores should increase monotonically with height.
\end{enumerate}
With these two conditions, we can now construct an objective function that should be minimized during training.
For training a body part regression model, a self-supervised approach is used. Therefore, we do not need ground truth labels, and we can use a large and diverse dataset. In the self-supervised approach, we gain a learning incentive from the intrinsic order of consecutive axial slices in a three-dimensional CT volume. Through the relative relation to other slices in the human body, the model can learn the relation between anatomy and slice score on its own without labeled data. 

In the following, we will try to find a general approach for a possible objective function, which considers the monotony condition and the independence condition. 
As a first approach for the objective function, the following equation can be stated
\begin{equation*}
    \begin{split}
    & \Phi = \sum_{i=1}^{N}\sum_{j=1}^{M_{i}-1}L_{\text{order}}(\Delta s_{ij}), \quad 
    \Delta s_{ij} = s_{ij+1} - s_{ij} 
    \end{split},
\end{equation*}
where $N$ is the number of available CT-volumes and $M_{i}$ the number of slices of volume $i$. 
The order loss $L_{\text{order}}$ should be designed such that it is high if $\Delta s_{ij} < 0 $ and low if $\Delta s_{ij} > 0$. It makes sure that the slice scores increase monotonically with slice height. In section \ref{sec:order-loss} two possible order loss functions will be derived. 
This objective function $\Phi$ can be seen as a potential, which we want to minimize to obtain good parameters $\boldsymbol{\theta}$ for our model $f_{\boldsymbol{\theta}}$, which will be a neural network in this work. Stochastic gradient descent with a mini-batch of size $B$ can be used to minimize the objective function. Assuming we are choosing $m$ ($m \ge 2$) equidistant random slices  per volume and $B$ random CT-volumes per update step, we need to calculate the following loss function, which is an expectation of our objective function, to calculate the gradients: 
\begin{equation*}
     L = \frac{1}{B} \sum_{i=1}^{B} \frac{1}{m-1}
    \sum_{j=1}^{m-1}L_{\text{order}}(\Delta s_{ij}).
\end{equation*}
The $L_{\text{order}}$ objective function is motivated by the monotony condition.
If the chosen neural network model has enough parameters $\boldsymbol{\theta}$, an easy way for the neural network to reduce the loss is to simply learn the slice score for each volume separately. Then the independence condition would be violated and anatomical  similar  regions from different patients would  not necessarily be mapped  to  the  same  slice  score. We need additional constraints to incorporate the second condition into the model as well. Loosely speaking, we want that for any transformation $g$ that does not change the assigned anatomical region of an image, the slice score mapping function should produce the same output. Let $G$ denote the family of transformations $g$ for which this assumption is true. We want that our slice to slice score mapping function $f$ is invariant under transformations $g \in G$: 
\begin{equation*}
   f_{\theta}(g(\mathbf{X}_{ij})) =
   f_{\theta}(\mathbf{X}_{ij})  = 
   s_{ij}.
\end{equation*}
In the machine learning setup, we can try to approximately implement this constraint by using transformations $g \in G$ as data augmentation. These data augmentations have to be applied independently to each individual sampled slice $\mathbf{X}_{ij}$. With slice-wise data augmentation, we try to prevent the network from overfitting. 

\begin{figure}
    \centering
    \includegraphics[width=\textwidth]{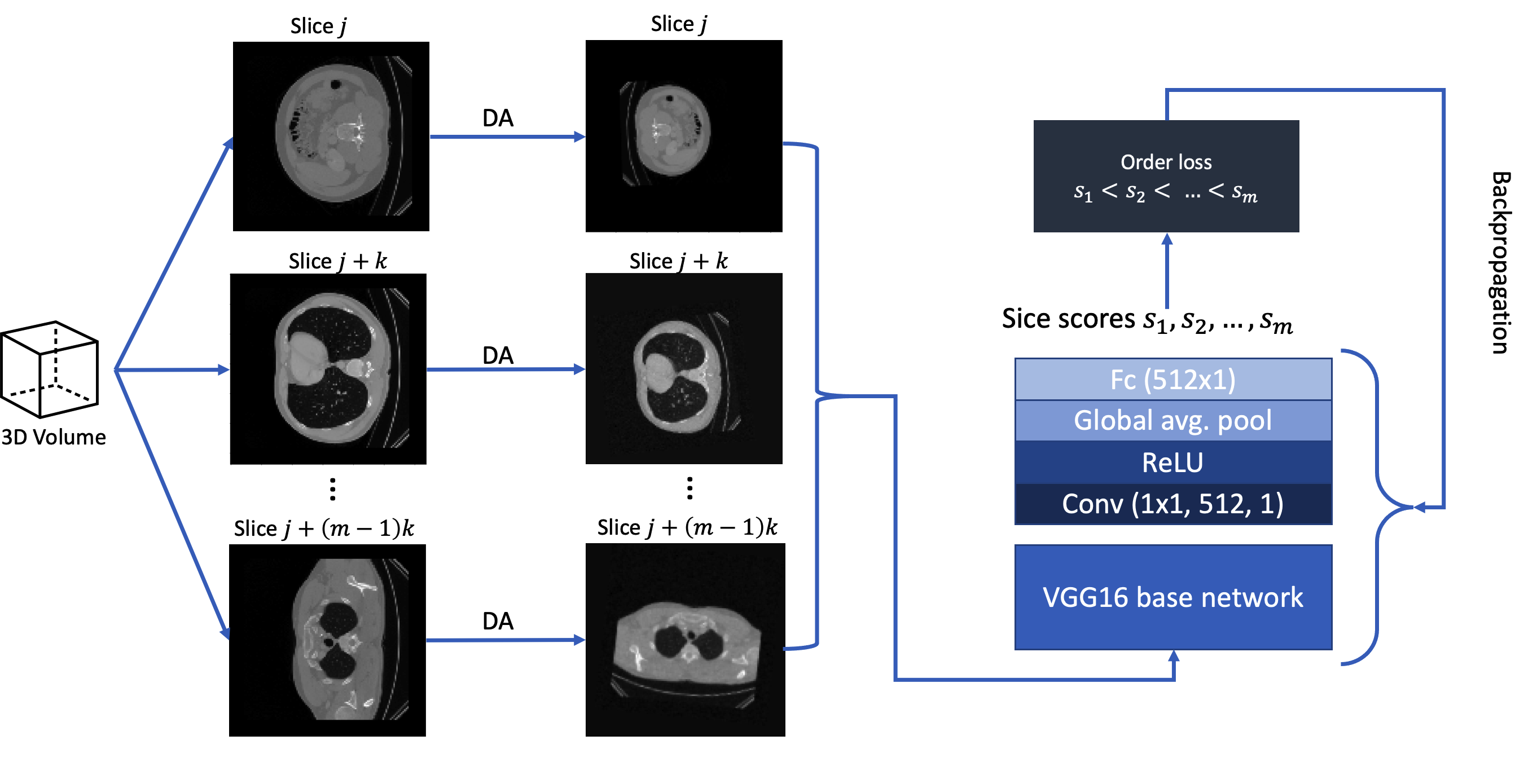}
    \caption{Overview of the learning framework for the self-supervised body part regression model. In this simplified illustration, the mini-batch size is equal to one. The abbreviation DA stands for diverse data augmentation techniques. The variable $k$ is the randomly sampled slice distance between the slices. The network architecture was adapted from the \ac{SSBR} model \cite{yan2018deep, yan2018unsupervised}. } 
    \label{fig:UAR}
\end{figure}

An example of the used network structure can be seen in figure \ref{fig:UAR}. The training procedure is as follows: first, $m$ equidistant slices are sampled from a random three-dimensional volume. The distance between the slices $\Delta h$ is sampled randomly from a $z$-range, e.g., from 5 mm to 100 mm. After sampling  $\Delta h$, the distance in mm is converted to a difference in slice indices $k$ with the help of the z-spacing $z$  by the equation
\begin{equation}
    k = \text{round}\left( \frac{\Delta h}{z} \right). 
\end{equation}
For each of the $m$ two-dimensional slices, data augmentation techniques are applied independently. After that, the slice scores are computed for each slice by a neural network. Then, based on the stack of predicted slice scores, the loss function is computed. Finally, the total loss is backpropagated to the neural network so that the model can learn from its mistakes, and the procedure starts all over again. 
The network architecture was adapted from the \ac{SSBR} model \cite{yan2018deep, yan2018unsupervised} with the following modifications: 
\begin{enumerate}
    \item In the \ac{SSBR} model, the slice-index difference $k$ is sampled directly. We will sample the distance $\Delta h$ in mm from the physical space between $[\Delta h_{\text{min}}, \Delta h_{\text{max}}]$. From this sampled distance $\Delta h$, the slice-index difference $k$ is computed. Because the distribution of the z-spacing in the training data is heterogeneous, the two approaches lead to different samplings strategies.  The sampling method was changed to obtain a more natural and realistic slice sampling and to be less dependent on the distribution of the z-spacing in the dataset. 
    \item In the \ac{SSBR} model, random cropping was used as the only data augmentation method \cite{yan2018deep}. This work will apply far more data augmentation techniques to prevent the model from overfitting on studies or patients. The used data augmentation techniques can be find in section \ref{sec:data-augmentation}.
    \item The original \ac{SSBR} model was only trained on body regions from the lower pelvis to the upper chest \cite{yan2018unsupervised}. In this work, the investigated body region was extended to the end of the head by using publicly available data. This makes the method applicable to more CT images. 
    \item In the \ac{SSBR} model, the classification order loss was used (see sec. \ref{sec:order-loss}). We will use the heuristic order loss. The derivations of the loss functions and the use of the heuristic order loss makes the used hyperparameters better interpretable (see sec. \ref{sec:order-loss}). 
    \item The original \ac{SSBR} model is validated by transferring the regression problem to a 3-class classification problem and measuring the accuracy. In this work, a metric is proposed that allows a more accurate evaluation of different models without simplifying the problem to a classification problem (see sec. \ref{sec:methods-evaluation-methods}). 
\end{enumerate}
\subsection{Order Loss}\label{sec:order-loss}
For self-supervised approaches, the loss function is crucial. The loss function of the body part regression model consists of an order loss and optionally an additional distance loss.
The order loss ensures the monotony condition, which states that slice scores should increase with height. In this section, two versions of the order loss are introduced. The classification order loss used by Yan et al. \cite{yan2018unsupervised} is mathematically derived, and a new variant, called heuristic order loss, is proposed. The classification order loss is based on transferring the order problem to a classification problem. For the heuristic order loss, we additionally incorporate the physical distance between sampled slices into the loss function. 
\begin{figure}
    \centering
    \includegraphics[width=0.65\textwidth]{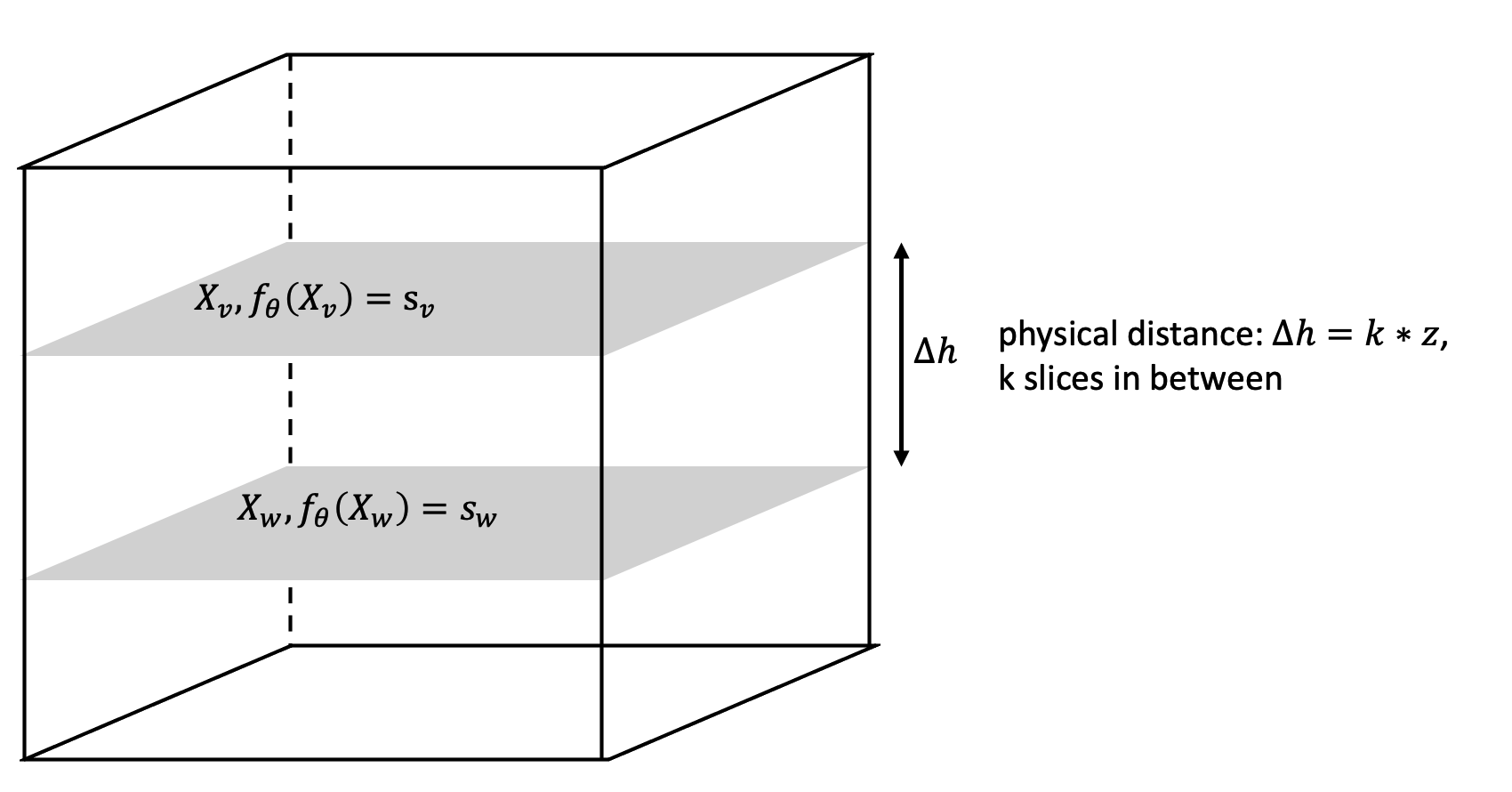}
    \caption{Sampled CT slices $\mathbf{X}_{v}, \mathbf{X}_{w}$, corresponding slice scores $s_{v}, s_{w}$ and physical distance $\Delta h$ of CT volume with z-spacing $z$.}
    \label{fig:orderloss-definitions} 
\end{figure} 
For the following calculations, we assume that we sample two two-dimensional slices $\mathbf{X}_{v}$ and $\mathbf{X}_{w}$ from a three-dimensional human CT-volume with the corresponding physical distance $\Delta h$, as it can be seen in figure \ref{fig:orderloss-definitions}. 

\textbf{Classification Order Loss: }
Because no mathematical derivation for the loss function was given in the literature, the loss function will be derived in this subsection. The key idea behind this order loss function is to transfer the order problem to a classification problem. 
Therefore, the derivation will be quite similar to the derivation of the Cross-Entropy loss (see sec. \ref{sec:loss-functions}). 
The ordering problem is converted to a classification problem by defining two classes $c \in \{0, 1\}$: 
\begin{equation*}
    c_{vw} = \begin{cases}
    1 & \text{if } v > w,\\ 
    0 & \text{if } v \le w.\\ 
    \end{cases}
\end{equation*}
If  slice $ \mathbf{X}_{v}$ lies above  slice $\mathbf{X}_{w} $, we assign the class 1. Otherwise, the class 0 is assigned. The slices $ \mathbf{X}_{v}$ and $\mathbf{X}_{w} $ are randomly sampled from one patient. 
We assume that the class labels $c$ are Bernoulli distributed 
\begin{equation*}
    \begin{split}
    & c_{vw} \sim \mathcal{B}(\pi_{vw}), \\ 
    & P(c=1|\mathbf{X}_{v},\mathbf{X}_{w}) = \pi_{vw}, \\
    & P(c=0|\mathbf{X}_{v}, \mathbf{X}_{w}) = 1 - \pi_{vw}, \\
    & \Rightarrow P(c|\mathbf{X}_{v}, \mathbf{X}_{w}) = \pi_{vw}^{c_{vw}} \cdot (1-\pi_{vw})^{1-c_{vw}}.
    \end{split}
\end{equation*}
Provided that we have a perfect body part regression model, which fulfills the \textit{monotony condition} (see sec. \ref{sec:architecture}), we can model the probability $\pi_{vw}$ based on the predicted scores $s_{v}$ and $s_{w}$. We assume that slice $\mathbf{X}_{v}$ lies above slice $\mathbf{X}_{w}$, if the score difference $\Delta s_{vw} = s_{v} - s_{w}$ is greater than zero. If the score difference $\Delta s_{vw}$ is less than zero, we assume, that slice $\mathbf{X}_{v}$ lies under slice $\mathbf{X}_{w}$. For $\Delta s_{vw}=0$, we expect a chance of 50 \% that $\mathbf{X}_{v}$ lies above slice $\mathbf{X}_{w}$. The probability $\pi_{vw}$ can be represented with the help of the sigmoid function $\sigma$: 
\begin{equation}\label{eq:pi}
    \pi_{vw} = \sigma(\Delta s_{vw}).
\end{equation}
The sigmoid function $\sigma$ is given by equation \ref{eq:sigmoid}. 
The greater $s_{v}$ compared to $s_{w}$, the higher is the probability $ P(c_{vw}=1)$.
To train a model, which fulfills the \textit{monotony condition}, we can use the maximum likelihood approach. 
With this approach, we find the model parameters $\boldsymbol{\theta}$ under which the observed slice orderings are most likely. 

\begin{figure}
    \centering
    \includegraphics[width=0.7\textwidth]{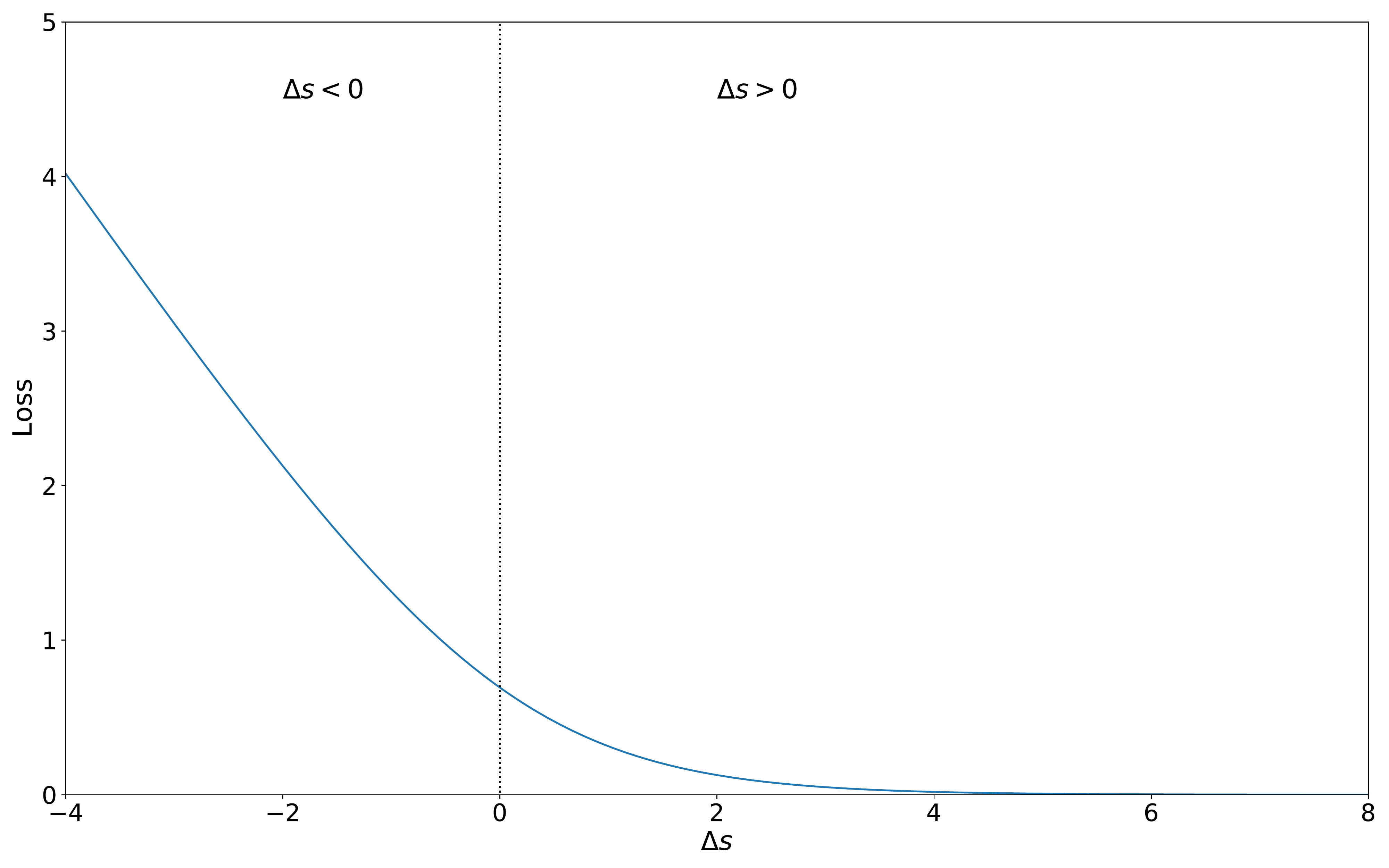}
    \caption{Trend of the classification order loss $L_{\text{order}}(\Delta s)$ from equation \ref{eq:order-loss-ubr} for $B=1$ and $m=2$. For $\Delta s >> 0$, the loss converges to zero. For $\Delta s < 0$ the loss and the gradient is high, the \textit{monotony condition} is not met.}
    \label{fig:ubr-order-loss}
\end{figure}

We can influence the observed slice orderings by the sampling strategy of the input volumes. 
By sampling a random start index $j$ and a random index difference $k$, we can define 
\begin{equation*}
    v=j+k,\quad w=j. 
\end{equation*}
Therefore, we know that slice $\mathbf{X}_{v}$ always lies above slice $\mathbf{X}_{w}$ and that the observed class label is always one. If we sample from $N$ independent volumes, the observed data is given by
$(\textbf{X}_{i,j}, \textbf{X}_{i,j+k}, c_{i}=1)_{i=1...N}$. Moreover, the likelihood function is equal to 
\begin{equation}\label{eq:likelihood-ce}
    \mathcal{L}({\boldsymbol{\theta}}) = p(\{c_{ij}=1\}_{i=1...N}|\boldsymbol{\theta}) = 
    \prod_{i=1}^{N} \sigma(f_{\boldsymbol{\theta}}(\mathbf{X}_{i,j+k}) - f_{\boldsymbol{\theta}}(\mathbf{X}_{ij})) = 
    \prod_{i=1}^{N} \sigma(\Delta s_{ij}), 
\end{equation}
where the slice score difference $\Delta s_{ij}$ is defined by 
\begin{equation*}
     \Delta s_{ij} = s_{i,j+k} - s_{ij},
    \quad f_{\boldsymbol{\theta}}(\mathbf{X}_{ij}) = s_{ij}, \quad f_{\boldsymbol{\theta}}(\mathbf{X}_{i,j+k}) = s_{i,j+k}.
\end{equation*}
By taking the logarithm of equation \ref{eq:likelihood-ce} we obtain the log-likelihood 
\begin{equation*}
    \text{ln }\mathcal{L}({\theta}) = \sum_{i=1}^{N} \text{ln}\left(\sigma(\Delta s_{ij})\right). 
\end{equation*}
The order loss is given by an estimate of the normalized negative log-likelihood  d
\begin{equation}\label{eq:order-loss-ubr}
    L_{\text{order}} = - \frac{1}{B}\sum_{i=1}^{B} \text{ln}\left(\sigma(\Delta s_{ij})\right), 
\end{equation} 
where $B$ is the batch size. This is equivalent to the classification order loss from the \ac{SSBR} model \cite{yan2018unsupervised} for $m = 2$. 
For $m>2$, we sample $m$ slices from a volume by choosing a random slice difference $k$ and a random start index. We obtain the following normalized order loss function of the \ac{SSBR} model 
\begin{equation} \label{eq:calssification-order-loss}
    L_{\text{order}}^{c} = -\frac{1}{B}\frac{1}{m-1}\sum_{i=1}^{B}\sum_{j=1}^{m-1} \text{ln} \ \sigma(\Delta s_{ij}), \qquad \Delta s_{ij} = s_{i,j+1} - s_{ij}.
\end{equation} 
Figure \ref{fig:ubr-order-loss} shows that the loss and the gradients are small if $\Delta s$ is greater than zero, and it is high if $\Delta s$ is smaller than zero. One weakness of this loss is, that for slice pairs with small physical distance $\Delta h < $ 5 mm we would assume that the slice scores are nearly the same, because the images are looking quite similar. Although, we expect $s_{ij} \simeq s_{i,j+1}$, the gradient of the loss at $\Delta s = 0$ is still high. This weakness can be overcome by introducing a minimum physical slice distance $\Delta h_{\text{min}}$ for sampling $\mathbf{X}_{i,j}$ and $\mathbf{X}_{i,j+1}$. 

\textbf{Heuristic Order Loss: } In the previous paragraph, we derived the classification order loss used by Yan et al. to train the \ac{SSBR} model \cite{yan2018unsupervised}. We will now investigate a completely different and new approach for handling the order problem. The main reason behind the introduction of the heuristic order loss is to incorporate the physical distance $\Delta h$ between slices in a meaningful way. In the classification order loss, the physical distance has no impact on the loss function. If two slices lie physically close to each other, they would be treated in the same way as two slices that lie far apart from each other. 
In the heuristic order loss, we concentrate on building a heuristic for the following probability: 
\begin{equation*}
    p(c_{vw}=1|\Delta s_{vw}) = \pi_{vw}, 
    \quad \Delta s_{vw} = s_{v} - s_{w}.
\end{equation*}
The probability $\pi_{vw}$ is the probability that slice $\mathbf{X}_{v}$ lies above slice $\mathbf{X}_{w}$, based on the predicted slice scores $s_{v}$ and $s_{w}$. 
We can model the predicted probability $\hat{\pi}_{vw}$ through the network with the sigmoid function as already derived in equation \ref{eq:pi}: 
\begin{equation*}
    \sigma(\Delta s_{vw}) = \hat{\pi}_{vw}, 
\end{equation*} 
In general, the probability $\hat{\pi}_{vw}$ is high if the difference between the predicted slice scores is as well high. 
Next, we want to construct a heuristic for the target probability $\pi_{vw}$. To build this pseudo-label, we assume that it should be easier to decide whether slice $\mathbf{X}_{v}$ lies above slice $\mathbf{X}_{w}$ if the physical slice distance $\Delta h$ between both slices is high. 
Let us consider two scenarios in which the goal is to order two slices from a human CT image: 
\begin{enumerate}
    \item One slice lies in the lung and the other one in the abdomen region. The physical slice difference between both slices is equal to $\Delta h =$ 200 mm. 
    \item The two slices are consecutive slices from the abdomen region, with a physical slice difference of  $\Delta h = $1 mm.
\end{enumerate}
A human can order the slices from scenario 1 with nearly a hundred percent certainty, whereas in scenario 2, a human would be less sure about the order of the slices.
Based on these considerations, we propose as a proxy for $\pi_{vw}$ 
\begin{equation*}
    \sigma(\beta \Delta h_{vw}) = \pi_{vw}^{\star},  \quad \Delta h_{vw} = h_{v} - h_{w}, 
\end{equation*}
where $\beta$ is a fixed hyperparameter. 
We assume high certainty for large $\Delta h$ (around 100 \%) and low certainty for small $\Delta h$ (around 50 \%). Moreover, we assume that the scores are continuous. 
For training the model, we can try to minimize the  distance between $\pi_{vw}^{\star}$ and $\hat{\pi}_{vw}$ through the least square error method \cite{durstewitz2017advanced}. If we chose for every volume $i$ in a batch a random start index $j$ and a random slice index difference $k$ with the corresponding physical slice distance $\Delta h_{i}$, we gain the following least square error
\begin{equation}\label{equation:order-loss-normal}
    L_{\text{order}}^{h} = \frac{1}{B} \sum_{i=1}^{B}||\pi_{ij}^{\star} - \hat{\pi}_{ij}||^{2}_{2} = 
   \frac{1}{B} \sum_{i=1}^{B} ||\sigma(\beta \Delta h_{i}) - \sigma(\Delta s_{ij})||^{2}_{2}, 
\end{equation}
where $B$ is the mini batch size. 
\begin{figure} 
    \centering
    \includegraphics[width=0.85\textwidth]{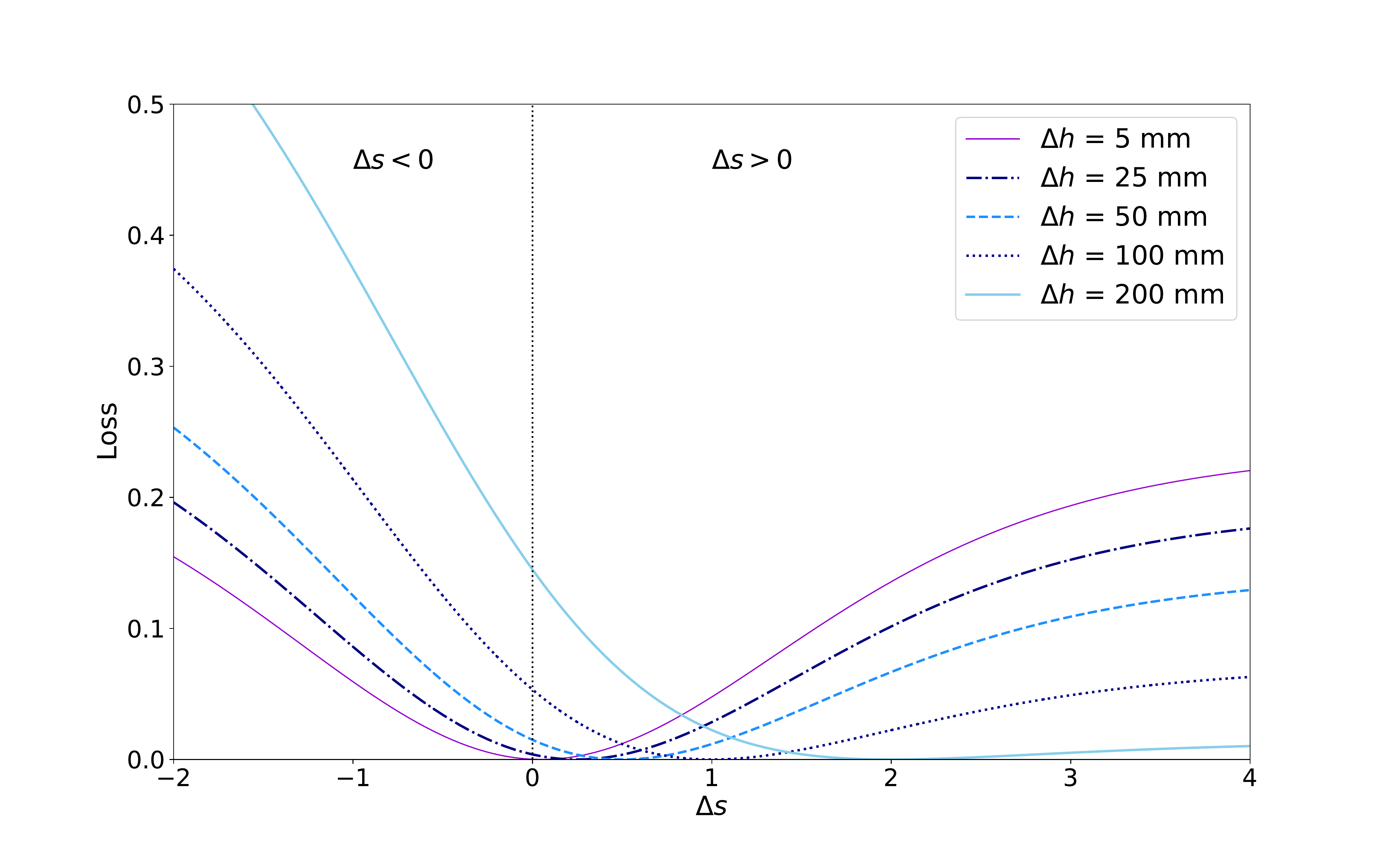}
    \caption{Order loss from equation \ref{equation:order-loss-normal} for different $\Delta h$, $B=1,$ and $\beta=\text{0.01/mm}$.}
    \label{fig:order-loss-normal}
\end{figure} 
For $m>2$, we sample $m$ slices from a volume by choosing a random start index and slice index difference $k$. We obtain the following loss function: 
\begin{equation}\label{equation:heuristic-order-loss}
    L^{h}_{order} = 
    \frac{1}{B(m-1)}\sum_{i=1}^{B}\sum_{j=1}^{m-1}
    ||\sigma(\beta \Delta h_{i}) - \sigma(\Delta s_{ij})||^{2}_{2},
    \quad \Delta s_{ij} = s_{i,j+1} - s_{ij}.
\end{equation}
The main advantage of this loss function is a meaningful integration of $\Delta h$. Figure \ref{fig:order-loss-normal} shows that the network can now learn something if $\Delta s$ is too high for small $\Delta h$: For small $\Delta h$, we want $\Delta s$ to be small as well because the visible slices look similar. If $\Delta s$ is too big, the network obtains the feedback that the predicted slice scores are too far apart. While we assume a linear relation between $\Delta s$ and $\Delta h$ in the range of small $\beta \Delta h$, the relation disappears for large $\beta \Delta h$, and the loss becomes more similar to the classification order loss. 
The order loss vanishes at the minimum for $\Delta s_{ij}$ equal to 
\begin{equation}\label{eq:proportionality-beta}
    L_{\text{order}} = 0 \quad \Rightarrow \quad 
    \sigma(\beta \Delta h_{i}) = \sigma(\Delta s_{ij})
    \quad  \Rightarrow \quad 
    \Delta s_{ij} = \beta\Delta h_{i}.
\end{equation}
To understand the influence of parameter $\beta$ better, it is helpful to look at the curvature of the minimum. At the minimum $\Delta s_{ij} = \beta \Delta h_{i}$ for $B=1$ and $m=2$, the curvature is given by 
\begin{equation}\label{eq:linearity-enforcement}
    L_{\text{order}}^{''}|_{\Delta s_{ij} = \beta \Delta h_{i}} = \frac{2 e^{-2\beta \Delta h_{i}}}{(1 + e^{-\beta \Delta h_{i}})^{4}}.
\end{equation}
The curvature is the second derivative of the loss function with respect to the score difference $\Delta s_{ij}$. 
If the curvature is high, the change in the gradients around the minimum is big, and the linear constraint is enforced more strongly by gradient descent. For $\beta\Delta h$ around zero, the curvature lies near $\frac{1}{8}$ and for $\beta\Delta h \rightarrow \infty$ the curvature goes to zero. This illustrates that the linear relation between $\Delta h$ and $\Delta s$ is only enforced by gradient descent for small $\beta\Delta h$ and disappears for large $\beta \Delta h$. For large $\beta \Delta h$, the curvature around the minimum is approximately zero, and we are in a flat region. If $\Delta s$ is large enough, the gradients are nearly zero, and the neural network has only a particular weak incentive to find the analytical minimum. 

\begin{figure}
    \centering
    \includegraphics[width=\textwidth]{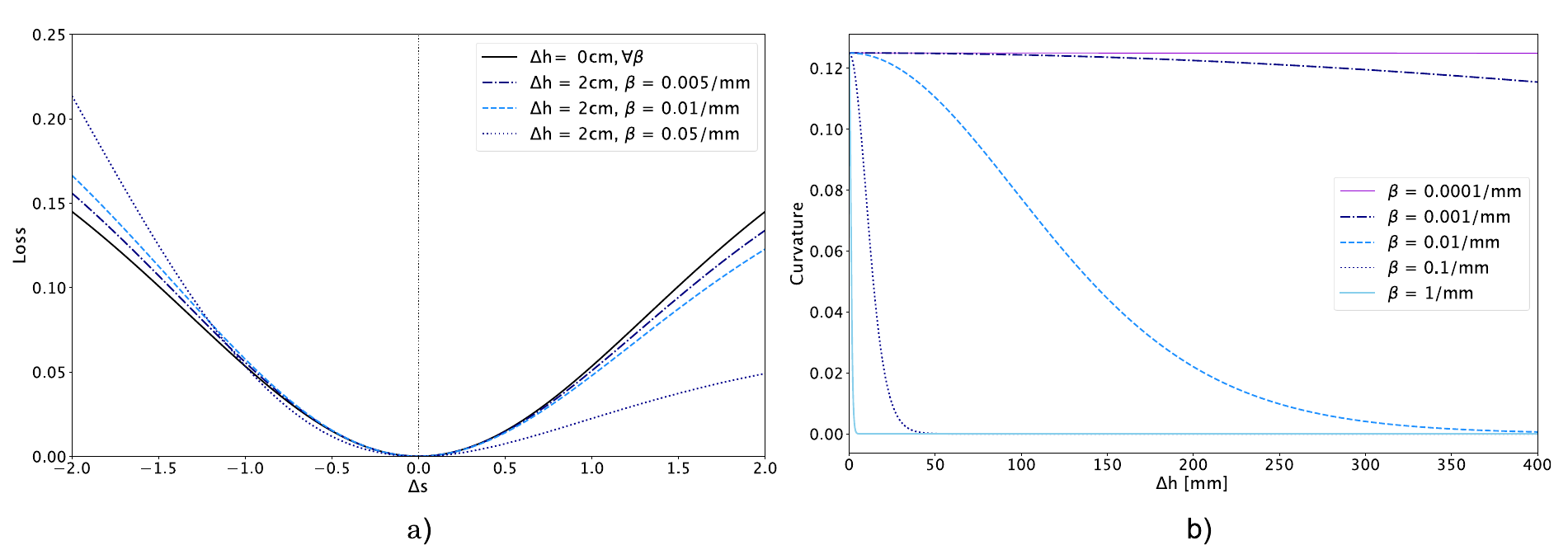}
    \caption{Analysis of parameter $\beta$ in heuristic order loss. In figure a), the heuristic order loss for different $\beta$ is visible. The graph was shifted to the left by the minimum $\Delta s = \beta \Delta h$ so that the minimum lies at $\Delta s = 0$ for all curves.
    In figure b), the curvature of the heuristic order loss  for different $\beta$ at the minimum $\Delta s = \beta \Delta h$ is visible.}
    \label{fig:alpha-beta-heuristic-order-loss}
\end{figure}

Figure \ref{fig:alpha-beta-heuristic-order-loss} b) visualizes the curvature at the minimum for different $\beta$. For $\Delta h = 0 \ \text{mm}$, the curvature is equal for all $\beta$. For $\Delta h >0 \ \text{mm}$, the $\beta$ parameter influences how quickly the curvature decreases. In figure \ref{fig:alpha-beta-heuristic-order-loss} a), we observe as well that for $\Delta h = 0 \ \text{mm}$ the different $\beta$ have no impact. For $\Delta h = 2 \ \text{cm}$, the curvature is smaller for bigger $\beta$. Therefore, we can say that the $\beta$ parameter influences how long the linearity condition $\Delta s \propto \Delta h$ persists. 
\subsection{Data Augmentations}\label{sec:data-augmentation}
Data augmentations, applied to each slice independently, play a key role in the learning procedure. The data augmentation techniques take care that the independence condition is met (see sec. \ref{sec:architecture}). 
Table \ref{tab:data-augmentation} summarizes the used data augmentations and parameters. In the following, the data augmentation methods applied throughout the experiments are explained.  The hyperparameters for the data augmentations were set manually to gain more variability in the image quality, appearance, and intensity values in the input image without receiving unrealistic CT slices. 

\textbf{Gaussian Noise: }
Gaussian noise was randomly applied to the images. The maximum possible standard deviation of the Gaussian noise $\sigma_{max}=0.04$ corresponds to $50 \text{HU}$, because the input volumes are rescaled from (-1000 HU, 1500 HU) to (-1,1): 
\begin{equation}\label{eq:hu}
    0.04 \cdot 2500 \text{ HU} /2 = 50 \text{ HU}.
\end{equation} 
This data augmentation was applied randomly with a chance of $p=50\%$ to the images in the dataset.

\textbf{Gaussian Blur: }
Gaussian blur was randomly applied with a chance of $p=50\%$ to the images. The maximum kernel size for blurring the input image is randomly chosen from three to seven. The maximum standard deviation for the blurring filter is set to 0.5.

\textbf{Flip, Scale and Rotate: }
In 50 \% of the cases, the image is randomly flipped horizontally or vertically. The images are transposed with a probability of 50 \%. By these two operations, mirroring and rotating the image by 90 degrees are covered. Furthermore, the images are randomly scaled by a scale factor between 80 \% and 120 \% and randomly rotated between zero and ten degrees. 

\textbf{Contrast and Brightness: }
The contrast is changed by linearly scaling the pixel intensities by the scaling factor, which is randomly chosen between 80 \% and 120\%. If the image extends the minimum or maximum possible value of -1 and 1, the extreme values are clipped. The brightness of the image is changed by randomly adding a constant between zero and 0.08. The maximum shift limit is equal to $100 \text{ HU} = 0.08 \cdot 2500 \text{ HU}/2 $ (see eq. \ref{eq:hu}). Contrast and brightness adaption are applied randomly with a chance of $p=50\%$ to the images in the dataset.

\begin{figure}
    \centering
    \begin{subfigure}[b]{0.49\textwidth}
        \centering
        \includegraphics[width=0.6\textwidth]{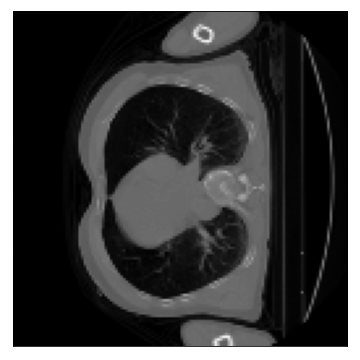}
    \end{subfigure}
    \begin{subfigure}[b]{0.49\textwidth}
        \centering
        \includegraphics[width=0.6\textwidth]{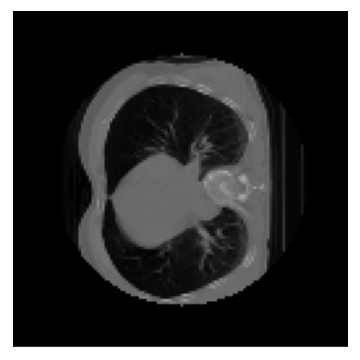}
    \end{subfigure}
    \caption{Example of Add Frame data augmentation. In the left figure, the original slice is visible. In the figure on the right, the Add Frame data augmentation technique was applied by adding a circular mask to the image. As a result, the CT table and the arms are not anymore visible.  }
    \label{fig:add-frame}
\end{figure}
\textbf{Add Frame: } In the Add Frame data augmentation technique, a circular mask is added to the input image. 
The transformation is applied randomly with a probability of $p=25\%$ to the images in the dataset. 
The circle has a radius of about 48 pixels. Figure \ref{fig:add-frame} visualizes an example of this data augmentation method based on a chest CT slice. Moreover, this figure visualizes that the data augmentation method helps ignore arms and CT tables in CT slices. Some patients have their arms positioned next to their thighs and others over their heads. In addition, the CT table appearance differs from device to device. With this data augmentation, the arms and the CT table are sometimes cut off, which helps the network to ignore these features for predicting the slice scores. 

\begin{table} 
    \centering
    \caption{Documentation of used data augmentation techniques and corresponding parameters for training a body part regression model. Several augmentation techniques were adopted from the python package Albumentation \cite{albumentation}. }

    \begin{tabular}{lll}
        \hline
         Data augmentation & Parameters & Implementation \\
         \hline
         GaussianNoise &std\_min=0, std\_max=0.04, p=0.5 & custom\\
         GaussianBlur & blur\_limit=(3,7), sigma\_limit=0.5, p=0.5& Albumentation \cite{albumentation} \\
         ShiftScaleRotate  &\makecell[l]{shift\_limit=0, scale\_limit=0.2, \\ rotate\_limit=10, border\_mode=reflect} & Albumentation \cite{albumentation}\\ 
         Flip  & p=0.5 & Albumentation \cite{albumentation}\\
         Transpose & p=0.5 & Albumentation \cite{albumentation}\\
         Contrast & scale\_delta=0.2, p=0.5 & custom\\
         Brightness &shift\_limit=0.08, p=0.5& custom\\
         Add Frame & diameter=0.75, p=0.25 & custom\\
         \hline
        \end{tabular}
    \label{tab:data-augmentation}
\end{table}
\newpage
\section{Evaluation Metrics}\label{sec:methods-evaluation-methods}
Evaluating a body part regression model is non-trivial. Because of a lack of ground truth labels, it is not easy to compare model performances. The mapping between anatomy and score is learned implicitly in an unsupervised manner, resulting in different slice score scales for each model. For example, one model may map the beginning of the pelvis to -5 and the end of the head to 6, and another model may map the beginning of the pelvis to -100 and the end of the head to 90. 
Even for a human, it is hard to define a slice-wise anatomy-to-score mapping that fulfills the monotony and independence condition from sec. \ref{sec:architecture}. 
One reason for this is the inter-patient anatomical variability. For example, the liver can start at the end of the kidneys or the end of the pelvis bones. The lungs can be bigger or smaller based on the breathing cycle
\cite{vauthey2002body, abdalla2004total}.
A good evaluation metric is necessary to measure how well a body part regression model fulfills the independence condition while allowing for cross-model comparison. In the following, two different metrics are described, which are used to evaluate the model performance of the body part regression model. The proposed 
Landmark Mean Square Error (LMSE) is introduced in section \ref{sec:lmse}. Moreover, the setup for the accuracy metric, which was already used for evaluation by Yan et al. \cite{yan2018unsupervised}, is explained in section \ref{sec:accuracy}. 

Before calculating the defined metrics, we have to compute the model specific \textbf{slice score reference table}, which establishes a correspondence between each evaluation landmark defined in table \ref{tab:landmark-definitions} and its slice score pseudo-label $\bar{s}$. 
The pseudo-labels $\bar{s}$ are calculated for each of the twelve defined evaluation landmarks by averaging across all annotated landmark positions of the training set
\begin{equation}
    \bar{s}_{l_{k}} = \frac{1}{N_{l_{k}}}\sum_{i=1}^{N_{l_{k}}} s_{il_{k}}, \qquad
    \forall l_{k} \in \{l_{1}, ..., l_{12}\}, 
\end{equation}
where $N_{l_{k}}$ indicates the number of CT volumes where the landmark $l_{k}$ was annotated.  The pseudo-labels are calculated with an annotated subset of the training dataset to be independent of the validation and test dataset.

\subsection{Landmark Mean Square Error} \label{sec:lmse}
The Landmark Mean Square Error (LMSE) $\bar{\phi}$ is used as the primary metric to compare different models. It is a normalized version of a Mean Square Error (MSE) with the difference that we do not have ground truth labels and therefore create pseudo-labels based on the annotated landmarks. As pseudo-labels $\bar{s}_{l_{k}}$, we are using the values of the slice score reference table.
Further, an additional normalization is applied to enable comparing different models with different slice score scales. 
The landmark scores $s$ are transformed into $s^{\star}$ such that the pseudo-label $\bar{s}_{l_{1}}$ of landmark $l_{1}$, which refers to the start of the pelvis, maps to 0 and the pseudo-label $\bar{s}_{l_{12}}$ of landmark $l_{12}$, which refers to the end of the eye socket, maps to 100. The transformation is: 
\begin{equation}\label{eq:transform-scores}
    s \mapsto s^{\star}=\frac{100 \cdot (s - \bar{s}_{l_{1}})}{\bar{s}_{l_{12}} - \bar{s}_{l_{1}}}
    = \frac{(s - \bar{s}_{l_{1}})}{d},
    \qquad d = \frac{1}{100}(\bar{s}_{l_{12}} - \bar{s}_{l_{1}}).
\end{equation}
In the following, the constant $d$, will be referred to as the \textbf{normalization constant}. 
The Landmark Mean Square Error per volume is defined as 
\begin{equation}
    \begin{split}
    \phi_{i}
    &= \frac{1}{|L_{i}|}\sum_{l_{k} \in L_{i}}  
    (\bar{s}^{\star}_{l_{k}} - s^{\star}_{il_{k}})^{2}\\
    &= \frac{1}{|L_{i}|}\sum_{l_{k} \in L_{i}}  
    \left(\frac{\bar{s}_{l_{k}} - \bar{s}_{l_{1}} - (s_{il_{k}} - \bar{s}_{l_{1}})}{d}\right)^{2}\\
    &=  \frac{1}{|L_{i}|}\sum_{l_{k} \in L_{i}}  
    \left(\frac{\bar{s}_{l_{k}} - s_{il_{k}}}{d}\right)^{2}, 
    \end{split}
\end{equation}
where $L_{i}$ is the set of annotated landmark positions for volume $i$. The Landmark Mean Square Error for the whole dataset $\bar{\phi}$ is the mean of the LMSEs per volume $\phi_{i}$
\begin{equation}
    \begin{split}
    \bar{\phi} = \frac{1}{N}\sum_{i=1}^{N} \phi_{i},
    \end{split}
\end{equation}
where $N$ is the number of volumes in the annotated dataset. 
The average over the LMSEs per volume is taken, because we are interested in how well the model generalizes to new volumes \cite{holland2020drawing}. 
The standard error of the LMSE  $\sigma_{\bar{\phi}}$ can be estimated by  
\begin{equation}
    \sigma_{\bar{\phi}} = \frac{\sigma_{\phi}}{\sqrt{N}} 
    = \frac{1}{\sqrt{N}}\cdot \sqrt{\frac{1}{N-1}\sum_{i=1}^{N}(\phi_{i} - \bar{\phi})^{2}}. 
\end{equation}
The LMSE is specialized in evaluating the independence condition, but it is insensitive to the monotony condition. 
For example, the LMSE can be zero if all slices are mapped to the same slice score. Therefore, we need to look at the predicted scores of a model to be sure that the model also fulfills the monotony condition. 

\textbf{Comparing Landmark Mean Square Errors}: 
In the following, the choice of significance test to compare different model performances based on the LMSE will be explained. 
If the number of annotated volumes $N$ is high enough, a normal distributed LMSE value can be assumed $\mathcal{N}(\bar{\phi}, \sigma_{\bar{\phi}})$ (based on the central limit theorem \cite{DeepVision}). 
Because of that, a standard two-sided z-test to compare model $x$ with model $y$ can be performed  \cite{durstewitz2017advanced}.
To test if model $x$ performs significantly different to model $y$ we can state the following null hypothesis $H_{0}$ and alternative hypothesis $H_{1}$: 
\begin{enumerate}
    \item[$H_{0}$:] $\mu_{\phi_{x}} = \mu_{\phi_{y}}$ ,
    \item[$H_{1}$:] $\mu_{\phi_{x}} \neq \mu_{\phi_{y}}$.
\end{enumerate} 
In this thesis, the significance level will be set for all z-tests to 5\%. 
The test statistic $t$ is computed by calculating the $\sigma$-deviation to zero of the difference between the performances $\bar{\phi}_{x}$ and $\bar{\phi}_{y}$ \cite{durstewitz2017advanced}:
\begin{equation}\label{eq:teststatistics}
    t = \frac{\bar{\phi}_{x} - \bar{\phi}_{y}}{\sqrt{\sigma_{\bar{\phi}_{x}}^{2} + \sigma_{\bar{\phi}_{y}}^{2}}} \sim \mathcal{N}(0, 1). 
\end{equation}
The test statistic is standard normal distributed if the null hypothesis $H_{0}$ is true. 
If the absolute value of the test statistic is greater than 1.96, the null hypothesis $H_{0}$ can be rejected based on the significance level, and we assume that model $x$ and model $y$ perform significantly different. 

\subsection{Accuracy}\label{sec:accuracy}
Because of the easy interpretability of the accuracy and its usage in previous work, we will also use the accuracy $\psi$ as an evaluation metric. For calculating the accuracy, five classes were defined: pelvis, abdomen, chest, neck, and head. Table \ref{tab:5class-accuracy} lists the class boundaries based on the annotated landmarks. For evaluation, the body range between the lowest landmark pelvis-start and the highest landmark eyes-end is considered. 
The accuracy $\psi$ compares the predicted classes $\hat{c}$ with the ground truth classes $c$ and is defined in equation \ref{eq:accuracy}. 
The ground truth class $c_{ij}$ for volume $i$ and slice index $j$ is determined by finding the start and end landmarks from table \ref{tab:5class-accuracy} that are surrounding slice $j$. Only if both boundary slice indices of a class are defined within a CT volume, the ground truth label was assigned. Otherwise, the class assignment would not be unique. 

The predicted class label $\hat{c}_{ij}$ for volume $i$ and slice index $j$ is determined by using the predicted slice score $s_{ij}$ of the slice $j$. With the help of the slice score reference table, the surrounding landmark slice score boundaries from table \ref{tab:5class-accuracy} are found and the 
corresponding body part class $\hat{c}_{ij}$ is assigned. 
 
\begin{table}
    \centering
    \caption{Definition of five body parts to determine the accuracy.}
    \begin{tabular}{lll}
        \hline
        Body part & Start landmark & End landmark  \\
        \hline
        pelvis  & pelvis-start & L5 \\
        abdomen & L5 & Th11 \\
        chest  & Th11 & Th2\\ 
        neck  & Th2 & C1\\ 
        head & C1 & eyes-end \\
        \hline
    \end{tabular}
    \label{tab:5class-accuracy}
\end{table}
\newpage
\section{Application and Deployment}\label{sec:methods-application-deployment}
In this thesis, three use cases (UC) for the body part regression model will be discussed:  
\begin{enumerate}
    \item[\textbf{UC1}] \textbf{Estimate Examined Body Part:} the body part regression model can be used for a better estimate of the \textit{BodyPartExamined} DICOM tag \cite{fedorov2016dicom}.  This information is beneficial for sorting and filtering medical datasets \cite{yan2018deep,gueld2002quality}. This use case will be discussed in section \ref{sec:methods-bpe}.

    \item[\textbf{UC2}] \textbf{Known Region Cropping:}
    Deep learning methods in the medical imaging domain are typically designed for one specific body region. For a robust transfer of deep learning models into the clinic, it is important to only select images which fit the scope of the algorithms. 
    Selecting and cropping images to the scope of an algorithm can be done by the body part regression model. Moreover, the body part regression model can be used as a post-processing step to detect false-positive predictions of a medical algorithm, which lie out of scope. This application use case is demonstrated by using medical segmentation algorithms as a representative for medical computer vision algorithms. The use case is explained in section \ref{sec:known-region-cropping}. 

    \item[\textbf{UC3}] \textbf{Data Sanity Checks: } 
The body part regression model can be used for basic data sanity checks, such as validating the z-spacing and the ordering of the z-axis. This use case will be discussed in section \ref{sec:method-dsc}. 
\end{enumerate}
In this thesis, the trained body part regression model is specialized on CT images. Therefore, the use cases will be evaluated on CT images. 
To make the use cases easier applicable, the methods were deployed through a python package and integrated into  Kaapana (see sec. \ref{sec:deployment}).

\begin{figure}
    \centering
    \includegraphics[width=\textwidth]{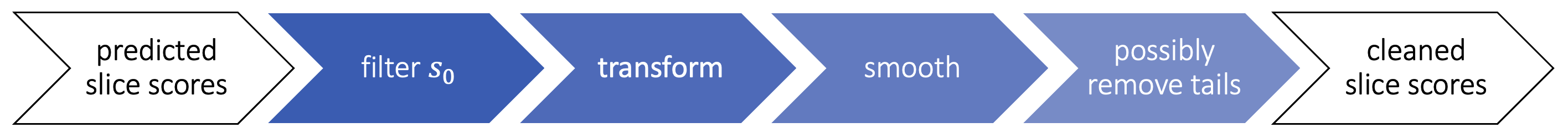}
    \caption{Slice score post-processing steps before using the slice scores for application purposes.}
    \label{fig:slice-score-postprocessing}  
\end{figure}

Predicted slice scores are post-processed by filtering empty slices, linearly transforming, and smoothing scores, and removing outliers. 
 The different steps are summarized in figure \ref{fig:slice-score-postprocessing}. 
To obtain the cleaned  slice scores from figure \ref{fig:slice-score-postprocessing}, the following steps need to be taken: 
\begin{enumerate}
    \item \textbf{Filter $s_{0}$}: For the body part regression model, the slice score prediction $s_{0}$ of an empty slice needs to be determined in advance. In empty slices, the pixels have all the HU value, which is equal to $-1024$ HU. No human and no CT table is visible. 
    Score predictions of empty slices $s=s_{0}$ are filtered out. 
    \item \textbf{Transform: } As described in section \ref{sec:methods-evaluation-methods}  the \textbf{slice score reference table} is employed to transform predicted slice scores to a common scale. The calculation of the table is done as in section \ref{sec:methods-evaluation-methods} with the difference that the data foundation are the annotated volumes from the training and the validation set to obtain a better estimate of the expected slice scores for each landmark. 
    With the help of the computed slice score reference table, the slice scores are linearly transformed so that the slice of the pelvis-start landmark maps to 0 and the landmark from eyes-end maps to 100 (see eq. \ref{eq:transform-scores}). This makes working with the slice scores more standardized and manageable.
    \item \textbf{Smooth: } The slice scores are smoothed to remove local discontinuities. The smoothing is done by applying a discrete convolution with a Gaussian filter kernel
    \begin{equation*}
    \begin{split}
         & s(h) \mapsto (f \ast g)(h), \\
         & g(h) = \frac{1}{\sqrt{2 \pi} \cdot \sigma} \cdot e^{\frac{-h^{2}}{2\sigma^{2}}},
    \end{split}
    \end{equation*}
    where $f(h)$ is our neural network which maps the slice at height $h$ of a patient to a slice score and where $g(h)$ is the Gaussian filter kernel. The standard deviation $\sigma$ of the Gaussian filter is set manually to $\sigma=\text{10 mm = 1 cm}$. This means that approximately a range of $\pm \text{1 cm}$ around a score has an influence on the smoothed score and should be in a comparable slice score range. Through this step, we take advantage of the assumption that the function should be continuous. 
    \item \textbf{Possibly remove tails: } Remove tails of the slice score curve before 0 or after 100 if tangential slopes $m_{t}$ are atypical.
    The tangential slope at index $i$ is calculated by 
    \begin{equation}
        m_{ti} = \frac{s_{i+1} - s_{i}}{h_{i+1} - h_{i}}. 
    \end{equation}
     For the body part regression model, a minimum valid tangential slope $m_{t, min}$ and a maximum valid tangential slope $m_{t, max}$ are set. 
     If a tangential slope $m_{t}$ before 0 or after 100 does not lie in the valid range, the tails until this score are removed from the slice score curve. 
     To obtain $m_{t, min}$ and  $m_{t, max}$ for a given model the distribution of the tangential slopes $m_{t}$ for the training dataset was calculated. For this, the score curve for each volume from the training dataset is transformed and smoothed as described in step 1 till step 3, step 4 is ignored. 
     The 0.5 \% quantile and the 99.5 \% quantile of the distribution was set as $m_{t, min}$ and  $m_{t, max}$. 

\end{enumerate}
For the data sanity checks, the mean slope $\bar{m}_{s}$ of the slice score curves needs to be found. The difference between the previously defined tangential slope $m_{t}$ and the slice score curve slope $m_{s}$ is that $m_{s}$ represents the overall slope of a slice score curve for one volume whereas the $m_{t}$ is the slope on one point.
To obtain the mean slice score curve slope $\bar{m}_{s}$, we are iterating through all volumes from the training data. For each volume, 
the cleaned slice scores were predicted through steps 1 till 4. To each slice score curves, a linear line is fitted
\begin{equation*}
    s_{i} = m_{s, i}\cdot z\cdot i + b, 
\end{equation*}
where $z$ is the z-spacing of the volume. The slope $m_{i, s}$ is saved. In the end, the mean $\bar{m}_{s}$ of the slopes $m_{i, s}$ across all training volumes $i$ is used for the data sanity checks.  

\subsection{Estimate Examined Body Part} \label{sec:methods-bpe}
To estimate the examined body part of a CT volume $X$ with $N$ slices, the cleaned slice scores and the slice score reference table are used. The estimated examined body part can be specified as a dictionary, called \textbf{body part examined dictionary} $d$, or it can be specified as a single tag as in the meta-data entry \textit{BodyPartExamined} in a DICOM file. This tag will be referred to as \textbf{body part examined tag} $t$ in the following section. 

\textbf{Body Part Examined Dictionary: }
Six different core body regions were defined to categorize the slices within a CT volume: legs, pelvis, abdomen, chest, neck-and-shoulder, and head. In table \ref{tab:body-parts}, the defined body parts with the corresponding start and end landmarks can be seen. The start and end landmarks were manually set through visual inspection of the results.
Some body regions are overlapping, as for example, abdomen and chest. In the abdomen body part, the abdominal organs should be visible as the liver or the kidneys. The chest body part was defined from the lowest Thoracic spine to the highest Thoracic spine. 
The body region \textit{leg} has no start landmark and the body region \textit{head} has no end landmark. If head slices are in an image, the slices until the end of the volume are assigned as head body region. If leg slices lie inside a CT volume, all slices from index zero to the score $\bar{s}_{pelvis-start}$ are assigned as leg body region. 
The body part examined dictionary $d$ summarizes the slice indices for each body region: 
\begin{equation}
\begin{split}
    d = \{ &\text{legs: }\{ i | \infty < s_{i} < \bar{s}_{\text{pelvis-start}} \wedge i \in [0, .., N] \},  \\
    &\text{pelvis: } \{ i | \bar{s}_{\text{pelvis-start}} < s_{i} < \bar{s}_{\text{pelvis-end}} \wedge i \in [0, .., N]\}, \\
    &\text{abdomen: } \{ i | \bar{s}_{\text{L5}} < s_{i} < \bar{s}_{\text{Th8}} \wedge i \in [0, .., N]\}, \\
    &\text{chest: } \{ i | \bar{s}_{\text{Th12}} < s_{i} < \bar{s}_{\text{Th1}} \wedge i \in [0, .., N]\}, \\
    &\text{neck-and-shoulder: } \{ i | \bar{s}_{\text{Th3}} < s_{i} < \bar{s}_{\text{C2}} \wedge i \in [0, .., N]\}, \\
    &\text{head: } \{ i | \bar{s}_{\text{C5}} < s_{i} < \infty \wedge i \in [0, .., N]\}\}.
\end{split}
\end{equation}
With the body part examined dictionary, a detailed overview of the imaged body parts is easily accessible. 

\begin{figure}
    \centering
    \includegraphics[width=0.9\textwidth]{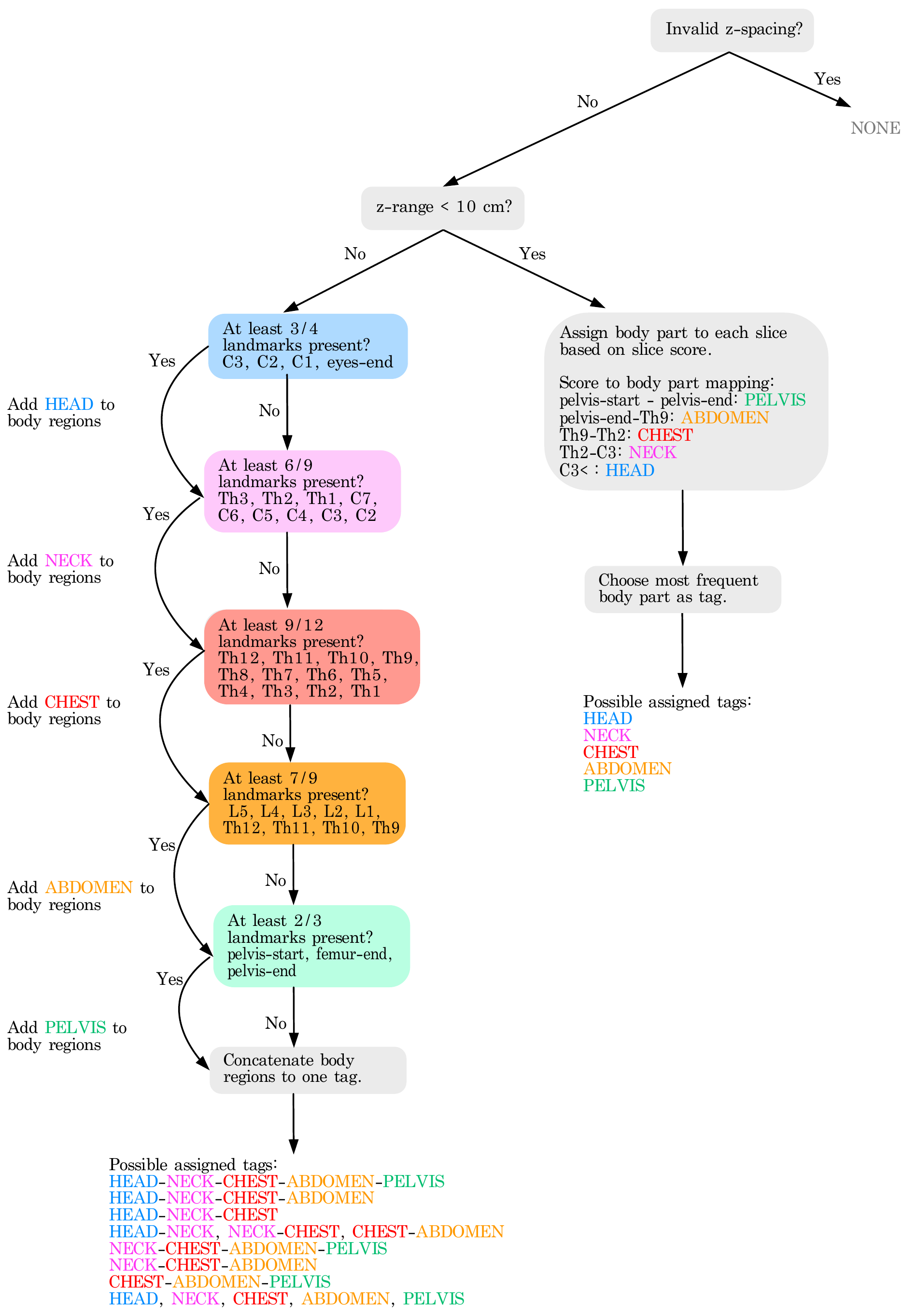}
    \caption{Decision tree to obtain the \textit{body part examined tag} $t$.}
    \label{fig:decision-tree}
\end{figure}

\begin{table}
    \centering
    \caption{Defined start and end landmarks for each body region. The expected start and end slice scores $\bar{s}$ are used from the \textit{slice score reference table}.}
    \begin{tabular}{lllll}
        \hline
         \makecell[l]{Body \\part} &\makecell[l]{Start \\landmark} & \makecell[l]{End \\landmark} & \makecell[l]{Start\\slice score} & \makecell[l]{End\\slice score}\\
         \hline
         legs & - & pelvis-start & $-\infty$ & $\bar{s}_{pelvis-start}$  \\
         pelvis & pelvis-start & pelvis-end & $\bar{s}_{pelvis-start}$  & $\bar{s}_{pelvis-end}$ \\
         abdomen & L5 & Th8 & $\bar{s}_{L5}$ & $\bar{s}_{Th8}$ \\
         chest & Th12 & Th1 & $\bar{s}_{Th12}$ & $\bar{s}_{Th1}$  \\
         neck-and-shoulder & Th3 & C2 & $\bar{s}_{Th3}$  & $\bar{s}_{C2}$ \\
         head & C5 & - & $\bar{s}_{C5}$  & $+\infty$\\
         \hline
    \end{tabular}
    \label{tab:body-parts}
\end{table} 
\textbf{Body Part Examined Tag}: For the body part examined tag $t$, one tag per CT volume is assigned. The assigned body regions are PELVIS, ABDOMEN, CHEST, NECK, and HEAD. If multiple body regions are visible, the tags are concatenated by a hyphen. The legs are excluded from the tag because they are out of the scope of the algorithm. 
If the data sanity check of section \ref{sec:method-dsc} outputs that the CT volume seems to be invalid based on the slope of the slice score curve, the resulting tag is NONE. For all other volumes, we differentiate between volumes with a small z-range of less than 10 cm and volumes with a bigger z-range of more than 10 cm. An overview of the decision tree can be found in figure \ref{fig:decision-tree}. 

For volumes with small z-ranges (z-range < 10 cm), no body region will likely be completely visible. Therefore, for every slice, a body region will be assigned based on the predicted cleaned slice score. After that, the most frequent body tag will be assigned to the whole CT volume. The slice score reference table is used to obtain the score boundaries for each class. The defined landmark boundaries for each body part class can be found in figure \ref{fig:decision-tree}. 

For volumes with bigger z-ranges, it is more likely that multiple body ranges were scanned. 
For each body part, a list of landmarks was defined through visual inspection. If a volume includes at least a manually defined minimum number of landmarks, the body part is declared as visible. 
A landmark $l_k$ is defined as visible if we find values in the cleaned slice scores which are bigger and smaller than the expected landmark slice score $\bar{s}_{l_{k}}$.
For each body part, the related landmarks and the minimum number of required landmarks can be found in figure \ref{fig:decision-tree}. In the end, the visible body parts are concatenated in descending order to one tag $t$ \mbox{(see fig. \ref{fig:decision-tree})}. 
\subsection{Known Region Cropping}\label{sec:known-region-cropping}
Because of a limitation in labeled data, medical image algorithms are often trained on a small dataset. These datasets normally only show the relevant CT body part for the prediction task.
For example, a CT kidney segmentation network is only trained on abdomen CT images and has never seen head CT scans during training \cite{hellerkits19}. This can lead to miss-classifications in head regions during test time. Therefore, clinical deep learning software can benefit from a basic understanding of the visible anatomy in a CT volume to filter false-positive predictions in irrelevant body areas.

\begin{figure}
    \centering
    \includegraphics[width=0.9\textwidth]{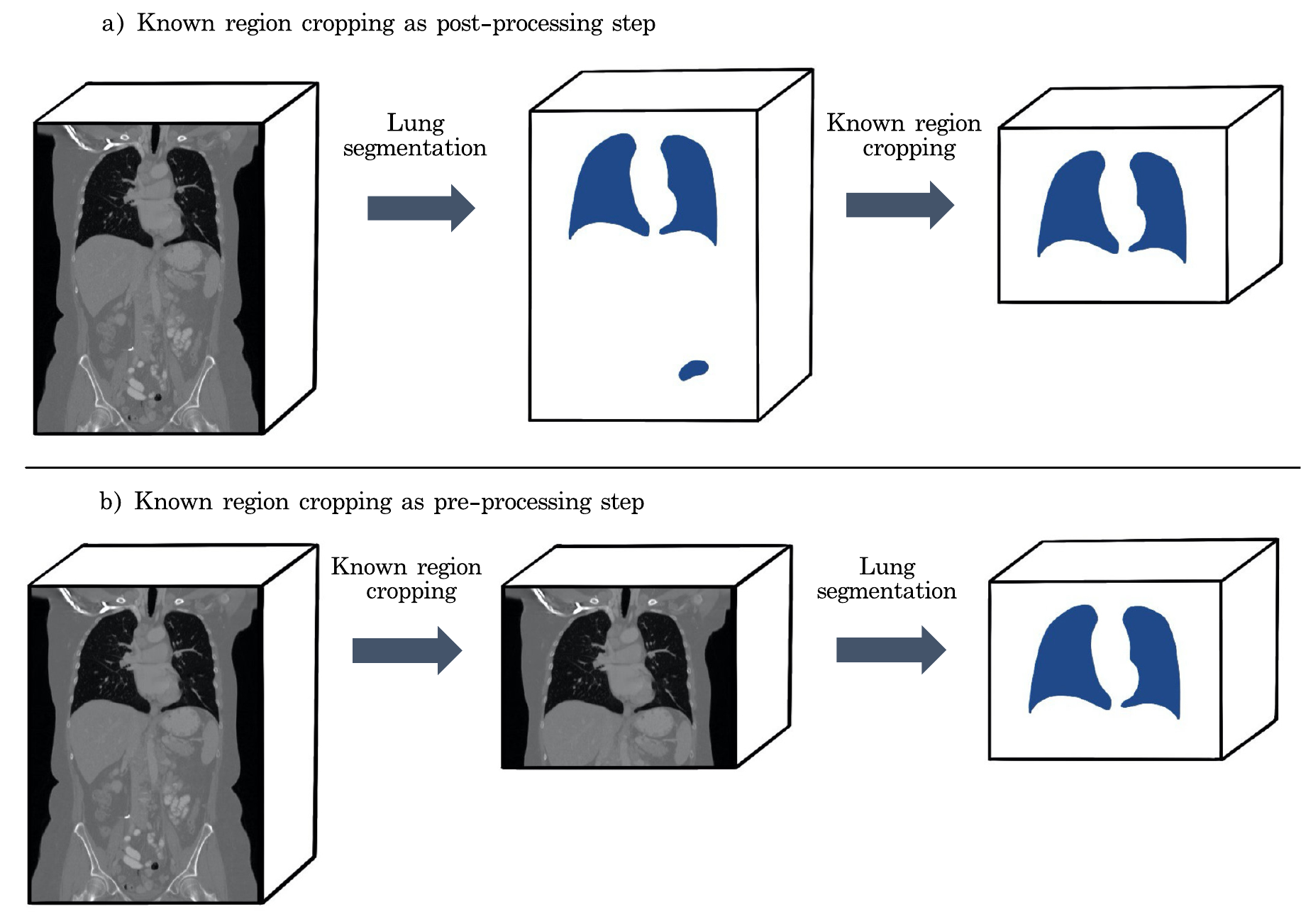}
    \caption{Visualization of known region cropping based on a lung segmentation algorithm. False-positive predictions can arise if the algorithm needs to process images that are out of the algorithm's scope.   }
    \label{fig:known-region-cropping-process}
\end{figure}

The body part regression model can be used as a preprocessing step of clinical end-to-end deep learning software to avoid false-positive predictions and to reduce the run-time. 
Moreover, \textit{known region cropping}  can be used as a post-processing step of a deep learning algorithm to catch false-positive predictions. 
The \textit{known region cropping}  process is independent of the used clinical algorithm. 
The different applications of \textit{known region cropping} as pre- or post-processing step are visualized in figure \ref{fig:known-region-cropping-process}. 
In the figure, a lung segmentation algorithm is used as an exemplary clinical computer vision task.

Before \textit{known region cropping} can be applied, we need to identify the known region of a medical algorithm. The known region is defined by the \textbf{minimum known score} $s_{\text{min}}$ and the \textbf{maximum known score} $s_{\text{max}}$. 
There are two main approaches to identify the score boundaries:
\begin{enumerate}
    \item Through the \textbf{data-driven approach}, the known range of an algorithm is determined with the used training data of the algorithm. 
    In this approach, for each volume of the training dataset, the cleaned slice score curve is predicted (see fig. \ref{fig:slice-score-postprocessing}), and the minimum and maximum slice scores are collected in two distinct lists. 
    In the end, the $q_{\text{min}}$-quantile of the list with the minimum slice-scores is used as the minimum known score $s_{\text{min}}$. The $q_{\text{max}}$-quantile of the list with the maximum slice scores is used as the maximum known score $s_{\text{max}}$. 
    The minimum quantile $q_{\text{min}}$ and the maximum quantile $q_{\text{max}}$-quantile are hyperparameters, which are manually set.
    \item Through the \textbf{manual approach}, the known range of an algorithm is determined manually based on the medical prediction task of the algorithm. It can be used if the training data of a deep learning algorithm is unknown. The minimum known score and maximum known score can be chosen based on the slice score reference table of the model and the prediction task. For example, for a kidney segmentation algorithm, the minimum known score can be set to the 
    expected score at the landmark femur-end, and the maximum known score can be set to the expected score of the landmark Th12. 
\end{enumerate} 
If we chose the difference  $\Delta s = s_{\text{max}} - s_{\text{min}}$ too high, it would lead to fewer detected false-positives. If we chose it too small, it would lead to the truncation of correct predictions. We need to find a trade of between the behaviors. In conclusion, for preparing \textit{known region cropping} for a model, the hyperparameters $q_{\text{min}}$ and $q_{\text{max}}$ needs to be set if the data-driven approach is used. Additionally, we need to find $s_{\text{min}}$ and $s_{\text{max}}$ for the model with the manual approach or the data-driven approach.
To apply the \textit{known region cropping} method to a CT volume $X$, we need to follow the steps: 
\begin{enumerate}
    \item Predict slice scores of CT volume $X$. 
    \item Calculate cleaned slice scores (see fig. \ref{fig:slice-score-postprocessing}). 
    \item Crop prediction or volume $X$, where the score is outside the known region defined by $s_{\text{min}}$ and $s_{\text{max}}$ (see fig. \ref{fig:known-region-cropping-process}).
\end{enumerate}

\subsection{Data Sanity Checks}\label{sec:method-dsc}
The height of women and men are approximately normal distributed with 
$h_{\text{women}} = 165 \pm 7 $ cm and $h_{\text{men}} = 184 \pm 8$ cm \cite{heights}. Therefore, the mean slopes $\bar{m}_{s}$ of the slice score curves should be as well approximately normal distributed. As discussed in section \ref{sec:methods-application-deployment},
the mean slope  $\bar{m}_{s}$ is estimated from the training data. With the help of the mean slope, $\bar{m}_{s}$ basic data sanity checks can be applied. On the one hand, it can be checked if the z-spacing is meaningful. On the other hand, it can be checked if the z-axis order is reversed. 
For the validation of the z-spacing a new variable, the slope ratio, $r_{m}$ and the expected relative error $\hat{\gamma}$ is defined. 
The slope ratio $r_{m}$ is the fraction between the observed slope $m_{X}$ for volume $X$ and the expected slope $\bar{m}_{s}$:  
\begin{equation}\label{eq:zratio}
    r_{m} = \frac{m_{X}}{\bar{m}_{s}}. 
\end{equation}
The expected relative error is given by 
\begin{equation}
    \hat{\gamma} = |1 - |r_{m}||.
\end{equation}
The expected relative error $\hat{\gamma}$ is, therefore, the absolute value of the relative error between the observed slope $m_{X}$ to the expected slope  $\bar{m}_{s}$. 
To decide if a slope ratio is valid, a threshold $\theta$ needs to be defined. The threshold will be found data-driven based on the validation set. The largest threshold $\theta$, for which for all volumes of the validation set $\hat{\gamma} < \theta$ is \mbox{chosen as threshold}. 

\begin{figure}
    \centering
    \includegraphics[width=0.6\textwidth]{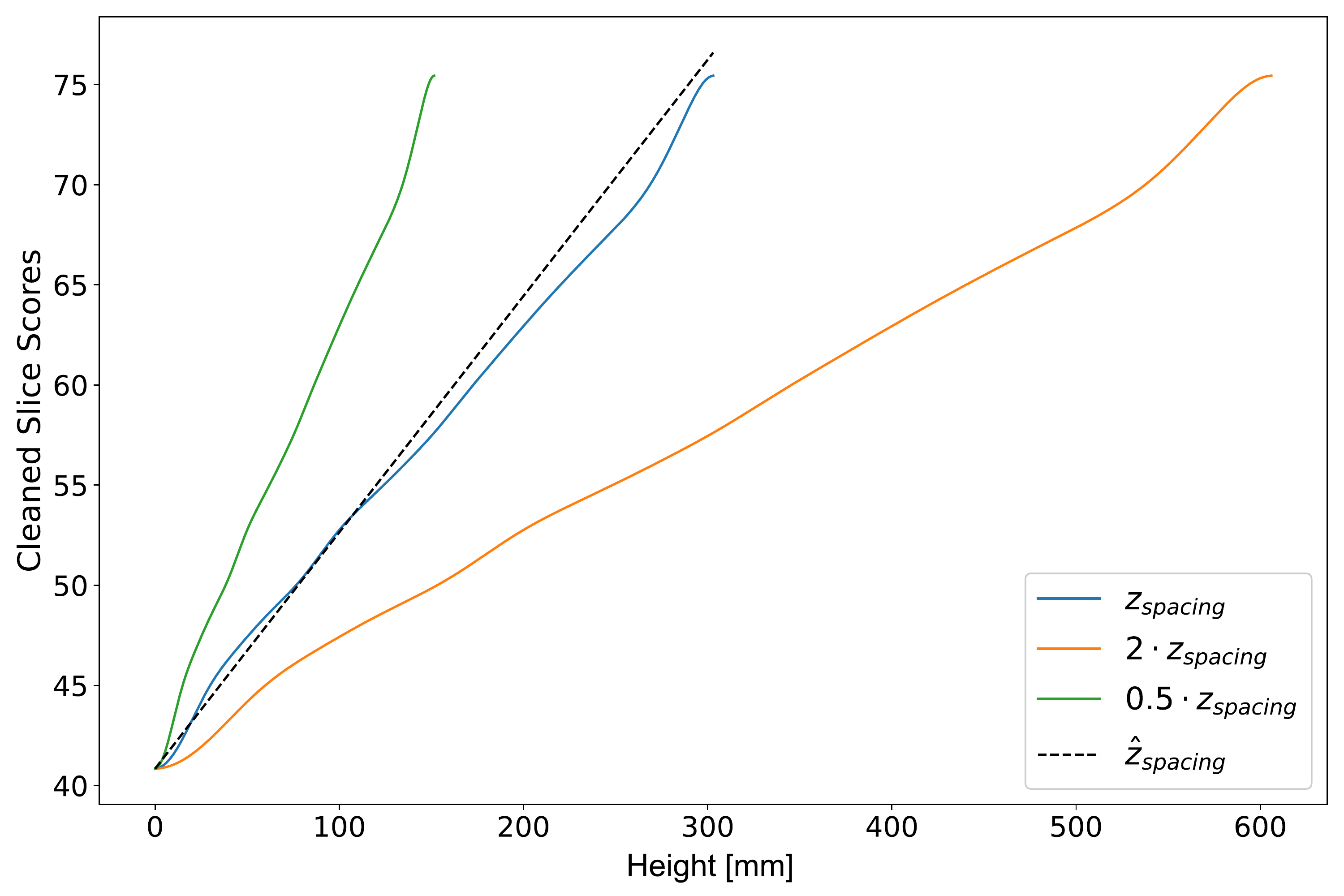}
    \caption{Slice score slopes for an exemplary volume and different z-spacings. The relation between z-spacing and the slope is visible. The dotted line has the expected slope $\bar{m}_{s}$. Relative errors in the z-spacing lead to a deviation in slope. }
    \label{fig:zspacing-slope}
\end{figure}

The expected z-spacing $\hat{z}$ can be calculated based on the slope of the slice score curve and the expected slope $\bar{m}_{s}$. For a volume $X$, we distinguish between two slopes. The slope $m^{\prime}_{X}$ depending on the slice index and the slope $m_{X}$ depending on the height $h$ in mm. The slopes are defined by fitting a linear line to the cleaned slice score curve. Both slopes can be transformed into each other through 
\begin{equation*}
    m_{X} = \frac{1}{z_{X}} \cdot m^{\prime}_{X}. 
\end{equation*}
The expected slope $\bar{m}_{s}$ is declared in dependence of the slice height $h$. Therefore, the expected z-spacing
$\hat{z}_{X}$ for volume $X$ can be calculated via the equation 
\begin{equation}\label{eq:expected-zspacing}
    \hat{z}_{X} =  \frac{m_{X}^{\prime}}{\bar{m}_{s}}.
\end{equation}
Figure \ref{fig:zspacing-slope} shows for an exemplary volume $X$ slice score curves of different z-spacings. Moreover, the linear line for the expected z-spacing is visible. It can be observed that the slope of the expected z-spacing and the observed z-spacing are quite similar. The slopes of the slice score curves with twice or half of the z-spacing can be visually detected as outliers. The slope of the slice score curve is an indicator of the meaningfulness of the z-spacing. 
To apply both data sanity checks for an arbitrary CT-volumes $X$ during test-time, we need to follow the steps: 
\begin{enumerate}
    \item Predict slice scores for CT-volume $X$ with body part regression model 
    \item Obtain cleaned slice scores (see fig. 
    \ref{fig:slice-score-postprocessing}). 
    \item Fit linear line to cleaned slice score curve. 
    \item Check if slope is negative  
    \begin{equation*}
    \begin{split}
        & \text{if: } m_{X} > 0 \Rightarrow
        \text{Natural z-ordering. Indices increase with patient height.} \\ 
        &  \text{if: } m_{X} < 0 \Rightarrow
        \text{Reverse z-ordering. Indices decrease with patient height.} \\ 
    \end{split}
    \end{equation*}

    \item Check if the expected relative error $\hat{\gamma}_{X}$ is valid
    \begin{equation*}
        \begin{split}
        &\text{if: } \hat{\gamma}_{X} > \theta \Rightarrow \text{z-spacing is invalid}\\
        &\text{if: } \hat{\gamma}_{X}  < \theta \Rightarrow \text{z-spacing is valid}
        \end{split}
    \end{equation*}

\end{enumerate}
\subsection{Deployment}\label{sec:deployment}
To make the use cases UC1, UC2, and UC3 better applicable for the clinical community and the research community, a public python package was deployed, and the model was integrated into the medical platform toolkit Kaapana (see sec. \ref{sec:background-deployment}). 

\textbf{Main Functionality: } For the deployment, a main function for inference was implemented. The function creates and saves a body part meta-data file
for CT images in the  \ac{NIfTI} data format. The meta-data files are saved in the 
\ac{JSON} data format. Table \ref{tab:json-meta-data} summarizes the information inside the JSON-file. The threshold $\theta$ for the data sanity check can be customized as well as the body part boundaries for calculating the \textit{body part examined dictionary} and the \textit{body part examined tag}.
\begin{table}
    \centering
    \caption{Explanation of the tags in the JSON body part meta-data file.}
    \begin{tabular}{lll}
        \hline
        Tag & Format & Description  \\
        \hline
         cleaned slice scores & list & cleaned slice scores from figure \ref{fig:slice-score-postprocessing}\\
         unprocessed slice scores & list & transformed unprocessed slice scores \\
         z & list & height in mm\\
         body part examined & dictionary & body part examined dict from section \ref{sec:methods-bpe}\\
         body part examined tag & string & body part examined tag from section \ref{sec:methods-bpe} \\
         look-up table & dictionary & slice score reference table from section \ref{sec:methods-application-deployment}\\
         reverse z-ordering & bool & data sanity check from section \ref{sec:method-dsc} \\
         valid z-spacing & bool & data sanity check from section \ref{sec:method-dsc} \\
         expected slope & float & expected slope $\bar{m}_{s}$ from section \ref{sec:method-dsc} \\
         observed slope & float & observed slope $m_{X}$ from section \ref{sec:method-dsc}\\
         slope ratio & float & $r_{m}$ from equation \ref{eq:zratio} \\
         expected z-spacing & float & $\hat{z}_{X}$ from equation \ref{eq:expected-zspacing}\\
         z-spacing & float & z-spacing of volume\\
         settings & dictionary & used settings to generate the JSON-file \\
         \hline
    \end{tabular}
    \label{tab:json-meta-data}
\end{table} 

 \textbf{Public Model: }To publish the body part regression model the model was trained on a subset of the original training data from section \ref{sec:dataset}. The data from the \textit{whole body CT} study and study \textit{Task051} was excluded based on the user restriction for the data. The validation data and the test data stayed the same.  In total, 2091 volumes were used for training, 100 volumes for validation and 100 volumes for testing. A subset of 42 volumes from the training data was annotated. 

\textbf{Python Package:} For the python package a README was created to describe how to install and use the package. Moreover, jupyter notebook tutorials were provided to explain how to train a body part regression model with the help of the python package and how to use the package for inference of CT volumes in different data formats. 

\textbf{Kaapana: }
For Kaapana, the public model was used. The main function of creating a JSON meta-data file for a NIfTI-volume was built into a Docker container with the help of a Dockerfile. To integrate the Docker container into the Kaapana toolkit, an operator was build upon the Docker container. Additionally, a \ac{DAG} was defined, which concatenates the following operators (see sec. \ref{sec:background-deployment}): 
\begin{enumerate}
    \item get-input-data
    \item dcm-converter
    \item bodpartregression
    \item workflow-cleaner 
\end{enumerate}
All operators, apart from the bodypartregression operator, already existed in Kaapana and are used to load the DICOM files, convert the DICOM files to NIfTI-files, and to clean the temporary data of the workflow. 
The DAG can be triggered through the Kaapana dashboard. The workflow uses the body parts predicted from the DICOM volumes to extend the meta-data database of the platform. This way, even more fine-grained exploration and filtering of data can be achieved.

\chapter{Experiments and Results}
In section \ref{sec:experiments-bpr}, we investigate the effect of different loss functions and data augmentation techniques on the model performance. 
Moreover, the best-performing body part regression model is presented. In section  \ref{sec:results-deployment}, the usefulness of the trained body part regression model is demonstrated through several use cases. Additionally, the deployment of the body part regression model is explained. 

\section{Body Part Regression}\label{sec:experiments-bpr}
Multiple experiments were carried out to choose the best loss function (see sec. \ref{sec:order-loss-results}) and to understand the model behavior better. All experiments were conducted on the same dataset. The dataset consists of 2192 CT volumes for training, 100 CT volumes for validation, and further 100 CT volumes for testing (see sec. \ref{sec:dataset}). Hyperparameters were optimized based on the model performance on the validation set. Final statements about the generalization to unseen data were made based on the model performance on the test set. After running all experiments, the model with the best model performance on the validation set was chosen and carefully evaluated. 

During the experiments, the data augmentations and some parameters were fixed to compare the experiments better and keep the search space small. 
The selected fixed hyperparameter for the data augmentations can be found in table \ref{tab:data-augmentation}. 
For optimization, the Adam optimizer was used, which runs quite robustly \cite{kingma2014adam}. Additionally, a learning rate of $10^{-4}$ was used.
Moreover, the sampled slices per batch and the total sampled slices per CT volume were fixed so that every model sees the same amount of two-dimensional slices per batch and during the whole training procedure.  Each mini-batch contains about 256 axial slices. The sampled slices per batch are given by
\begin{equation*}
    \text{sampled slices per batch} = \text{batch-size} \cdot m, 
\end{equation*}
where $m$ is  the number of sampled slices per volume. From each CT volume about 1920 slices are randomly sampled over the whole training period. The total sampled slices per volume are given by
\begin{equation*}
    \text{total sampled slices per volume} = m \cdot \text{epochs}. 
\end{equation*}
The random seed was fixed to zero. Further, the range from which $\Delta h$ was sampled, was fixed to [5 mm, 100 mm].
Table \ref{tab:fixed-parameters} summarizes the selected hyperparameter settings. 
Training a body part regression model took between 1.5 and 2.5 hours on a Nvidia RTX 2080 ti GPU with 11 GB memory. 
\begin{table}
    \centering
    \caption{Fixed parameters for the body part regression experiments. }
    \begin{tabular}{ll}
         \hline
         Fixed parameter & Value \\
         \hline
         learning rate & $10^{-4}$ \\
         total sampled slices per volume & 1920 \\
         sampled slices per batch & 256 \\
         $\Delta h$-range & [5mm, 100mm] \\
         random seed & 0 \\
         \hline
    \end{tabular}
    \label{tab:fixed-parameters}
\end{table}
\subsection{Loss Functions}\label{sec:order-loss-results}
\begin{figure}
    \centering
    \includegraphics[width=\textwidth]{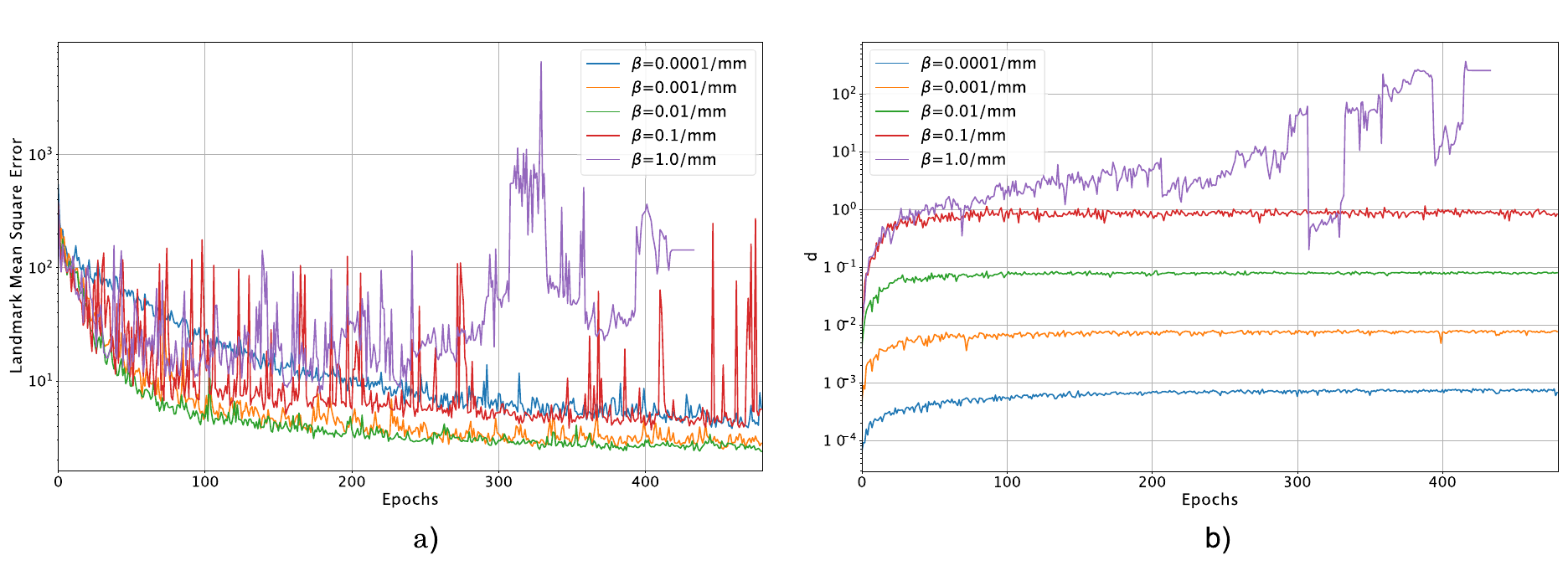}
    \caption{Results for the heuristic order loss experiment. In figure a) the influence of $\beta$ on the training convergence can be seen.The training convergence is measured by the LMSE on the validation set. In figure b) the procedure of the normalization constant $d = \frac{1}{100}(\bar{s}_{l_{12}} - \bar{s}_{l_{1}})$ during training is visible.} 
    \label{fig:loh-experiment-mse}
\end{figure}

In the following, the classification order loss (see sec. \ref{sec:order-loss}) and the heuristic order loss (see sec. \ref{sec:order-loss}) will be compared. Further, we will investigate the distance loss.
To be able to compare both order losses, first a good parameter $\beta$ for the newly proposed heuristic order loss needs to be found. The optimal $\beta$ was  searched on a logarithmic grid. Table \ref{tab:order-loss-beta} contains the search range, along with the other hyperparameters for this experiment. 
\begin{table}
    \centering
    \caption{Experiment setup for finding a good parameter $\beta$ for the heuristic order loss.}
    \begin{tabular}{lll}
         \hline
         Parameter & Description & Value[s]  \\
         \hline
         m & sampled slices per volume & 4 \\
         batch-size & data-items in one mini-batch & 64 \\
         epochs & training epochs & 480 \\
         $\beta$ & parameter for heuristic order loss & 0.0001, 0.001, 0.01, 0.1, 1 \\
         \hline
    \end{tabular}
    \label{tab:order-loss-beta}
\end{table} 
As discussed in section \ref{sec:order-loss}, the parameter $\beta$ is an additional constraint parameter, which influences how long the linearity condition between $\Delta h$ and $\Delta s$ holds. Thus, a greater $\beta$ leads to a weaker constraint. 
Figure \ref{fig:loh-experiment-mse} a) shows the \ac{LMSE} curves during training for the different values of $\beta$. For $\beta=1/\text{mm}$, the LMSE deteriorates, the constraint is too low.
Additionally, the absolute slice score predictions are drifting apart during training.
For $\beta=0.0001/\text{mm}$ the linear constraint is too high, therefore the LMSE converges rather slowly. This would result in longer training times. Moreover, the parameter $\beta$ is the 
proportionality factor between $\Delta s$ and $\Delta h$ (see eq. \ref{eq:proportionality-beta}). Therefore, we can observe in figure \ref{fig:loh-experiment-mse} b) that the normalization constant $d$ increases with increasing $\beta$. 
Table \ref{tab:loh-experiment} shows a summary of the evaluation metrics for all trained models. It can be seen that $\beta=0.01/\text{mm}$ performs best for the analyzed metrics. This parameter is used for further experiments.
\begin{table}
    \centering
    \caption{
    Influence of the hyperparameter $\beta$ on the model performance evaluated with the LMSE metric and the accuracy defined in section \ref{sec:methods-evaluation-methods}.}
    \begin{tabular}{lll}
        \hline
         $\beta$ &$\bar{\phi}_{\text{val}} \pm \sigma_{\bar{\phi}}$ &  Accuracy in \% \\
         \hline
         0.0001 & 6.3 $\pm$ 0.7  &  89.4 \\	
         0.001 & 3.01 $\pm$ 0.27 &   93.9 \\ 
         0.01 & \textbf{2.39 $\pm$ 0.23} &      \textbf{94.0}  \\  
         0.1 & 6.6 $\pm$ 1.3  &       91.7  \\ 
         1   & -- &       -- \\
         \hline
    \end{tabular}
    \label{tab:loh-experiment}
\end{table} 
Next, we compare the heuristic order loss with the classification order loss. The heuristic order loss and classification order loss are defined in equation \ref{equation:heuristic-order-loss} and equation \ref{eq:calssification-order-loss}.

\textbf{Instability of the Classification Order Loss: }During initial experiments, an instability of the classification order loss was observed. The gradients and the loss sometimes became "NaN". The reason for the instability is that the classification order loss is not defined if $\sigma(\Delta s) = 0$. Mathematically, this can not happen, because the sigmoid function can not map to zero. But numerically $\sigma(\Delta s)$ can be rounded to zero, if $\Delta s \ll 0$. To stabilize the training, $\sigma(\Delta s) = 0$ was ignored and the updated order loss function is given by 
\begin{equation} \label{eq:updated-calssification-order-loss}
    \begin{split}
    L_{\text{order}}^{c} &= -\frac{1}{N^{\star}}\sum_{i=1}^{B}\sum_{j=1,\text{if }\sigma(\Delta s_{ij}) \neq 0}^{m-1} \text{ln} \ \sigma(\Delta s_{ij}),  \\
    N^{\star} &= \sum_{i=1}^{B}\sum_{j=1}^{m-1} [\sigma(\Delta s_{ij}) \neq 0].
    \end{split}
\end{equation}
For comparing both loss functions, the following loss construction was used
\begin{equation}\label{eq:loss-constitution-alpha}
    L = L_{\text{order}} + \alpha \cdot L_{\text{dist}}, 
\end{equation}
with the classification order loss (eq. \ref{eq:updated-calssification-order-loss}) or the heuristic order loss (eq. \ref{equation:heuristic-order-loss}) as order loss function and the distance loss function (eq. \ref{eq:distance-loss}) as additional constraint. 

\textbf{Experimental Setup: } To ensure a fair comparison of the methods, we will first tune the hyperparameters.
Table \ref{tab:order-loss-experiment-setup} describes the experimental setup. Based on the fixed total sampled slices, slices per batch, and chosen $m$, the corresponding batch size and epoch were calculated (see tab. \ref{tab:fixed-parameters}). For the classification order loss, Yan et al. used $\alpha=1$ for their analysis \cite{yan2018deep}, therefore apart from the experiment with $\alpha=0$, the search space was centered around one. For the sampled slices per data instance $m$, Yan et al. chose $m=8$. Therefore, the search space for $m$ was centered around 8. 
An additional experiment was carried out without any order loss and only with the distance loss to understand the impact of the distance loss better.

\begin{figure}
    \centering
    \includegraphics[width=\textwidth]{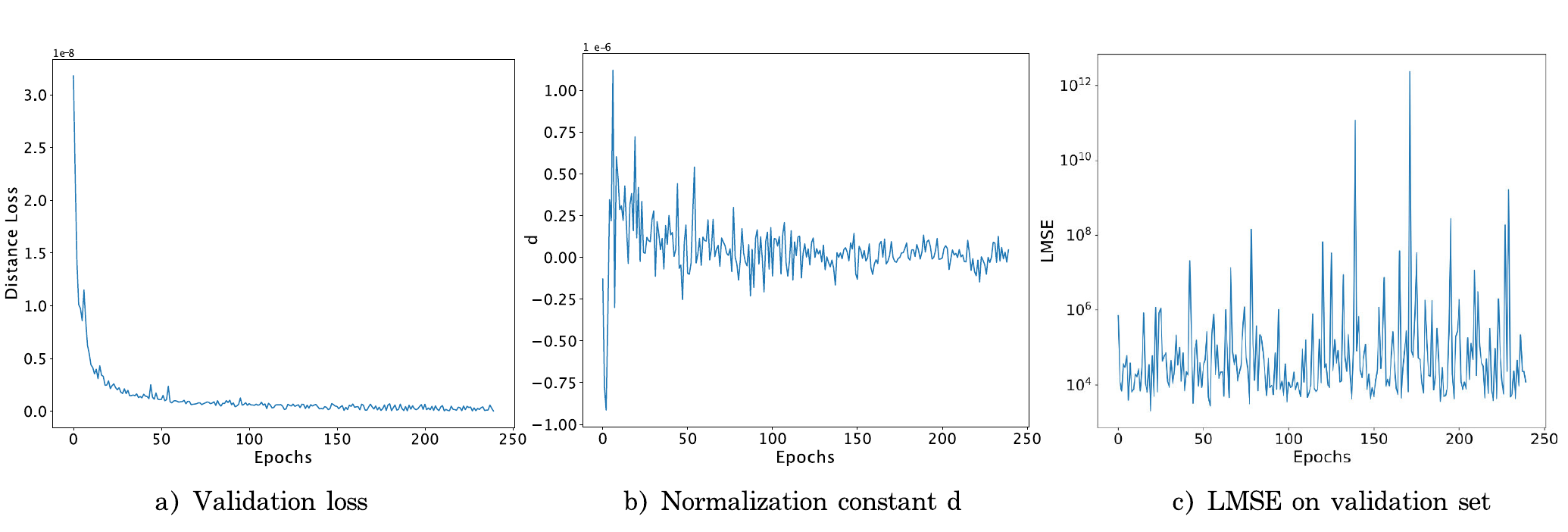}
    \caption{Training behavior of model with distance loss as only loss function and with the hyperparameters $\alpha=1$ and $m=8$. The normalization constant $d$ is given by $d = \frac{1}{100}(\bar{s}_{l_{12}} - \bar{s}_{l_{1}})$.}
    \label{fig:ldist}
\end{figure}

\begin{table}
    \centering
    \caption{Experimental setup for comparing the performances between the classification order loss and the heuristic order loss. The parameters $\alpha$, which controls the distance loss and the sampled slices per updated step $m$ were varied. }
    \begin{tabular}{lll}
         \hline
         Order loss & Parameters & Values  \\
         \hline
         heuristic order loss & (m, batch-size, epoch) &
         (4, 64, 480), (8, 32, 240), (12, 21, 160) \\
         & $\alpha$ & 0, 0.01, 0.1, 1\\
         & $\beta$ & 0.01 \\
         \hline
         classification order loss &  (m, batch-size, epoch) &
         (4, 64, 480), (8, 32, 240), (12, 21, 160) \\
         & $\alpha$ & 0, 0.8, 1.0, 1.2 \\
         \hline
         no order loss & (m, batch-size, epoch) & (8, 32, 240) \\
         & $\alpha$ & 1 \\
         \hline 
    \end{tabular}
    \label{tab:order-loss-experiment-setup}
\end{table} 
\textbf{Experimental Results: }In table \ref{tab:order-loss-experiment-results} the LMSE and the accuracy for each trained model is shown. The model which only uses the $L_{\text{dist}}$ as loss function is by far the worst. Figure \ref{fig:ldist} shows the learning procedure of this model. Based on figure \ref{fig:ldist} b), we can observe that the model predicts for all landmarks nearly the same value.
Moreover, in figure \ref{fig:ldist} c) we observe that the LMSE has a high variance and does not decrease in the course of training, although the validation loss $L_{\text{dist}}$ decreases.
We can conclude that the model performance is so bad because an easy minimum for $L_{\text{dist}}$ is to predict all slices to the same value. Without an additional order loss, nothing prevents the model from converging to this state. The distance loss $L_{\text{dist}}$ needs to be seen as an additional constraint, which encourages a linear behavior between slice index and slice scores. However, as an independent loss, it is not suitable for body part regression. It is interesting to notice that the best model for the heuristic order loss includes no additional distance loss $\alpha=0$. This supports the hypothesis that the heuristic order loss has the linear constraint already incorporated, and therefore the additional constraint through the distance loss is useless. 
Based on the complete hyperparameter tuning results in table \ref{tab:order-loss-experiment-results}, the following research questions are addressed: 
\begin{enumerate}
    \item[] \textbf{RQ1:} Is the best model trained with the heuristic order loss significantly better than the model trained with the classification order loss? 
    \item[] \textbf{RQ2:} Do the parameters $m$ and $\alpha$ play a significant role in the model performance? 
\end{enumerate}
\textbf{Influence of Different Order Losses on Performance:} To address \textbf{RQ1}, the performance on the test set for the best model for each order loss was calculated (see tab. \ref{tab:order-loss-test}).
Based on the LMSE values $\bar{\phi}_{test,13}$ of model 13  and $\bar{\phi}_{test,6}$ of model 6, we can perform the z-test as described in section \ref{sec:lmse}: 
\begin{equation*}\label{eq:test-significance}
    \frac{3.3 - 2.65}{\sqrt{0.4^{2} + 0.28^{2}}} \approx 1.3 \ < 1.96. 
\end{equation*}
Therefore, for a z-test with a significance level of 5\% (corresponds to a standard deviation of 1.96), the models perform not significantly different from each other. We can conclude that the heuristic order loss model performs equally well as the combination of classification order loss and distance order loss. Both approaches are valid. In the further analysis, we will proceed with the heuristic order loss because the local linear constraint is theoretically better motivated through the parameter $\beta$ than for the distance loss. Moreover, the heuristic order loss incorporates the physical distance $\Delta h$ into the constraint. Therefore, it should be less dependent on the sampling strategy of $\Delta h$. 

\begin{figure}
     \centering
     \includegraphics[width=\textwidth]{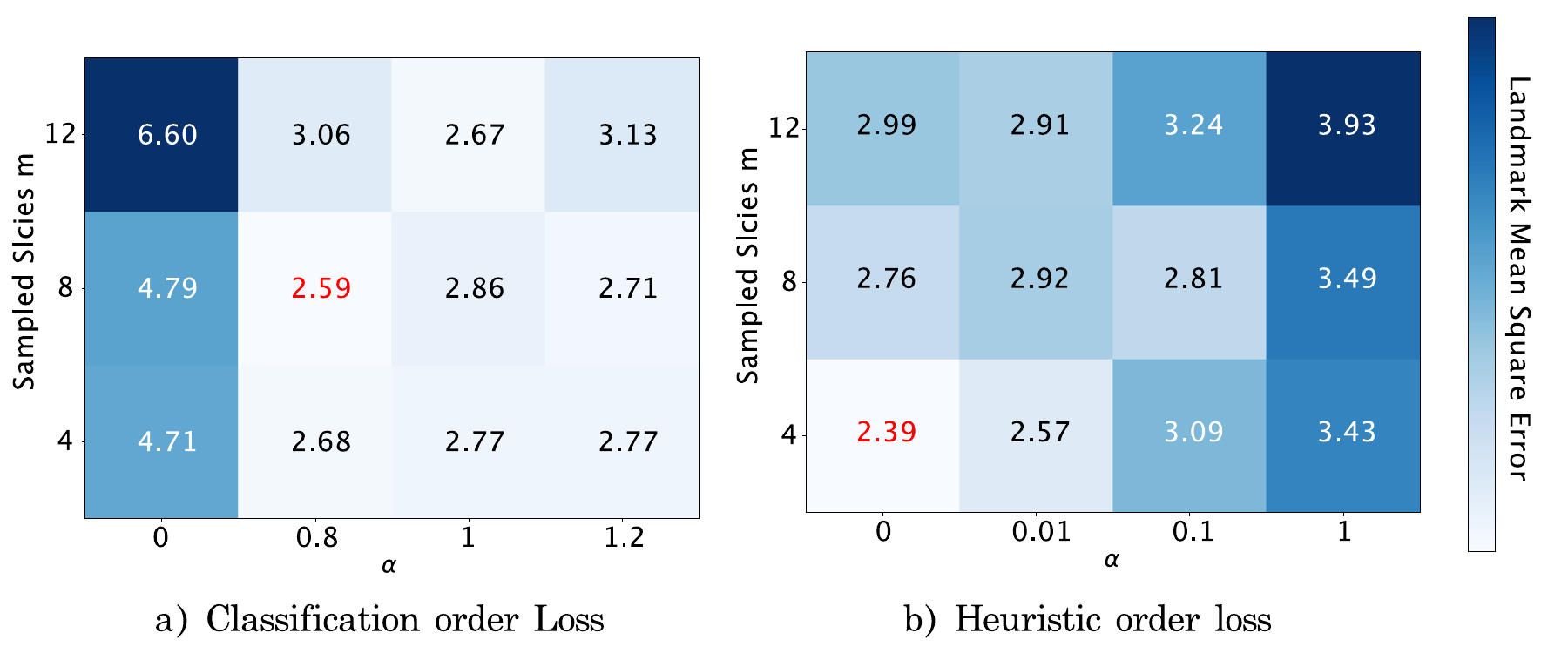}
     \caption{Visualization of influence of hyperparameter $\alpha$ and the sampled slices $m$ on the model performance for the heuristic order loss and the classification order loss. The model performance is given by the Landmark Mean Square Error. The red numbers represent the model with the best LMSE score.}
    \label{fig:m-alpha-analysis}
 \end{figure}

\textbf{Influence of Hyperparameters on Performance:} To address \textbf{RQ2} and investigate the influence of the parameters $m$ and $\alpha$ on the model performance, figure \ref{fig:m-alpha-analysis} visualizes  the different model performances in dependence of $m$ and $\alpha$. In figure b) it can be seen that the sampled slices $m$ have an impact on the model performance. For $m=4$ the model performances are better than for $m=12$. 
 The sampled slices $m$ influence the compilation of a batch and the sampled physical distances $\Delta h$. The ideal parameter $m$ may be dependent on the used training data. For the classification order loss, an impact of $m$ on the model performance is not clearly visible. 
Furthermore, figure \ref{fig:m-alpha-analysis} shows that the parameter $\alpha$ has a big impact on the model performance for models trained with the classification order loss. If $\alpha$ is too small, the linear constraint is too weak, and the model performance deteriorates. For $\alpha=0$ the model performance is much worse than for $\alpha \in [0.8, 1, 1.2]$. For the heuristic order loss, the data confirm that the parameter $\alpha>0$ is not beneficial for training. Big values of $\alpha$ lead to bad model performances.
This strengthens the hypothesis that the linear constraint approach is already incorporated in the heuristic order loss. Therefore, adding a distance loss does not benefit the model. Therefore, the distance loss will not be used for the heuristic order loss in the following sections. 

\textbf{Choice of Evaluation Metric:} Another exciting fact about table \ref{tab:order-loss-experiment-results} is that the model ranking resulting from the LMSE metric is not the same as the order resulting from the accuracy metric. For example, model 13 performs significantly better than model 23 based on the LMSE (z-test with a significance level of 5 \%), but model 23 has higher accuracy on the validation set than model 13.  In table \ref{tab:compare-lookups} the LMSE per landmark is shown for model 23 and model 13 on the validation set. 
Model 23 has an overall better performance on the class boundary landmarks, which are used to compute the accuracy. This behavior leads to a better performance of model 23 than model 13 in terms of the accuracy metric, although the overall performance on the different landmarks is not better. 
In the following, the LMSE is used as the primary evaluation metric because it accounts for all landmarks. 
\begin{table}[t]
    \centering
    \caption{Comparing LMSE $\bar{\phi}$ per landmark on validation dataset for model 13 and model 23. For each landmark, the better LMSE value is visualized in bold letters. Through the checkmarks, the landmarks which are used for calculating the accuracy are marked.}
    \begin{tabular}{lrrl}
        \hline
         Landmark name& $\bar{\phi}_{\text{23}} \pm \sigma_{\bar{\phi}}$ & 
         $\bar{\phi}_{\text{13}} \pm \sigma_{\bar{\phi}}$ & \makecell[l]{Class \\ boundary} \\
         \hline
            pelvis-start &  \textbf{2.70 $\pm$ 0.53} &  3.10 $\pm$ 0.51 &      \checkmark \\
            femur-end    &  \textbf{0.47 $\pm$ 0.07} &  0.56 $\pm$ 0.11 &            \\
            L5           &  \textbf{1.53 $\pm$ 0.29} &  2.43 $\pm$ 0.46 &      \checkmark \\
            L3           &  3.88 $\pm$ 0.56 &  \textbf{2.84 $\pm$ 0.39} &            \\
            L1           &  4.67 $\pm$ 0.67 &  \textbf{2.59 $\pm$ 0.46} &            \\
            Th11         &  4.79 $\pm$ 0.99 &  \textbf{2.58 $\pm$ 0.61} &      \checkmark \\
            Th8          &  4.49 $\pm$ 0.99 &  \textbf{2.79 $\pm$ 0.69} &            \\
            Th5          &  3.75 $\pm$ 0.87 &  \textbf{2.06 $\pm$ 0.43} &            \\
            Th2          &  \textbf{2.13 $\pm$ 0.37} &  2.14 $\pm$ 0.48 &      \checkmark \\
            C6           &  1.71 $\pm$ 0.33 &  \textbf{1.31 $\pm$ 0.25} &            \\
            C1           &  1.55 $\pm$ 0.49 &  \textbf{1.16 $\pm$ 0.33} &      \checkmark \\
            eyes-end     &  \textbf{1.48 $\pm$ 0.35} &  4.82 $\pm$ 0.82 &      \checkmark \\
            \hline
            \hline
            mean &     2.76 & \textbf{2.37} & \\
            \makecell[l]{mean for class \\ boundary landmarks}
            & \textbf{2.36} & 2.71 & \\
            accuracy & \textbf{94.2 \%} & 94.0 \% & \\
            \hline
    \end{tabular}
    \label{tab:compare-lookups}
\end{table} 
\begin{table}
    \centering
    \caption{Summary of model performances for best heuristic order loss model and best classification order loss model.}
    \begin{tabular}{llllll}
        \hline
         Model name&  Order loss& $m$ & $\alpha$ & $\bar{\phi}_{\text{val}} \pm \sigma_{\bar{\phi}}$ & $\bar{\phi}_{\text{test}} \pm \sigma_{\bar{\phi}}$\\
         \hline
         model 13 &  $L_{\text{order}}^{h}$   & 4 & 0   & 2.39 $\pm$ 0.23 & 2.65 $\pm$ 0.28 \\ 
         model 6  &  $L_{\text{order}}^{c}$   & 8 & 0.8 & 2.59 $\pm$ 0.27 & 3.3 $\pm$ 0.4 \\
         \hline 
    \end{tabular}
    \label{tab:order-loss-test}
\end{table} 

\begin{table}
    \centering
    \caption{LMSE $\bar{\phi}$ with the standard error $ \sigma_{\bar{\phi}}$  and accuracy $\psi$ for order loss experiment on the validation set}
\begin{tabular}{llllll}
\hline
Model &       Order-loss &  $m$ &  $\alpha$ & $\bar{\phi}_{\text{val}} \pm \sigma_{\bar{\phi}}$ & Accuracy in \% \\
\hline
model 1  &         $L_{\text{order}}^{c}$&           4 &   0.00 &  4.7 $\pm$ 0.4 &        93.5 \\
model 2  &         $L_{\text{order}}^{c}$&           4 &   0.80 &  2.68 $\pm$ 0.26 &        94.8 \\
model 3  &         $L_{\text{order}}^{c}$&           4 &   1.00 &   2.8 $\pm$ 0.3 &        94.6 \\
model 4  &         $L_{\text{order}}^{c}$&           4 &   1.20 &  2.77 $\pm$ 0.26 &        94.5 \\
model 5  &         $L_{\text{order}}^{c}$&           8 &   0.00 &  4.8 $\pm$ 0.5 &        92.7 \\
model 6  &         $L_{\text{order}}^{c}$&           8 &   0.80 &  \textbf{2.59 $\pm$ 0.27} &        94.6 \\
model 7  &         $L_{\text{order}}^{c}$&           8 &   1.00 &  2.9 $\pm$ 0.3 &        94.5 \\
model 8  &         $L_{\text{order}}^{c}$&           8 &   1.20 &  2.71 $\pm$ 0.28 &        94.7 \\
model 9  &         $L_{\text{order}}^{c}$&          12 &   0.00 &   6.6 $\pm$ 0.7 &        92.1 \\
model 10 &         $L_{\text{order}}^{c}$&          12 &   0.80 &  3.06 $\pm$ 0.3 &        94.0 \\
model 11 &         $L_{\text{order}}^{c}$&          12 &   1.00 &  2.7 $\pm$ 0.28 &        94.4 \\
model 12 &         $L_{\text{order}}^{c}$&          12 &   1.20 &  3.1 $\pm$ 0.3 &        94.3 \\
\hline 
model 13 &         $L_{\text{order}}^{h}$&           4 &   0.00 &  \textbf{2.39 $\pm$ 0.23} &       94.0 \\
model 14 &         $L_{\text{order}}^{h}$&           4 &   0.01 &  2.57 $\pm$ 0.22 &        94.4 \\
model 15 &         $L_{\text{order}}^{h}$&           4 &   0.10 &  3.09 $\pm$ 0.28 &        93.9 \\
model 16 &         $L_{\text{order}}^{h}$&           4 &   1.00 &  3.4 $\pm$ 0.4 &        93.4 \\
model 17 &         $L_{\text{order}}^{h}$&           8 &   0.00 &  2.76 $\pm$ 0.27 &        94.6 \\
model 18 &         $L_{\text{order}}^{h}$&           8 &   0.01 &  2.9 $\pm$ 0.3 &        94.0 \\
model 19 &         $L_{\text{order}}^{h}$&           8 &   0.10 &   2.8 $\pm$ 0.3 &        94.5 \\
model 20 &         $L_{\text{order}}^{h}$&           8 &   1.00 &  3.5 $\pm$ 0.4 &        94.0 \\
model 21 &         $L_{\text{order}}^{h}$&          12 &   0.00 &  2.99 $\pm$ 0.27 &        93.8 \\
model 22 &         $L_{\text{order}}^{h}$&          12 &   0.01 &  2.91 $\pm$ 0.29 &        94.3 \\
model 23 &         $L_{\text{order}}^{h}$&          12 &   0.10 &  3.2 $\pm$ 0.3 &        94.2 \\
model 24 &         $L_{\text{order}}^{h}$&          12 &   1.00 &  3.9 $\pm$ 0.4 &        93.6 \\
\hline
model 25 &               ---     &           8 &   1.00 & 11100 $\pm$ 800 & 5.0 \\
\hline
\end{tabular} 
    \label{tab:order-loss-experiment-results}
\end{table} 

\subsection{Data Augmentations}\label{sec:data-augmentation-results}
\begin{table}
    \centering
    \caption{Grouping of used data augmentation techniques from table \ref{tab:data-augmentation}.}
    \begin{tabular}{ll}
        \hline
         Data augmentation group & Augmentations  \\
         \hline
         physical transforms & Flip, Transpose, ShiftScaleRotate \\
         intensity transforms & Contrast, Brightness\\
         image quality transforms & GaussianNoise, GaussianBlur\\
         add frame transform & AddFrame\\
         \hline
    \end{tabular}
    \label{tab:data-augmentation-groups}
\end{table} 
\begin{table}
    \centering
    \caption{Data augmentation results. The LMSE $\bar{\phi}$ was calculated on the validation set and the test set. The test statistic $t$ was calculated based on equation \ref{eq:teststatistics} and the significance test was calculated based on section \ref{sec:lmse}.  }
    \begin{tabular}{llllll}
    \hline
        Model       & Unused data augmentations & $\bar{\phi}_{\text{val}} \pm \sigma_{\bar{\phi}}$& $\bar{\phi}_{\text{test}} \pm \sigma_{\bar{\phi}}$& $t$ & significant\\
        \hline
        model 1 &                 --- &  \textbf{2.39 $\pm$ 0.23} &      \textbf{2.653 $\pm$ 0.279}  & 0.0 & \\
        model 2 & add frame transform &  3.08 $\pm$ 0.26&      3.803 $\pm$ 0.402  & 2.4& \checkmark\\
        model 3 & physical transforms &  3.0 $\pm$ 0.3 &      3.835  $\pm$ 0.360 & 2.6 & \checkmark\\
        model 4 & image quality transforms &  3.5 $\pm$ 0.3 & 4.297 $\pm$ 0.484 & 2.9& \checkmark\\
        model 5 & intensity transforms &   3.5 $\pm$ 0.4 &    4.619 $\pm$ 0.552 & 3.2& \checkmark\\
        model 6 &                  all &  4.9 $\pm$ 0.6 &     6.496  $\pm$ 1.066  & 3.5 & \checkmark\\
        \hline
    \end{tabular}
    \label{tab:data-augmentation-results}
\end{table}
In this section, the influence of different data augmentation techniques on the model performance is investigated. For this purpose, a leave-one-out experiment was designed. First, the data augmentation techniques were grouped into data augmentation classes. Afterward, several models were trained where for each model, one group was left out. Table \ref{tab:data-augmentation-groups} summarizes the defined data augmentation groups. 
By leaving one group out of the training, we can analyze if the impact of the data augmentation group is crucial for the model performance. 
For conducting this experiment, the evaluated model from section \ref{sec:order-loss} was used, where $\alpha=0$ and $m=4$. 

For each of the data augmentation groups the model performance to the original model was compared with a z-test (see sec. \ref{sec:lmse}).
Under the significance level of 5 \% the tests were performed.  
The results of the experiment and the test statistic $t$ can be seen in table \ref{tab:data-augmentation-results}. The model performances $\bar{\phi}$ on the test set were used for comparison.
Based on table \ref{tab:data-augmentation-results} it can be seen, that for all models from 2 to 6, the model performance is significantly reduced compared to the original model 1. 
Especially, without any data augmentation technique, the model performance is significantly worse than with all the data augmentation techniques. The LMSE of the model without data augmentation techniques (model 6) is three times bigger than the performance of model 1. 
This suggests that data augmentations are essential for training a good body part regression model, which fulfills the independence condition.

\subsection{Model Evaluation}\label{sec:model-evaluation} 
The best model on the validation set with $\alpha=0$ and $m=4$ will be qualitatively and quantitatively evaluated in this section. The validation loss, training loss and validation LMSE during training can be found in figure \ref{fig:best-model-loss-curves}. 
On the validation set, the model has a final LMSE $\bar{\phi}$ with the standard error  $\sigma_{\bar{\phi}}$ and the accuracy $\psi$ of 
\begin{equation*}
    \bar{\phi}_{\text{val}} = 2.39 \pm 0.23, \quad \psi_{\text{val}} = 94.0 \%, 
\end{equation*}
and 
\begin{equation*}
    \bar{\phi}_{\text{test}} = 2.65 \pm 0.28, \quad \psi_{\text{test}} = 94.7 \%, 
\end{equation*}
on the test set. 
In table \ref{tab:slice-score-reference-table-train-val} the \textbf{slice score reference table} for all defined landmarks on the training and validation set can be found.
The slice scores were linearly transformed so that the landmark pelvis-start is mapped to 0 and the eyes-end landmark maps to 100. 
In section \ref{sec:architecture} we defined two criterions, which the body part regression function should fulfill: 
\begin{enumerate}
    \item \textbf{Independence condition: }The same anatomical region should map to the same score, independently of patient and study. 
    \item \textbf{Monotony condition: }The anatomical slice scores should increase monotonously with height.
\end{enumerate}
Table \ref{tab:slice-score-reference-table-train-val} shows that the standard deviations $\sigma_{s}$ of the predicted scores lie for the bone landmarks between 1 and 3.5. The small standard deviations compared to the overall slice score range suggests that the independence condition is fulfilled. Moreover, the mean slice scores $\bar{s}$ are increasing with increasing patient height. Therefore, the model fulfills both established conditions.  

\begin{figure}
    \centering
    \includegraphics{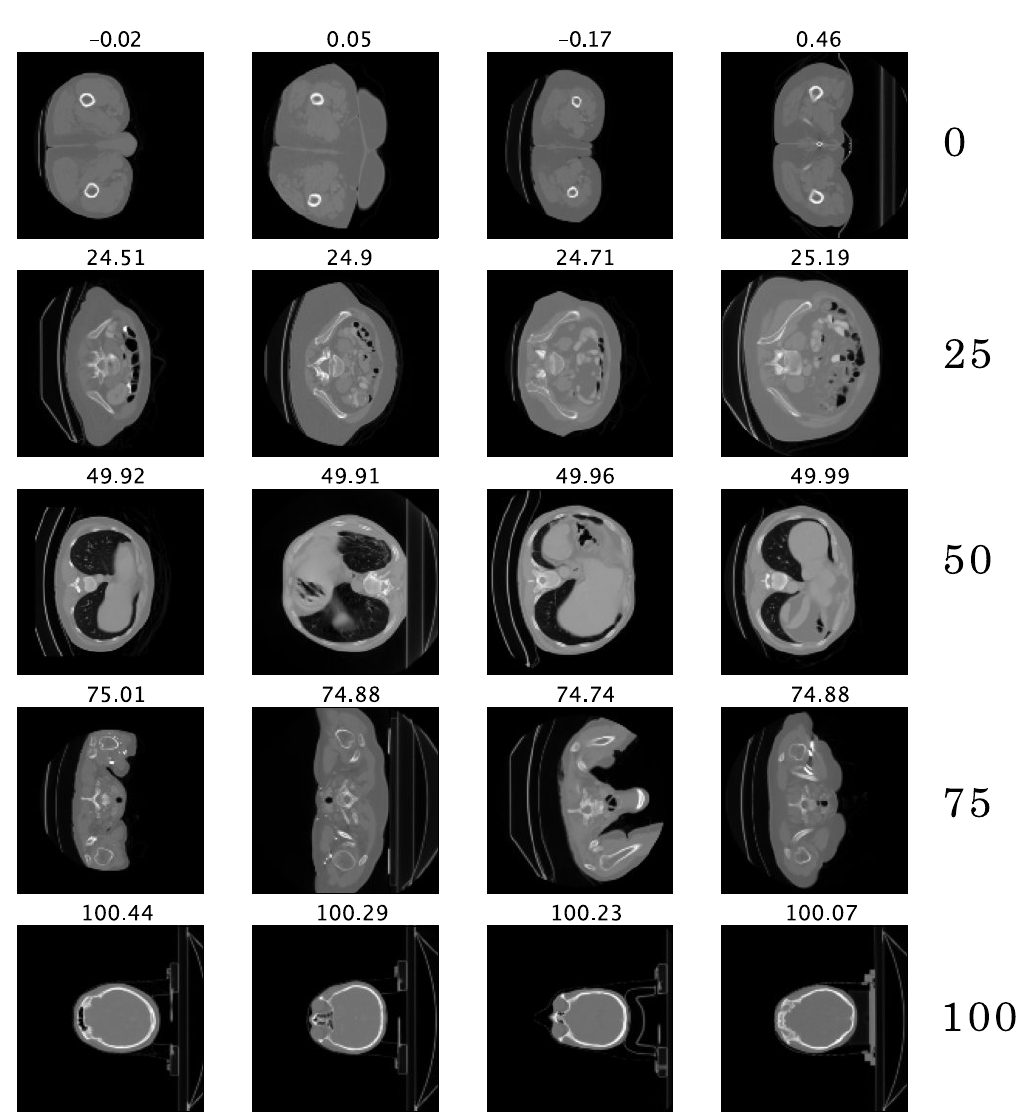}
    \caption{Sampled slices from the test dataset with scores around 0, 25, 50, 75 and 100.  From top to bottom, different scores are visualized, and from left to right, different example slices from the test set are shown. At the top of each image, the corresponding predicted slice score can be found.}
    \label{fig:model-evaluation-nearby-slices}
\end{figure}

\begin{figure}
    \centering
    \includegraphics[width=\textwidth]{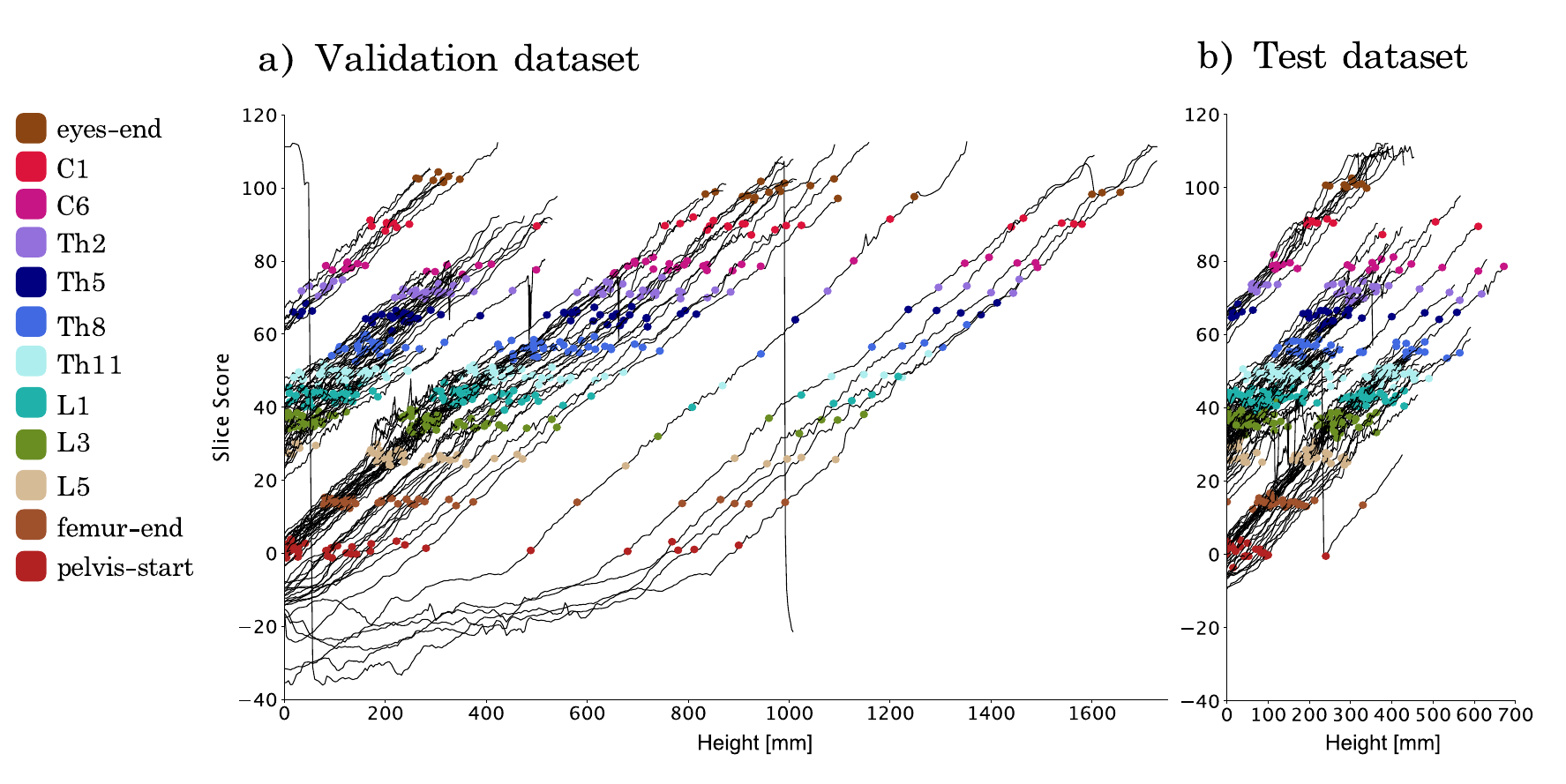}
    \caption{Slice score predictions for volumes from the test and validation set. The landmark positions are labeled with different markers.}
    \label{fig:evaluate-volumes}
\end{figure}

Table \ref{tab:landmarks-model-performance} depicts that the organ landmarks have a mean standard deviation of $\bar{\sigma_{s}} = 3.98$
and the bone landmarks have a mean standard deviation of $\bar{\sigma_{s}} = 2.43$. Compared to the organ landmarks, the standard deviation of the bone landmarks is smaller, this can be explained with the fact that bones have a lower variability across humans compared to soft tissue and organs. 
\begin{table}
    \centering
    \caption{Mean standard deviation  $\bar{\sigma}_{s}$ of score predictions for landmark groups.
    The means $\bar{\sigma}_{s}$ were calculated with the data of table \ref{tab:slice-score-reference-table-train-val}.}
    \begin{tabular}{lll}
        \hline
        Landmarks & $\bar{\sigma}_{s} \pm \sigma_{\bar{\sigma}_{s}}$\\
        \hline
        all landmarks & 2.64 $\pm$ 0.15\\
        evaluation landmarks & 2.30 $\pm$ 0.19\\
        bone landmarks & 2.43  $\pm$ 0.11 \\
        organ landmarks & 3.98 $\pm$ 0.53 \\
        \hline
    \end{tabular}
    \label{tab:landmarks-model-performance}
\end{table} 
In figure \ref{fig:model-evaluation-nearby-slices}, exemplary slices from the test set with predicted slice scores close to the values 0, 25, 50, 75 and 100 are visualized. Different scores are visualized from top to bottom, and from left to right, different example slices from the test set are shown.  Because each row shows similar body regions, we can qualitatively see that similar scores correspond to similar body parts. The visible body regions also match with the results of table \ref{tab:slice-score-reference-table-train-val}. The score 0 maps to the start of the pelvis,  the score 25 maps between the end of the femur bone and the end of the pelvis, score 50 corresponds to slices at the beginning of the lung, score 75 maps to the end of the lung and score 100 correspond to upper head slices.
In figure \ref{fig:evaluate-volumes} the slice score is plotted against the height for all CT volumes in the test and validation dataset. The slice scores which correspond to annotated evaluation landmarks are labeled with colored dots. The variance in y-intercepts reflects the different starting points of the imaged body range. Each curve's minimum and maximum value is an indicator for the body range seen inside the volume. As expected, it can be seen that the slice scores monotonously increase with height. Moreover, it can be observed that the slopes of the different curves are quite similar.  
The variance in slopes can be explained by variance in body heights. 
For some volumes, an unexpected drop or rise in the slice score curve can be observed before the start of the pelvis and after the landmark eyes-end. This can happen if empty slices without any imaged body part exist before or after the landmarks pelvis-start and eyes-end or if body parts outside the network's scope are visible, for example, the feet. 

\begin{figure}
    \centering
    \includegraphics[width=\textwidth]{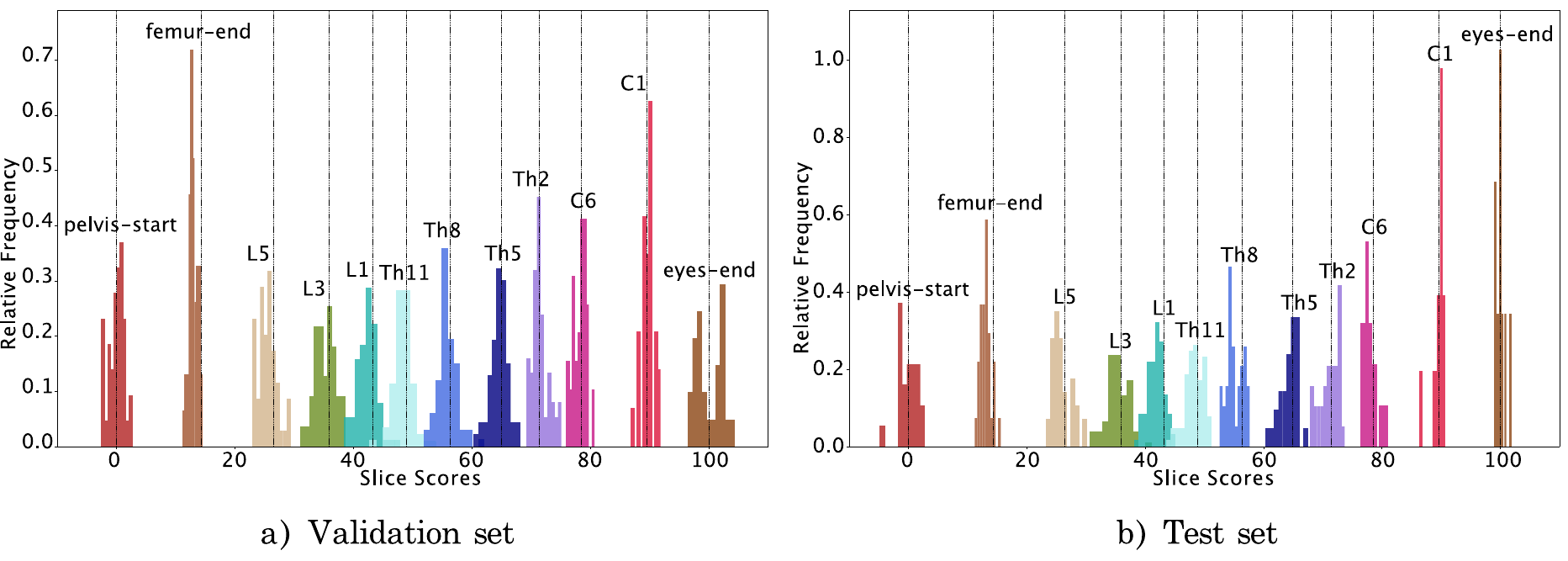}
    \caption{Relative frequency of slice scores on evaluation landmarks for the test and the validation dataset. Each histogram represents the distributions of the predicted slice scores at a specific landmark. The black lines represent the expected slice scores of the landmarks given by the slice score reference table.}
    \label{fig:my_label}
    \label{fig:evaluate-landmarks}
\end{figure}

Based on the landmarks in figure  \ref{fig:evaluate-volumes} it can be observed that similar landmarks are mapped to similar slice scores. This confirms our observations from table \ref{tab:slice-score-reference-table-train-val}. 
In the following, we will refer to predicted scores at a landmark as \textbf{landmark slice scores}. For an ideal model and no intrinsic anatomical variance across patients, the landmark slice scores for one landmark would be the same, and we would expect to have no variance in the landmark slice score. Therefore, we want to achieve a landmark slice score distribution with low variance. Additionally, we want to be able to distinguish the landmarks based on the landmark slice scores. Figure \ref{fig:evaluate-landmarks} shows that for most landmarks, we can separate the slice scores and that the landmark slice score histograms have a relatively low variance. Moreover, it can be observed that some landmarks have a higher landmark slice score variance than other landmarks. Apart from poor landmark annotations, the reason for a higher landmark slice score variance can be explained by higher anatomical variance at the landmark.

\subsection{State-of-the-Art Comparison}\label{sec:sota-comparison}
As discussed in the related work section, this thesis is built upon the \ac{SSBR} model introduced in 2018 by Yan et al. \cite{yan2018unsupervised}. In this section, we compare the \ac{SSBR} model from Yan et al. with the model trained in this thesis (referred as \textbf{$\text{SSBR}^{\star}$}). Furthermore, we compare the trained model with the BUSN model, which uses a 2-stage approach to improve the results of the \ac{SSBR} model. It was introduced by Tang et al. in 2021 \cite{tang2021body} (see sec. \ref{sec:busn}). 

\textbf{Original SSBR Model: }
In table \ref{tab:ssbr-parameters} the parameters which were used to run the grid search and to find the best parameters for the \ac{SSBR} model are shown. For the \ac{SSBR} models, the slices are sampled from the CT volumes based on a random slice index difference $k$. 
The parameters for $\alpha$ were chosen around 1 and the sampled slices $m$ around 8, because these were the final parameters used by Yan et al. \cite{yan2018unsupervised}. In short, the main differences between the \ac{SSBR} and the $\text{SSBR}^{\star}$ are the chosen loss function, the slice sampling strategy and the used data augmentation techniques.
As already discussed in section \ref{sec:order-loss-results} diverging patterns were observed by using the order-loss described by Yan et al.  \cite{yan2018unsupervised} (see fig. \ref{fig:illustration-uncorrected-order-loss}). Because of this incorrect behavior the corrected version of the classification order loss of equation \ref{eq:updated-calssification-order-loss} was used for this analysis. 
The results of the hyperparameter tuning experiment can be seen in table \ref{tab:ssbr-experiment}. It can be observed that model 9 with the parameters $m=12$ and $\alpha=1.2$ is the best performing model based on the LMSE. 
An overview of the model performances of the best \ac{SSBR} and the $\text{SSBR}^{\star}$  on the validation and the test set can be found in table \ref{tab:ssbr-test}.  
The test statistic of the z-test (see sec. \ref{sec:lmse}) between both LMSE values is equal to $t=3.9$. Therefore, our $\text{SSBR}^{\star}$ model performs significantly better than the \ac{SSBR} model (under a significance level of 5 \%).  
The original \ac{SSBR} model from Yan et al. \cite{yan2018unsupervised} was evaluated based on the accuracy metric. It can be observed that our  $\text{SSBR}^{\star}$ outperforms the \ac{SSBR} model as well in the accuracy metric on the test set (see tab. \ref{tab:ssbr-test}). 
\begin{table}
    \centering
    \caption{Overview of grid search for SSBR model. 
    For the data augmentation with random cropping the transformation function ShiftScaleRotate of the Albumentation package \cite{albumentation} was used. }
    \begin{tabular}{ll}
        \hline 
        Parameters & Values \\
        \hline 
         m, batch-size, epoch &  (4, 64, 480), (8, 32, 240), (12, 21, 160) \\
         $\alpha$ & 0.8, 1, 1.2 \\
         k & [2, 30] \\
         random cropping & shift-limit: 0.3, scale-limit: (0, 1.5), p: 1 \\
         \hline
    \end{tabular}
    \label{tab:ssbr-parameters}
\end{table} 
\begin{table}
    \centering
    \caption{Results of the \ac{SSBR} experiment for different hyperparameter configurations on the validation set.}
    \begin{tabular}{lllll}
    \hline 
    Model & m & $\alpha$ & $\bar{\phi}_{\text{val}} \pm \sigma_{\bar{\phi}}$& Accuracy in \% \\
    \hline 
    model 1 &   4.0 &    0.8 &  4.3 $\pm$ 0.5 &        93.3 \\
    model 2 &   4.0 &    1.0 &  4.3 $\pm$ 0.5 &        93.1 \\
    model 3 &   4.0 &    1.2 &  4.1 $\pm$ 0.4 &        93.7 \\
    model 4 &   8.0 &    0.8 &  4.1 $\pm$ 0.4 &        93.5 \\
    model 5 &   8.0 &    1.0 &   3.6 $\pm$ 0.4 &        93.6 \\
    model 6 &   8.0 &    1.2 &  7.8 $\pm$ 4.5 &        94.0 \\
    model 7 &  12.0 &    0.8 &  3.7 $\pm$ 0.4 &        93.3 \\
    model 8 &  12.0 &    1.0 &  3.7 $\pm$ 0.4 &        94.3 \\
    model 9 &  12.0 &    1.2 &  \textbf{3.56 $\pm$ 0.4} &        93.9 \\
    \hline
    \end{tabular}
    \label{tab:ssbr-experiment}
\end{table} 
\begin{table}[ht]
    \centering
    \caption{Performance comparison of the \ac{SSBR} and the $SSBR^{\star}$ model on the validation and test dataset. Model performance is measured based on the LMSE and the accuracy. } 
    \begin{tabular}{lllllll}
        \hline
         Model & $m$ & $\alpha$ & $\bar{\phi}_{\text{val}} \pm \sigma_{\bar{\phi}}$ &
         $\bar{\phi}_{\text{test}} \pm \sigma_{\bar{\phi}}$& $\psi_{\text{val}}$ & $\psi_{\text{test}}$\\
         \hline
         $\text{SSBR}^{\star}$  & 4 & 0   & 2.39 $\pm$ 0.23 & 2.65 $\pm$ 0.28 & 94.0 \% & 94.7\% \\ 
         SSBR  & 12 & 1.2 & 3.6 $\pm$ 0.4 & 4.9 $\pm$ 0.5 & 93.9 \% & 94.0 \% \\
         \hline
    \end{tabular}
    \label{tab:ssbr-test}
\end{table} 

\begin{figure}[t]
    \centering
    \includegraphics[width=\textwidth]{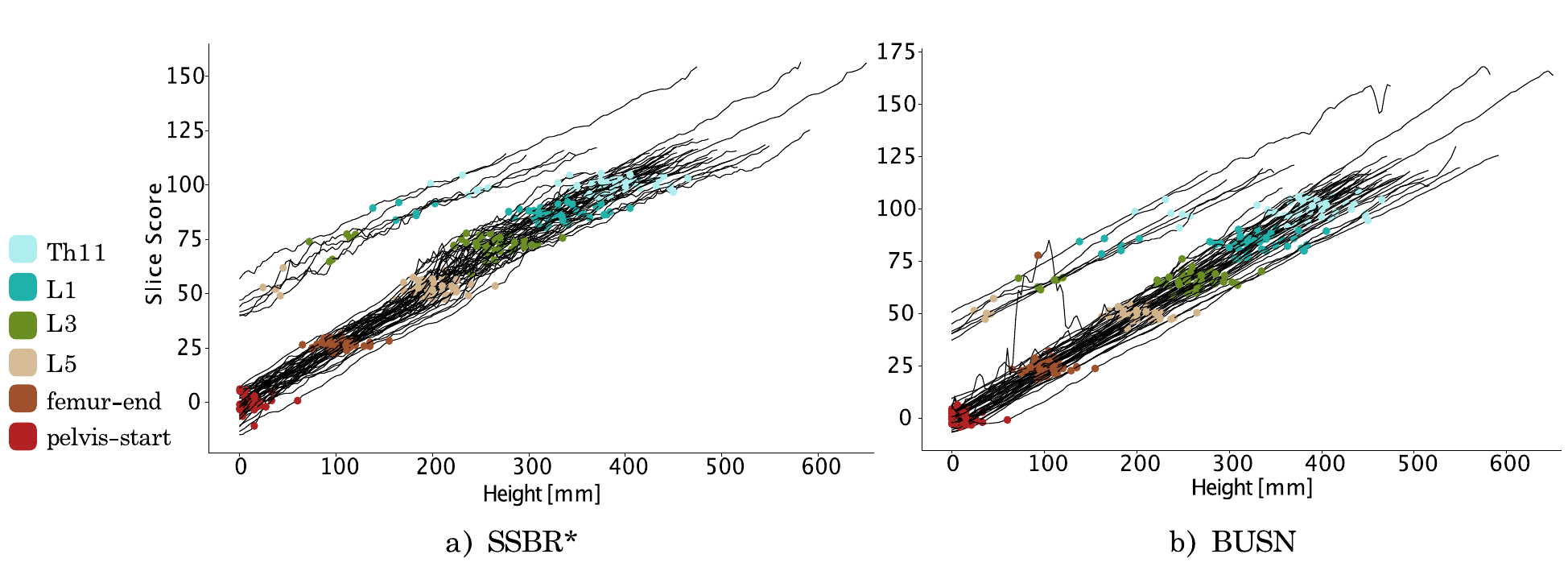}
    \caption{Predicted and linearly transformed slice score curves for all 50 volumes from the BTCV-dataset for the BUSN and the $\text{SSBR}^{\star}$ model. The annotated landmark positions are marked with dots.  }
    \label{fig:evaluation-slice-scores-busn}
\end{figure}

\textbf{BUSN Model: } The $\text{SSBR}^{\star}$ model performance is compared to the 2-stage BUSN model (see sec. \ref{sec:busn}) presented by Tang et al. \cite{tang2021body}, who kindly provided their predictions on their test dataset. The test dataset from the BUSN model was the BTCV dataset \cite{dataSynapse} which is publicly available. Ten of the fifty CT-volumes are already included in the test dataset and referred to as volumes from study Task017 (see tab. \ref{tab:test_data}). To be able to evaluate the BUSN model against the $\text{SSBR}^{\star}$ model as good as possible, the evaluation landmarks of the remaining 40 volumes were additionally annotated as described in section \ref{sec:annotation}. The volumes of this dataset mainly range between the landmark pelvis-start $l_{1}$ and Th11 $l_{5}$. To be able to compare the LMSE between the models, the normalization constant $d$ was redefined for this evaluation to 
\begin{equation*}
    d = \frac{1}{100}(\bar{s}_{l_{5}} - \bar{s}_{l_{1}}). 
\end{equation*} 
Because of this redefinition, the LMSE results of this paragraph are only comparable within this paragraph and not with other LMSE results. 
Moreover, the LMSE falls back to pseudo-labels which were calculated on an annotated subset of the training dataset defined in table \ref{tab:train_data}. This information is not available for the BUSN model. Therefore the pseudo-labels were calculated based on the mean landmark slice score predictions of the BTCV-dataset. The $R^{2}$-metric was calculated for each model by fitting a straight line to the slice score curves and using the linear fit as ground truth (see sec. \ref{sec:related-work-evaluation-methods}).
In table \ref{tab:busn-comparison} an overview of the performances of the BUSN model, the $\text{SSBR}^{\star}$ model and the SSBR model on the BTCV-dataset can be found. 
The BUSN model performs equally well to the  $\text{SSBR}^{\star}$model based on the $R^{2}$-metric. Nevertheless, based on the LMSE metric, the $\text{SSBR}^{\star}$ model performs superior.  
It is interesting to note that the BUSN model does not significantly outperform our $\text{SSBR}^{\star}$ model and also not the SSBR model based on the LMSE metric. 
From table \ref{tab:busn-landmark-LMSE} and figure \ref{fig:evaluation-slice-scores-busn} it can be seen that the high variance of the LMSE from the BUSN model results from an outliers at the landmark femur-end. Besides the femur-end landmark, both models perform equally well. 
For completeness, in figure \ref{fig:evaluation-landmarks-examples-busn}, the relative frequency distributions of the BUSN model and the $\text{SSBR}^{\star}$ model for different landmarks can be found.
\begin{table}[ht]
    \centering
    \caption{Comparison between BUSN, SSBR, and $\text{SSBR}^{\star}$ model based on the BTCV-dataset \cite{dataSynapse}.}
    \begin{tabular}{lll}
        \hline
        & $R^{2}$-metric& $\bar{\phi}_{\text{test}} \pm \sigma_{\bar{\phi}}$ \\
        \hline        
        BUSN              & \textbf{0.994  $\pm$ 0.004}  & 20 $\pm$ 10  \\
        $\text{SSBR}^{\star}$ & 0.9918 $\pm$ 0.0011 & \textbf{9.8 $\pm$ 1.2}\\
        SSBR              & 0.9797 $\pm$ 0.0021 & 19.3 $\pm$ 2.5 \\ 
        \hline
    \end{tabular} 
    \label{tab:busn-comparison}
\end{table} 
\begin{table}[t]
    \centering
    \caption{LMSE $\bar{\phi}$ for each landmark on the BTCV-dataset \cite{dataSynapse}.}
    \begin{tabular}{llll}
    \hline 
    & $\bar{\phi}_{\text{BUSN}}$ per landmark & $\bar{\phi}_{\text{SSBR}^{\star}}$ per landmark & \\
    \hline 
    pelvis-start &    \textbf{7.63 $\pm$ 1.52} &  12.21 $\pm$ 3.81 \\
    femur-end    &  70.84 $\pm$ 62.57 &   \textbf{3.37 $\pm$ 0.70} \\
    L5           &    \textbf{6.74 $\pm$ 1.56} &   9.99 $\pm$ 2.17 \\
    L3           &   \textbf{11.38 $\pm$ 2.58} &  14.74 $\pm$ 4.03 \\
    L1           &   13.79 $\pm$ 2.42 &   \textbf{8.93 $\pm$ 1.64} \\
    Th11         &   17.29 $\pm$ 2.77 &   \textbf{8.80 $\pm$ 1.60} \\
    \hline
    \end{tabular}
    \label{tab:busn-landmark-LMSE}
\end{table}
\section{Application and Deployment}\label{sec:results-deployment}
To be able to use the best performing body part regression model from section \ref{sec:model-evaluation} for the proposed use cases we have to compute same characteristics of the model. Thus, the mean slice score curve slope $\bar{m}_{s}$ and the mean of the tangential slope $\bar{m}_{t}$ were calculated for the model based on the training set predictions (see sec. \ref{sec:methods-application-deployment}). The mean of the tangential slope, the \mbox{99.5 \%-quantile}  and the 0.5 \%-quantile of the tangential slope distribution were given by 
\begin{equation*}
    \bar{m}_{t} = 0.113,\quad q_{99.5\%}(m_{t})=0.25, \quad q_{0.5\%}(m_{t})=-0.037. 
\end{equation*}
For the mean slice score curve slope, we obtain 
\begin{equation*}
    \bar{m}_{s} = 0.118, \ \sigma_{m_{s}} = 0.012. 
\end{equation*}
Moreover, for empty slices, the model predicts $s_{0}\approx110.83$. 
With these characteristics, the cleaned slice scores described in section \ref{sec:methods-application-deployment} can be calculated. 
 
\subsection{Estimate Examined Body Part}\label{sec:results-bpe}
The \textit{body part examined dictionary} and the \textit{body part examined tag} defined in  section \ref{sec:methods-bpe} were evaluated qualitatively based on three studies from the TCIA. All of these datasets were not used before in this thesis. The selected datasets are from the studies: 
APOLLO \cite{apollo, clark2013cancer}, TCGA-KIRC \cite{clark2013cancer,akin2016radiology} and CPTAC-LSCC 
\cite{clark2013cancer, cptaclssc}. The CPTAC-LSCC data was generated by the National Cancer Institute Clinical Proteomic Tumor Analysis Consortium (CPTAC) and the TCGA-KIRC data comes from the Cancer Genom Atlas (TCGA) \cite{cancergenom}. The APOLLO data were generated by the Applied Proteogenomics OrganizationaL Learning and Outcomes (APOLLO) Research Network, a Federal Precision Oncology and Cancer Moonshot Program of the Department of Defense, Department of Veterans Affairs, and National Cancer Institute.

For the analysis, only three-dimensional volumes with the size of:  
$\text{512 px} \times  \text{512 px} \times z$ and $z>10$ 
slices were used. 
In the APOLLO study, 38 volumes, in the TCGA-KIRC study 404 volumes and in the CPTAC-LSCC 151 volumes were analyzed.

\begin{figure}
    \centering
    \includegraphics[width=\textwidth]{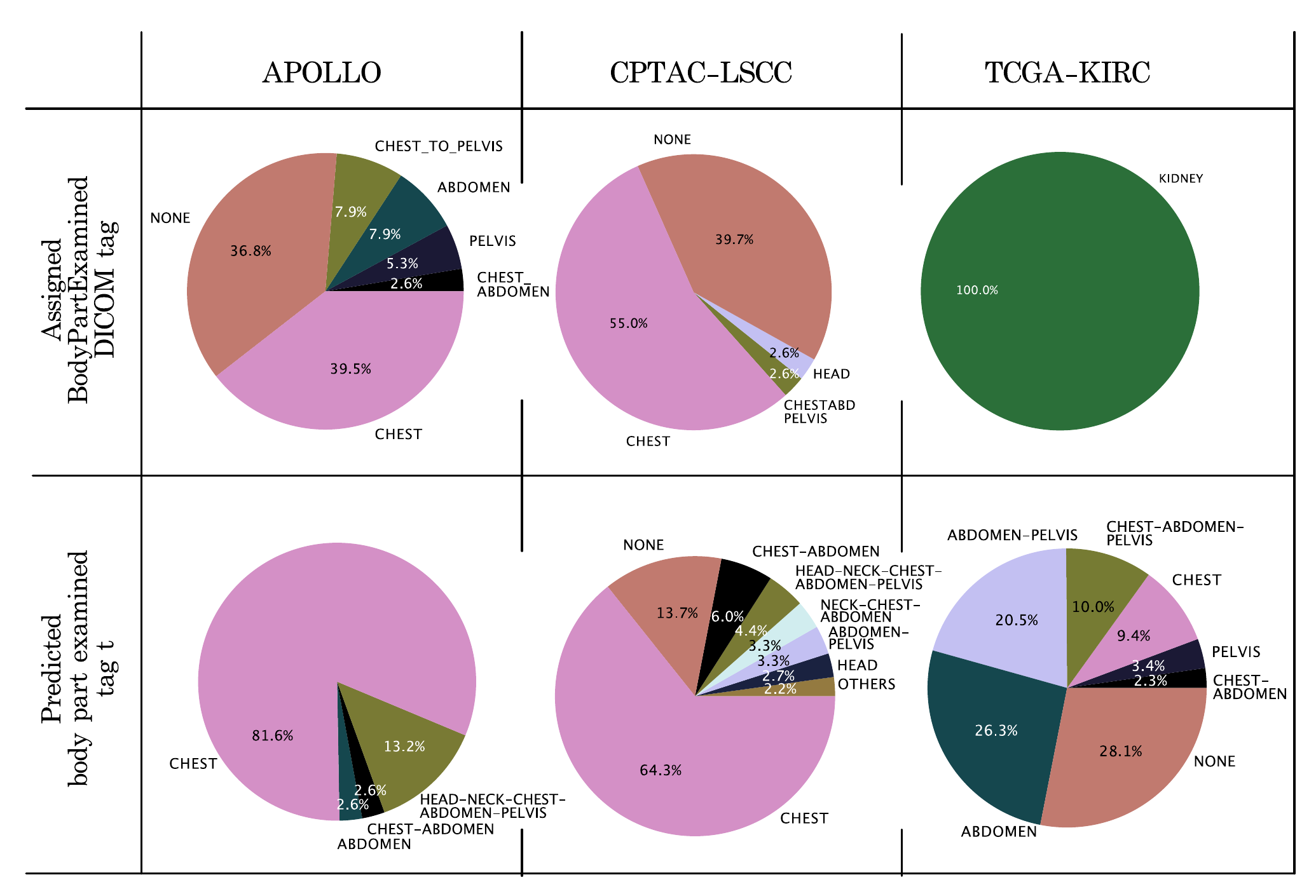}
    \caption{Distributions of the \textit{body part examined tag} and assigned \textit{BodyPartExamined} DICOM tag for the studies: APOLLO, CPTAC-LSCC and TCGA-KIRC. Tags which are less than 1\% represented in the data are summarized as OTHERS.
    }
    \label{fig:bpe-pie-charts}
\end{figure}

Figure \ref{fig:bpe-pie-charts} visualizes a pie chart for each study with the distribution of the \textit{BodyPartExamined} DICOM
tag and the predicted \textit{body part examined tag}.  Compared to the DICOM tags, we can observe that for the TCGA-KIRC study and the CPTAC-LSCC study, we gain a more accurate and fine-grained view of the scanned body ranges. Moreover, the missing values regarding the scanned body range were reduced for the CPTAC-LSCC study and the APOLLO study. 

\begin{figure}
    \centering
    \includegraphics[width=\textwidth]{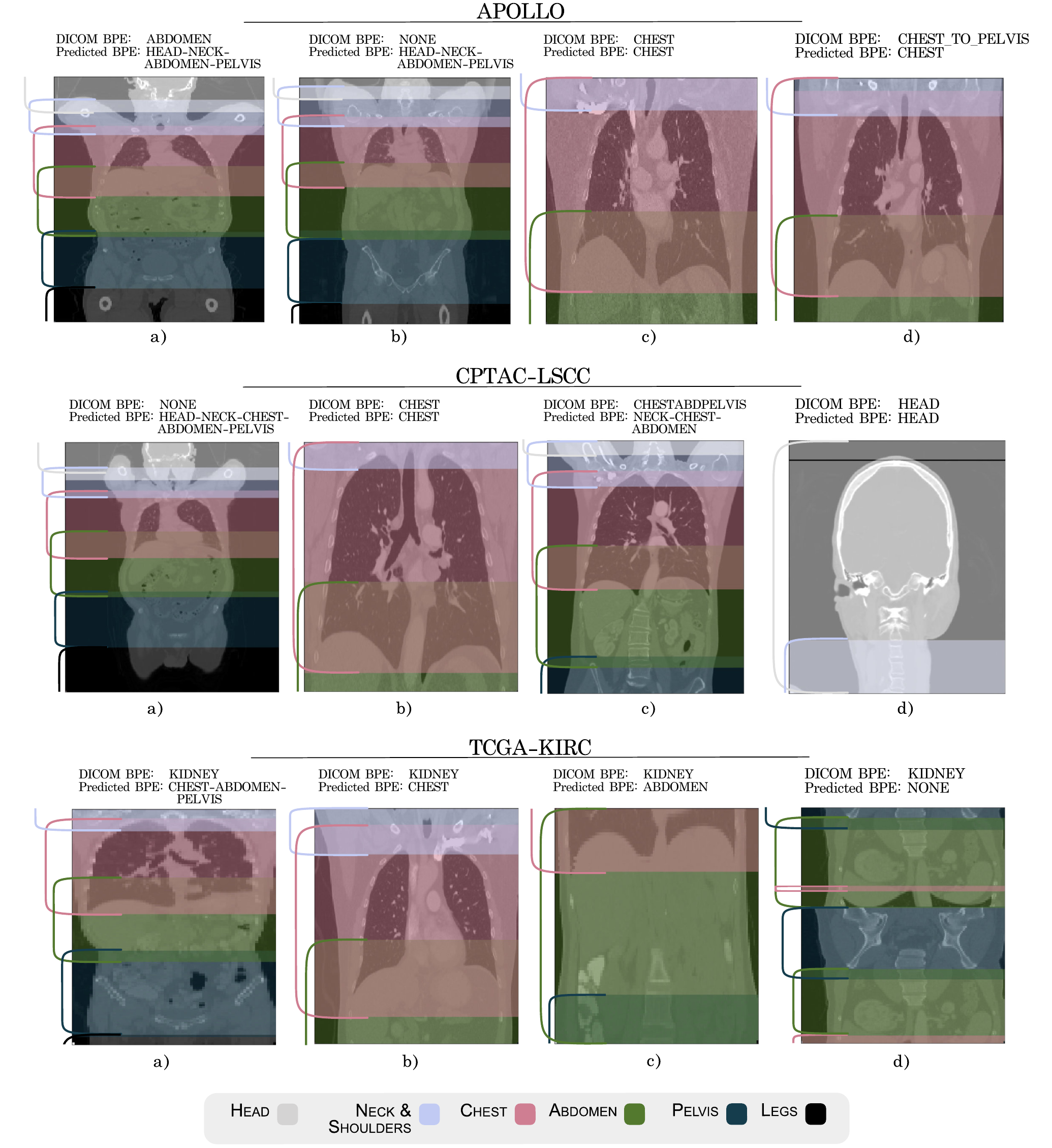}
    \caption{For each study (APOLLO, CPTAC-LSCC \& TCGA-KIRC) 4 example images were plotted together with the associated \textit{BodyPartExamined} DICOM tag (DICOM BPE) and the predicted \textit{body part examined tag} (Predicted BPE). Moreover, the \textit{body part examined dictionary} is visualized by the superimposed color mask. }
    \label{fig:bodypartexamined}
\end{figure}

For each of the studies, the \textit{BodyPartExamined} DICOM tag was compared to the predicted \textit{body part examined tag} $t$ and the \textit{body part examined dictionary} $d$ for 4 exemplary images, visible in figure \ref{fig:bodypartexamined} row 1. 
For all four images from the APOLLO study, the predicted \textit{body part examined dictionary} and the \textit{body part examined tag} fit well to the seen body parts. 
In figure a), the image contains more information than the DICOM tag ABDOMEN suggests. In figure b), the DICOM tag is missing. The \textit{body part examined tag} $t$ is equal to HEAD-NECK-ABDOMEN-PELVIS. In the last slice of volume b) the eyes and the nose is visible. Therefore, the predicted tag $t$ fits well to the volume. In figure c), the DICOM tag represents quite well the seen body range. Furthermore, for figure d), the DICOM tag is wrong, but the predicted tag $t$ is appropriate. 

For the CPTAC-LSCC study 4 exemplary images are visible in figure \ref{fig:bodypartexamined} row 2. It can be observed that the information of the DICOM tag, the predicted dictionary $d$, and the predicted tag $t$ fits well to the visible body parts. For the volume in figure a) no DICOM tag exists, the \textit{body part examined dictionary} can be used as a replacement.
It can be observed that the information of the \textit{body part examined dictionary}  is more precise. It contains as well information about where which body region is visible.

For the TCGA-KIRC study, the 4 exemplary images can be found in figure \ref{fig:bodypartexamined} row 3. For this study, all CT images were labeled with the KIDNEY tag. In figure  \ref{fig:bodypartexamined},  we can observe that although the kidney is visible in most of the figures, the DICOM label KIDNEY is not particularly precise. In most of the figures, more than just the kidneys are visible.  The \textit{body part examined dictionary} and the \textit{body part examined tag} provide a richer information than the KIDNEY DICOM tag. For figure b), the kidney label is wrong because the imaged body area is mainly the chest and not the abdomen. In figure d), we can find a corrupted image, where a CT scan of the abdomen is stacked to a pelvis-abdomen CT scan with a reverse z-ordering. Therefore, the predicted tag $t$ is equal to NONE. In general, we can automatically find corrupted files like this with the predicted \textit{body part examined tag}. 

\FloatBarrier
\subsection{Known Region Cropping} 
To evaluate the application use case of \textbf{known region cropping}, we will focus on medical image segmentation which serves as a representation for an arbitrary clinical deep learning algorithm. A strong baseline for medical segmentation is the nnU-Net \cite{isensee2021nnu}, which already leads to excellent segmentation performances. The known region cropping is applied after the segmentation algorithm as a post-processing step to fetch false-positive segmentations. 

In total, eight nnU-Net models were used for evaluation. Fabian Isensee kindly provided these models. Each of these eight nnU-Nets was trained on different datasets, which the body part regression model has not seen during training. Table \ref{tab:nnunets} summarizes the  nnU-Net segmentation models. For each model, the training data was available. Therefore, the \textbf{data-driven} approach was used to obtain the appropriate minimum known score $s_{\text{min}}$ and maximum known score $s_{\text{max}}$. The scores stake out the \textbf{known region}. The quantiles for the data-driven approach were manually chosen beforehand so that the results do not overfit on these hyperparameters. They were set to $q_{\text{min}} = 25 \%$ and $q_{\text{max}} = 75 \%$.
\begin{table}
    \centering
    \caption{Medical segmentation nnU-Nets with training data and segmentation \mbox{targets \cite{isensee2021nnu}}.}
    \begin{tabular}{llllll}
        \hline
         Models & Segmentation target & Data source & \# Train. data & $s_{\text{min}}$ & $s_{\text{max}}$   \\
         \hline 
         model 1 & lung cancer & Task006 & 63 & 40 & 75 \\ 
         model 2 & hepatic vessels, liver cancer& Task008 & 303 & 28 & 55\\ 
         model 3 & spleen & Task009 & 41& 2& 56\\ 
         model 4 & liver & Task017 & 30 & 0 & 56\\ 
         model 5 & \makecell[l]{pelvis organs\\ 
         (bladder, uterus, \\
         rectum, small bowel)}& Task018 & 30& -9 & 48\\ 
         model 6 & liver & Task046 & 90& 1 & 55\\ 
         model 7 & \makecell[l]{head tissue\\
         (eyes, lens, optical, nerves, \\ 
         optical chiasma, pituitary,  \\
         brain stem, temporal lobes, \\
         parotid glands, mandible, \\
         inner and middle ear, \\
         temporomandibular joint)} & Task049 &  50 & 52 & 110\\ 
         model 8 & kidney and kidney tumor& Task064 & 210 & 1 & 56\\ 
         \hline
    \end{tabular}
    \label{tab:nnunets}
\end{table} 
The segmentation algorithms were applied on the dataset Task055 (see tab. \ref{tab:test_data}) with $s_{\text{min}}=35$ and $s_{\text{max}}=89$. The Task055 dataset consists of 60 volumes. In some volumes, the neck and the head area are present, which are not present in most of the training sets from table \ref{tab:nnunets}. It can be assumed that model 1 till 8 leads to false-positive predictions on Task055 based on unknown body parts during runtime. In the Task055 dataset, the segmentation targets of model 1 till 8 were not masked, and therefore the ground truth segmentations were not available. 

\begin{figure}
    \centering
    \includegraphics[width=\textwidth]{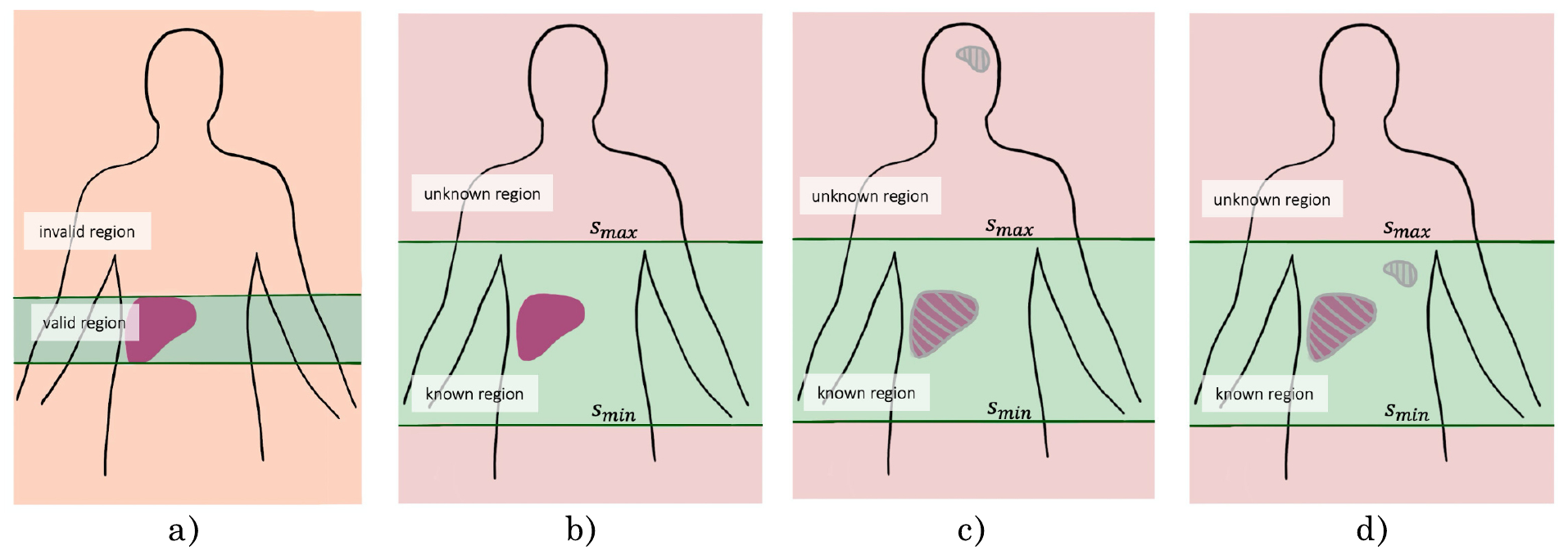}
    \caption{On the example of liver segmentation, the definition of the valid and invalid region can be seen in figure a). In figure b) the estimated known region based on the predicted scores $s_{\text{min}}$ and $s_{\text{max}}$ is visible. In figure c) and d) examples for possible false-positive predictions of the segmentation algorithm are visible. In figure c) the false prediction can be filtered through \textit{known region cropping}, because the false-positive segmentation lies within the unknown region of the segmentation model. The false-positive prediction of figure d) can not be detected, because the false prediction lies inside the known region of the segmentation algorithm.}
    \label{fig:intercepted-false-positives}
\end{figure}

In this experiment the following question was investigated:
\begin{enumerate}
    \item[] How many CT images with false-positive segmentations in \textbf{invalid regions} can be intercepted by \textbf{known region cropping}?  
\end{enumerate}
Slices, where the segmentation target is visible, are declared as \textbf{valid region}. All slices, where no segmentation target is present, are declared as \textbf{invalid region} (see fig. \ref{fig:intercepted-false-positives} a)). For each target, the valid region  was annotated manually on Task055.
For a perfect body part regression model, all misclassifications in the unknown region would be intercepted, and no true positive segmentation would be cropped
(see fig. \ref{fig:intercepted-false-positives} c)). 
All false-positive segmentations
that lie inside the known region of the model cannot be intercepted as it can be seen in figure \ref{fig:intercepted-false-positives} d).\ \\
The results for model 1 till 8 can be found in table \ref{tab:data-driven-slice-score-cutting}. The total number of volumes of the Task055 dataset with segmentations of model $x$ in invalid regions can be found in the column "false-positives". The column "intercepted false-positives" shows the number of volumes where false-positive segmentations in unknown regions were detected by the \textit{known region cropping} post-processing step. The "truncated segmentations" column shows the number of true positive predictions which were truncated by mistake through \textit{known region cropping}.  

Table \ref{tab:data-driven-slice-score-cutting} shows that through adding \textit{known region cropping} to the segmentation pipeline, on average 86.4 \% of volumes with false-positive predictions can be caught for a nnU-Net. The 13 volumes, which could not be caught through \textit{known region cropping} were manually investigated. For all these volumes, we found that the misclassifications lied in the known region of the model (case of figure \ref{fig:intercepted-false-positives} d)). The body part regression model behaved as expected. 
The 13 volumes with truncated segmentations can be traced back to a too small known region because of an inappropriate selection of $q_{\text{min}}$ and $q_{\text{max}}$ for these datasets. 
\begin{table}
    \centering
    \caption{Results of \textit{known region cropping} as a post-processing step for nnU-Nets. All models were applied on the dataset Task055. The number of volumes for each model with false-positive predictions in the invalid-range are shown in the column "False-positives".}
    \begin{tabular}{lllll}
        \hline
         \makecell[l]{nnU-Net\\model}& False-positives &
         \makecell[l]{Intercepted\\false-positives} & \makecell[l]{Truncated\\segmentations}& Accuracy     \\
         \hline 
         model 1 & 10 & 10 & 0 & 100    \% \\ 
         model 2 & 13 & 13 & 0 & 100    \% \\ 
         model 3 & 29 & 22 & 0 & 76     \% \\ 
         model 4 & 5 & 3 & 3 & 60       \% \\ 
         model 5 & 37 & 37& 0 & 100     \% \\ 
         model 6 & 14 & 14 & 10 & 100   \% \\ 
         model 7 & 60 & 60 & 0 & 100    \%\\ 
         model 8 & 9 & 5& 0 & 56        \% \\ 
         \hline 
         sum & 177 & 164 & 13 &  \\
         \hline
         \hline
         & global accuracy & 92.66 \% & mean accuracy & 86.40 \% \\
         \hline
    \end{tabular}
    \label{tab:data-driven-slice-score-cutting}
\end{table}

\FloatBarrier
\subsection{Data Sanity Checks}
To evaluate the proposed basic data sanity checks, two experiments were designed. 

\textbf{Evaluation of the z-Ordering Check: } The z-ordering of randomly picked 50 volumes  from the test dataset (see sec. \ref{sec:dataset}) were reversed. After that, the cleaned slice scores for all volumes were calculated, and the z-ordering test (see sec. \ref{sec:method-dsc}) was applied. With an accuracy of 100 \%, the z-ordering was correctly detected as reverse or rather not-reverse. The accuracy is this high because the slice score slope is quite stable, and the test concentrates only on the sign of the slope.

\textbf{Evaluation of the z-Spacing Check: } The threshold $\theta$, to define if a z-spacing is valid or not, was found data-driven by analyzing different $\theta$ on the validation set. As a result, the threshold was fixed at $\theta=28 \%$. For this threshold, the first time, no volume from the validation set was declared wrongly as a volume with invalid z-spacing. 

\begin{figure}
    \centering
    \includegraphics[width=0.7\textwidth]{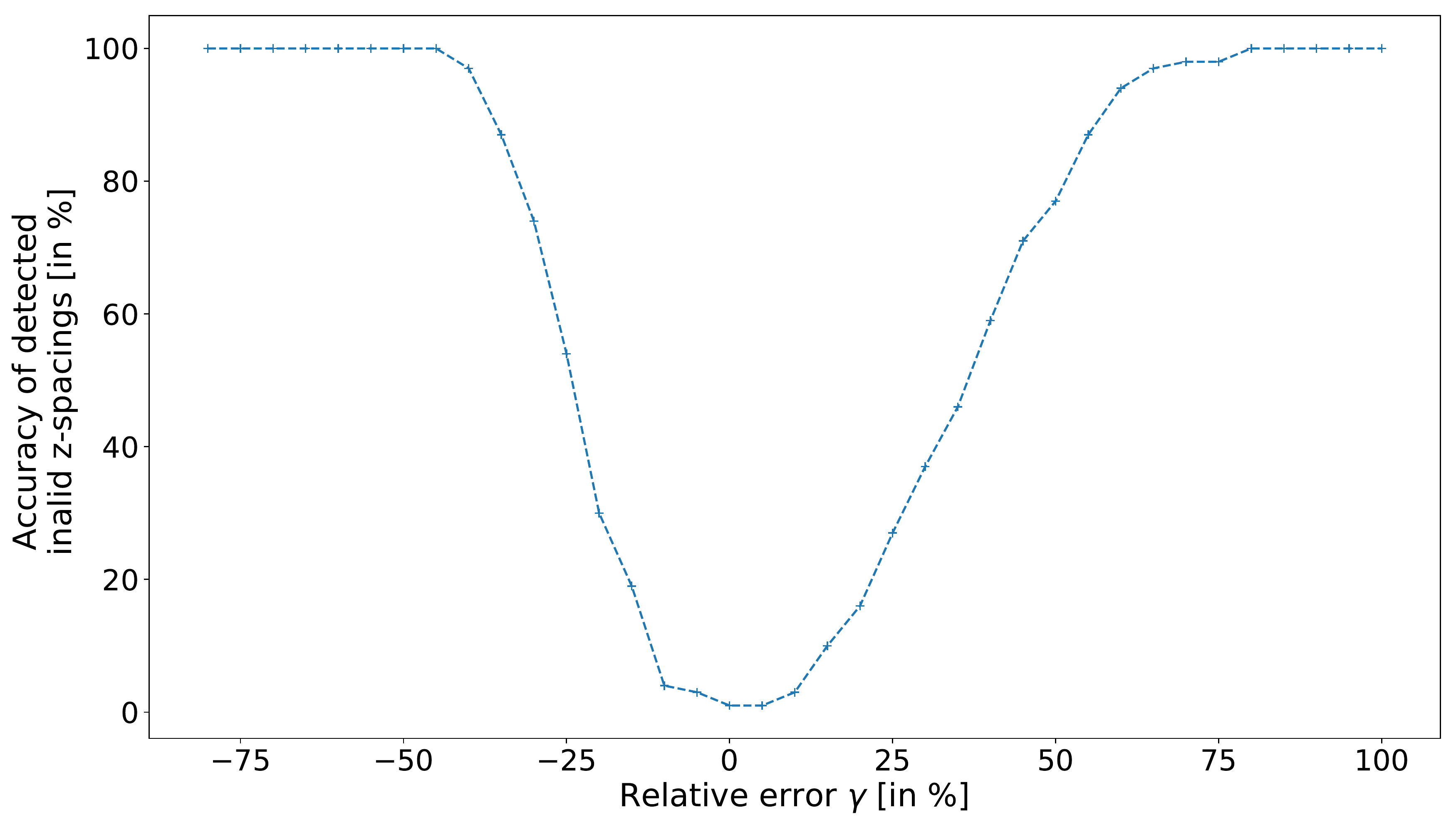}
    \caption{Relation between relative error $\gamma$ of z-spacing and accuracy of detected invalid z-spacings in the test dataset.}
    \label{fig:detected-invalid-zspacings}
\end{figure}

To analyze the performance of the z-spacing check, the volumes from the test dataset were used. Further, the z-spacings were artificially changed by a relative error $\gamma$ 
\begin{equation*}
    z^{\prime} = (1 + \gamma) \cdot z, \qquad \gamma \in [0, 1]. 
\end{equation*}
For each relative error $\gamma$, we evaluated how many volumes from the test set were declared as volumes with invalid z-spacing. 
For an ideal z-spacing check, the accuracy would always lie at 100 \%  for $\gamma \neq 0$. In figure \ref{fig:detected-invalid-zspacings} the result of the experiment can be found. 
For $\gamma=0$, one volume was incorrectly predicted as a volume with invalid z-spacing. For a relative error of over $+ 40 \%$ and a relative error of less than $- 25 \%$ about more than half of the volumes were declared as volumes with invalid z-spacing. 

\subsection{Deployment}
The python package was published on \href{https://github.com/MIC-DKFZ/BodyPartRegression}{GitHub}.
Moreover, the main functionality of the body part regression model was added to the Kaapana framework and is available within platforms via an extension.

\begin{figure}
    \centering
    \includegraphics[width=0.9\textwidth]{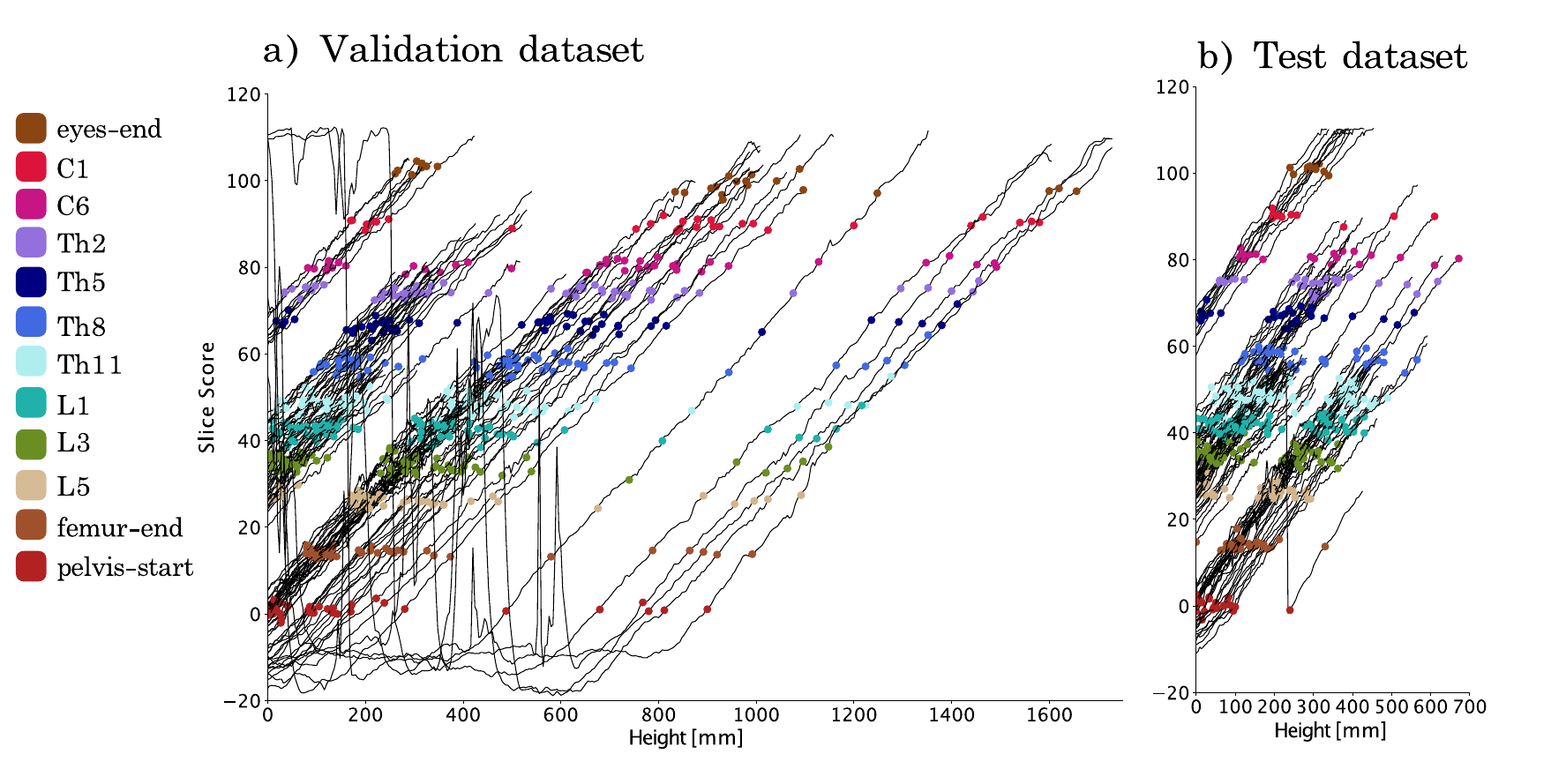}
    \caption{Slice score predictions from the public model for volumes from the test and validation set. The landmark positions were labeled with different markers.}
    \label{fig:public-slice-score-plot}
\end{figure}

\textbf{Public Model: }As discussed in section \ref{sec:deployment}, a body part regression model with a modified training dataset was trained to obtain a model which can be used in a public python package and in the Kaapana toolkit. We will refer to this model as \textbf{public model}. Respectively, we will refer the model evaluated in section  \ref{sec:model-evaluation} as \textbf{private model}. The public model has an LMSE $\bar{\phi}$ with the standard error $\sigma_{\bar{\phi}}$ and an accuracy $\psi$ of 
\begin{equation*}
    \bar{\phi}_{\text{val}} = 2.32 \pm 0.21, \quad \psi_{\text{val}} = 94.9 \% , 
\end{equation*}
on the validation set. On the test set, the model performance is equal to 
\begin{equation*}
    \bar{\phi}_{\text{test}} = 2.69 \pm 0.32, \quad \psi_{\text{test}} = 95.4 \%. 
\end{equation*}
The model performs equally well on the validation and the test set as the private model. To give a summary of the qualitative performance of this model, in figure 
\ref{fig:public-slice-score-plot} the predicted slice score curves from the validation and the test dataset can be found. For the \textit{whole body CT} study in the validation set, the model performs worse on the leg region compared to the private model. This can be easily explained by the fact that the public model has not seen this region during training. As a supplement to figure \ref{fig:public-slice-score-plot}, the distribution of the slice score predictions on the evaluation landmarks can be found in figure \ref{fig:public-evaluate-landmarks} in the appendix. The public model was made publicly available through zenodo: \\
\textit{\url{https://zenodo.org/record/5113483##.YQOl-VMzZQI}}. 
\textbf{Python Package: } 
The python package was published on GitHub and can be found at: 
\textit{\url{https://github.com/MIC-DKFZ/BodyPartRegression}}. 

\begin{figure}
    \centering
    \includegraphics[width=0.6\textwidth]{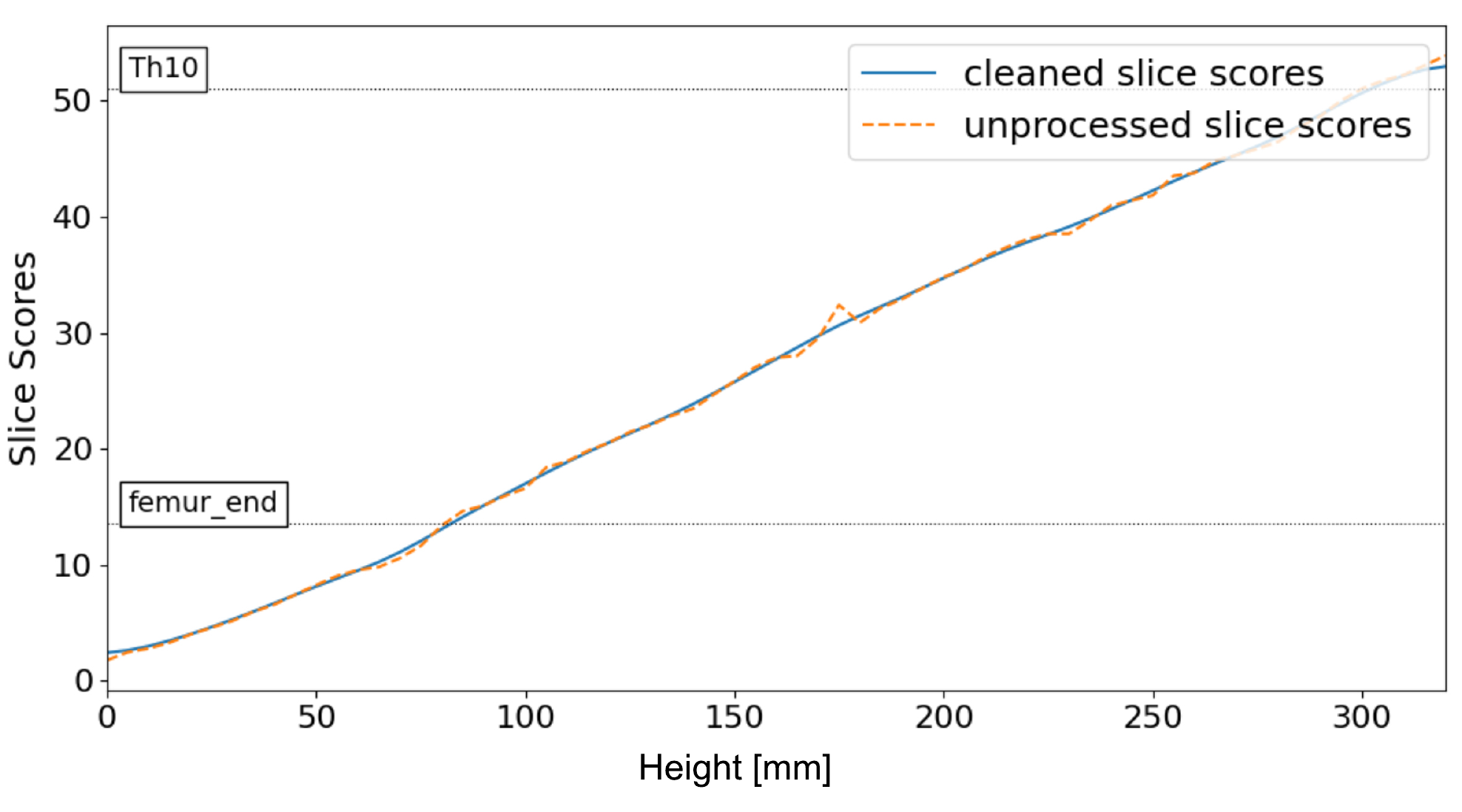}
    \caption{Example output slice score plot from \textit{bpreg\_predict} method. Source volume comes from the Task010 colon dataset. }
    \label{fig:bpreg-predict-slice-score-plot}
\end{figure}

The installation is explained in the README.md file and can be easily done through pip install. 
The main function can be triggered in the terminal by the command:
\begin{lstlisting}
bpreg_predict -i <input-path> -o <output-path>.
\end{lstlisting}
For every NIfTI-file in the input directory, an additional JSON-file is created in the output directory. Moreover, a file that explains the different entries of the JSON-file is saved in the output directory. Next to the input and the output path, several additional parameters can be set, for example --\textit{plot}. If the plot parameter is set to true, an image of the predicted unprocessed slice scores and the cleaned slice scores is visualized for every NIfTI-file and saved as a PNG file into the output directory (see fig. \ref{fig:bpreg-predict-slice-score-plot}). 
As the default model for predicting the slice scores, the public model is used. At the first run of \textit{bpreg\_predict}, the public model is downloaded automatically from zenodo.

\begin{figure}
    \begin{subfigure}[b]{\textwidth}
        \centering
        \includegraphics[width=\textwidth]{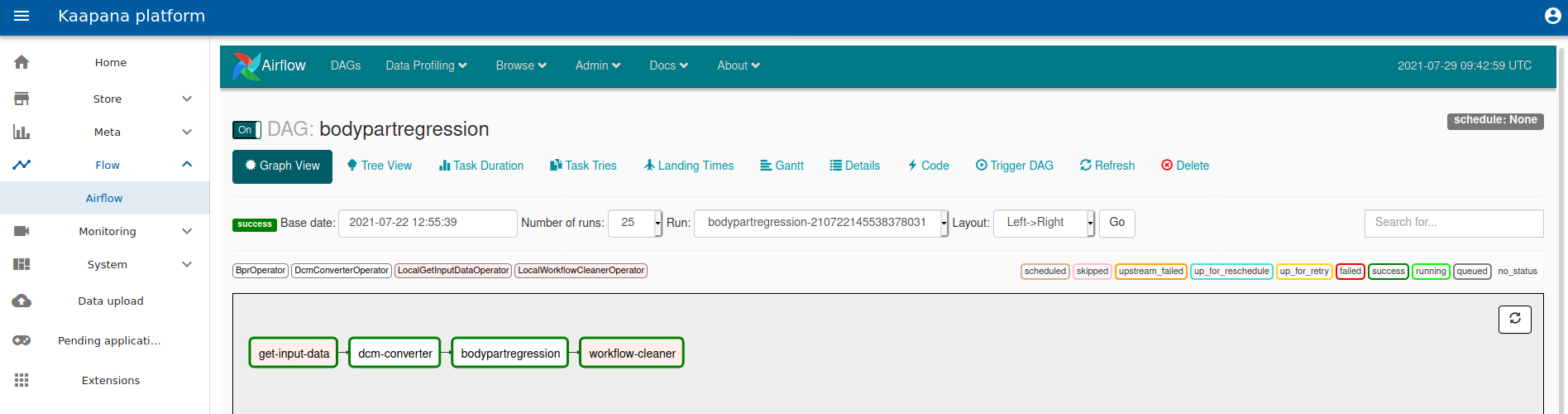}
        \caption{Visualization of a finished bodypartregression DAG processing pipeline through Airflow.}
        \label{fig:kaapana-dag}
    \end{subfigure}
    \par\bigskip 
    \begin{subfigure}[b]{\textwidth}
        \centering
        \includegraphics[width=\textwidth]{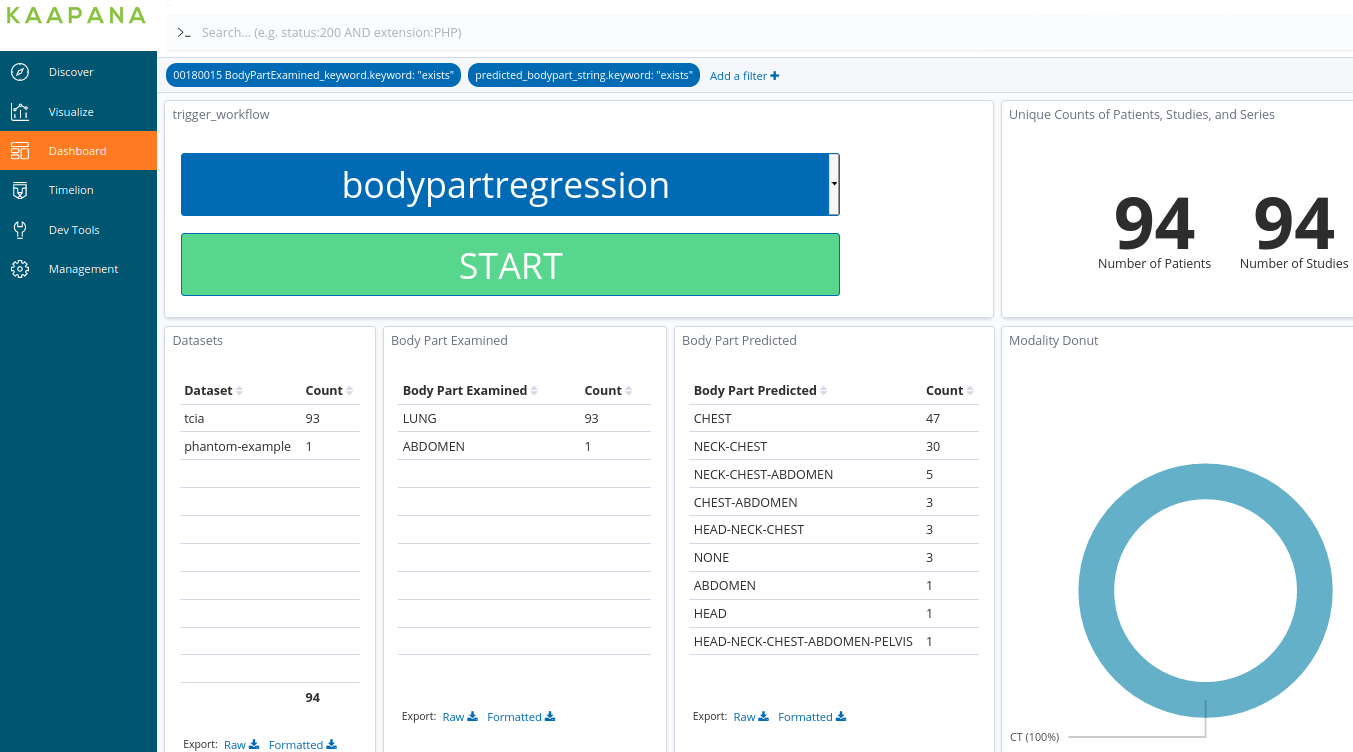}
        \caption{Visualization of the \textit{body part examined tag} from the JSON-file in the table \textit{Predicted Body Part} for some example CT images. Moreover, the bodypartregression DAG model can be triggered through selecting \textit{bodypartregression} at the button in the top left corner.}
        \vspace{0.5cm}
        \label{fig:kaapana-body-part-predicted}
    \end{subfigure}
    \caption{Screenshots of the Kaapana deployment.}
    \label{fig:kaapana-deployment}
\end{figure}

\textbf{Kaapana: } 
For the Kaapana integration, the public model was used and packed up into a Docker container. Figure \ref{fig:kaapana-dag} shows a screenshot of the \textit{bodypartregression} \ac{DAG}, which can be triggered through the dashboard and creates for each selected volume a JSON meta-data file
which is added to the meta-data knowledge-base of Kaapana. The \textit{body part examined tag} from the JSON meta-data file is visible at the Kaapana dashboard in a table called \textbf{Body Part Predicted} visible in figure \ref{fig:kaapana-body-part-predicted}. 
Since not all servers Kaapana is deployed on have a GPU available, the execution has been shifted to the CPU to ensure high compatibility. 

Depending on the device on which the model inference is made, the time, RAM usage, and GPU memory usage to create and save a JSON meta-data file for a NIfTI-volume is \mbox{different (see tab. \ref{tab:ram-time})}.

\begin{table}
    \centering
    \caption{Overview of GPU memory usage, RAM usage, end-to-end time $t_{\text{total}}$ and inference time $t_{\text{inference}}$ of the \textit{bpreg\_predict} method depending on the model inference device. The end-to-end time $t_{total}$ is the time needed for loading the NIfTI-file, computing the data for the JSON-file, and saving the JSON-file. The inference time $t_{\text{inference}}$ is only the time the model needs to compute the slice scores. 
    For the GPU prediction, a GeForce RTX 2080 with 11 GB memory was used. For the CPU and GPU computation, a CPU with about 62 GB RAM was used.  To compute the times, the method was applied to 100 random CT NIfTI-files. }
    \begin{tabular}{lllll}
        \hline
         Device &  GPU memory usage & RAM usage & $\bar{t}_{\text{total}} \pm \sigma_{t}$ & $t_{\text{inference}} \pm \sigma_{t}$\\
         \hline
         GPU & $\sim$ 6.4 GB & $\sim$ 5 GB & $2.0s \pm 1.6s$ & $0.14s \pm 0.11s$ \\
         CPU & -- & $\sim$ 5 GB & $5s \pm 4s$  & $3.2s \pm 2.6s$\\
         \hline
    \end{tabular}

    \label{tab:ram-time}
\end{table}

\chapter{Discussion}
\section{Body Part Regression}
The body part regression model $\text{SSBR}^{\star}$, trained in this thesis has an LMSE of $\bar{\phi}=2.65\pm 0.28$ and an accuracy of $94.7\%$ on the test set. 
The completely self-supervised approach shows robust and accurate performance across a variety of different datasets (see sec. \ref{sec:model-evaluation}). 
The performance in terms of accuracy of the trained  $\text{SSBR}^{\star}$ model is comparable with body part classification networks as the network from Roth et al., which was evaluated on five classes and achieved an accuracy of $94.1\%$ \cite{roth2015anatomy} (see sec. \ref{sec:related-work-bpr}). 
The $\text{SSBR}^{\star}$ model outperforms the state-of-the-art body part regression \ac{SSBR} model, which achieves an LMSE of $\bar{\phi}=4.9\pm 0.5$ and an accuracy of $94.0\%$  on the test set (see sec. \ref{sec:sota-comparison}). 
Moreover, it was shown that a 1-stage $\text{SSBR}^{\star}$ model can outperform the 2-stage BUSN approach if the $\text{SSBR}^{\star}$ model includes diverse data augmentation techniques (see sec. \ref{sec:sota-comparison}). The 2-stage BUSN model achieved an LMSE of $\bar{\phi}=20\pm 10$ and the $\text{SSBR}^{\star}$ an LMSE of $\bar{\phi}=9.8\pm 1.2$ on an independent test set with a separate definition of the normalization constant $d$.

\textbf{Self-Supervision: }
Usually, self-supervised methods are used as a pretraining for the actual problem with a lack of data (see sec. \ref{sec:learning-methodologies}). This was not the case for the body part regression model. 
Although the body part regression model is trained completely self-supervised, it shows accurate and robust performance across different studies. The self-supervised approach made it possible to train based on several diverse studies and over 2000 CT volumes, where only 10\% of the data was annotated for evaluation purposes. The heterogeneous dataset
facilitates generalization towards different diseases, anatomical conditions, study-related special features, and CT settings. Through the fully self-supervised training, the model learned completely on its own how the human anatomy should be mapped to slice scores and has therefore experienced less human bias during training. 

\textbf{Model Conditions: } At the beginning of this thesis, we have proposed two conditions which the trained body part regression model should fulfill: the \textit{monotony condition} and the \textit{independence} condition (see sec. \ref{sec:architecture}). The monotony condition is preserved during training because of the order loss, which is the main objective function of the deep learning task. The order loss enforces the model to order the slice scores monotonously within patients. The independence condition is measured by the LMSE and is preserved implicitly through the diverse data augmentation techniques. As it can be seen in the data augmentation experiment in section \ref{sec:data-augmentation-results}, the data augmentations are crucial for a good performance. In the previous literature, the impact of data augmentation techniques was underrated. For example, in the original paper, where the \ac{SSBR} model was introduced the first time \cite{yan2018unsupervised} no data augmentation techniques were used at all. In the update paper of the \ac{SSBR} model \cite{yan2018deep} random cropping was used as only data augmentation technique. 
In section \ref{sec:sota-comparison} we have seen that the $\text{SSBR}^{\star}$  performs significantly better than the \ac{SSBR} model. Together with the data augmentation experiments from section \ref{sec:data-augmentation-results} we can hypothesize that from all differences between both models, the lack of data augmentations is crucial for the significant difference in model performance. The performance of the \ac{SSBR} model ($\bar{\phi} = 3.6$)  is comparable with the model performances from table \ref{tab:data-augmentation-results}, where no image quality transform augmentations or no intensity transform augmentations were used to train the model. The random cropping data augmentation technique from the \ac{SSBR} model is a kind of physical transformation. Based on the fact that no image quality transforms and image intensity transform were applied as data augmentation, the \ac{SSBR} model may treat images of certain studies differently, based on overfitting on certain intensity values or the image quality. The BUSN model did not use data augmentation techniques at all \cite{tang2021body}. This could be a reason for the bad performance of the \ac{SSBR} model in the first stage of the BUSN model. 

\textbf{Model Parameters: }
Not all model parameters of the body part regression model were investigated in this thesis. The slice sampling $m$ per volume, the number of two-dimensional images in one batch, the $\Delta h$ sampling, and the model architecture were, for example, not investigated further.
While bigger $m$ allow the representation of larger contiguous regions per batch, smaller m generate batches with higher inter-patient anatomical variation.
The ultimate influence of $m$ on the training behavior is unknown and remains to be analyzed. In this thesis, the number of two-dimensional images per mini-batch stayed the same, with 256 slices per mini-batch. With this value, we had good experiences during training. For too few slices per mini-batch, we observed a high variance in convergence. This might be the case because the estimation of the gradients have more variance. The hyperparameter can, of course, be tuned and was not thoroughly analyzed in this thesis. The sampling of $\Delta h$ was fixed to a range of [5 mm, 100 mm] for most of the experiments. We sampled from a range of heights because the original paper sampled the distances from a list. It remains to be shown how the minimal possible sampled $\Delta h_{\text{min}}$ and the maximum possible sampled $\Delta h_{\text{max}}$ influences the performance of the model. Moreover, we do not yet know if it is necessary to sample $\Delta h$ from a range. Maybe it is enough to set $\Delta h$ to a fixed value for the whole training and to convert it for each volume individually to a distance in terms of slice index difference $k$ depending on the z-spacing. 

\textbf{Regularization: }
We have seen in the order-loss experiments that adding a linear constraint through a distance loss or the parameter $\beta$ of the heuristic order loss is beneficial for learning a body part regression model. In the order loss section \ref{sec:order-loss}, we derived that $\beta$ is a linear constraint parameter, which handles how long the linear constraint will be maintained. The linear constraint can be seen as a kind of regularization. The slice score mapping problem is an ill-posed problem with no unique solution. With the linear constraint, the solution space shrinks, and possible solutions are solutions, where the slice score increases approximately linear with $\Delta h$.  If $\beta$ is too high, the regularization is too low, and the slice score scale $\Delta s$ is unstable. 
Too small $\beta$ parameters lead to too strong regularization and the learning is prolonged. This behavior can be recognized in figure \ref{fig:loh-experiment-mse}. For $\beta=1/\text{mm}$, the LMSE deteriorates, the regularization is too low. For $\beta=0.0001/\text{mm}$, the regularization is too high, and the LMSE converges rather slowly. 
The linear constraint is no objective on its own because there exists no reason why the anatomy across patients should be linearly transformable into each other. 
If the anatomy of the human body would be linearly transformable across patients, then all differences in anatomy could be attributed to different patient heights. Two patients with the same height would then always have the same pelvis height, head height, leg height, etc. This is obviously not the case. The human anatomy is highly variable.  In the \ac{SSBR} paper \cite{yan2018deep, yan2016multi} the distance loss was treated as an additional objective function. In this thesis, we have seen that the distance loss is rather a kind of regularization that helps the model meet the independence condition and is not an objective on its own. 

\textbf{Linear Constraint: }
Two different versions of implementing a linear constraint were shown in this thesis. On the one hand, the linear constraint can be realized through an additional distance loss $L_{\text{dist}}$. On the other hand, it can be realized through the $\beta$ parameter in the heuristic order loss $L_{\text{order}}^{h}$. The distance loss constraints a linear relation between slice index and slice scores, whereas the heuristic order loss constraints a linear relation between slice height and slice scores. Therefore, the constraint of the distance loss is patient height independent, and the constraint of the heuristic order loss is not. 
In section \ref{sec:order-loss}, it is explained that the $\beta$ has an impact on the slice score scale. For small $\beta\Delta h$, it is the proportionality factor between $\Delta h$ and $\Delta s$:  $\Delta s \propto \beta \Delta h$. Moreover, a higher $\beta$ leads to a lower linear constraint, and vise versa (see sec. \ref{sec:order-loss}). Therefore, the impact of parameter $\beta$ on the predicted slice scores and the linear constraint is well understood. 
For the distance loss, it remains to be shown how the parameter $\alpha$ from equation \ref{eq:loss-constitution-alpha} influences the linear constraint and the slice score prediction. We were able to observe that big $\alpha$'s lead to smaller slice scores.  Additionally, big $\alpha$'s lead to more regularization, as it can be seen in the loss-experiments in section  \ref{sec:order-loss-results}. Because of the better theoretical understanding, the heuristic order loss was used to incorporate the linear constraint. Although, we have seen in the order loss experiments in section \ref{sec:order-loss-results} that both constraint methods are valid and lead to good experimental results.

\textbf{Recent Paper: }In a recent paper, Proskurov et al. trained a body part regression model with the following loss function \cite{proskurov2021fast}
\begin{equation}\label{eq:proskurov-loss}
    L = \frac{1}{B}\frac{1}{m-1} \sum_{i=1}^{B}\sum_{j=1}^{m-1}
    ||\Delta h_{i}  - \Delta s_{ij}||_{2}^{2}.
\end{equation}
This loss function is comparable to the heuristic order loss in the limit for small $\beta \ll 1$. For the limit, we need to take the tailor series of the sigmoid function into account. It is given by 
\begin{equation*}
     \sigma(x) \approx \frac{1}{2} + \frac{1}{4}x. 
\end{equation*}
For small $\beta \Delta h$ the linear constraint is particularly strong, and we will gain approximately 
\begin{equation*}
    \Delta s \approx \beta \Delta h. 
\end{equation*}
Therefore, if $\beta$ is small,  $\Delta s$ will become small as well. 
For the limit of small $\beta$ we obtain
\begin{equation}\label{eq:beta-0-limit}
\begin{split}
    \lim_{\beta \ll 1} \ L_{\text{order}}^{h}
    &=  \lim_{\beta \ll 1} \frac{1}{B}\frac{1}{m-1} \sum_{i=1}^{B}\sum_{j=1}^{m-1}
    ||\sigma(\beta \Delta h_{i})  - \sigma(\Delta s_{ij})||_{2}^{2} \\
    &= \frac{1}{B}\frac{1}{m-1} \sum_{i=1}^{B}\sum_{j=1}^{m-1} ||\frac{1}{4}(\beta\Delta h_{i}  - \Delta s_{ij})||_{2}^{2}.
\end{split}
\end{equation}
Equation \ref{eq:beta-0-limit} has an additional overall scaling factor of $\frac{1}{16}$ compared to equation \ref{eq:proskurov-loss}, which is not influencing the minimum. The minimum for both equations is equivalent expect for the scaling factor $\beta$ of $\Delta s$. The linear constraint is particularly strong and for all $\Delta h$ equally important. It would be interesting to see if the results from Proskurov et al. can be improved by using the heuristic order loss from this thesis with less regularization. Moreover, Proskurov et al. used a ResNet instead of a VGG network as base architecture. The ResNet architecture is a more recent architecture than the VGG and performs for multiple classification tasks better \cite{he2016deep}. Detailed analysis and comparison of different base network architecture for the body part regression network remains to be done. In this thesis, we showed that the VGG network as a base network already leads to outstanding results (see sec. \ref{sec:model-evaluation}).

\section{Evaluation Metrics}\label{sec:discussion-evaluation}
In this thesis, a modified version of the Mean Square Error, the \textbf{Landmark Mean Square Error} was used for evaluation. With the help of this method, we were able to evaluate and compare models. Based on the LMSE, we selected the model with the best performance. The selected model has outstanding results in qualitative and quantitative analysis, which speaks for a good selection metric (see sec. \ref{sec:model-evaluation}). Moreover, it allows comparing the slice scores of similar body regions via defined landmarks across different CT scans. 
However, the used metric also has limits and disadvantages. On the one hand, the LMSE can become zero if all slices map to the same score. Then, the model would not anymore fulfill the \textit{monotony condition} and is useless for body part recognition. Therefore, it is essential to have an additional look at the predicted slice score curves. 
In our case, we did not face this problem because the model was explicitly trained to optimize the \textit{monotony condition} with the help of the order loss function. On the other hand, the LMSE metric depends on the chosen landmarks. To obtain a precise metric, it is essential to choose landmarks that are well defined, easy to interpret and located in body regions with low inter-patient and intra-patient (e.g., independent of breathing cycle) variability. Further, it is essential to choose landmarks that are equally distributed along with the patient height such that different body regions have an equal contribution to the error metric. 
In the beginning, we agreed on using bone landmarks for evaluation which proved to show a smaller variance (see tab. \ref{tab:slice-score-reference-table-train-val}).
The variance of bone landmarks was on average equal to $\bar{\sigma}=2.43 \pm 0.11$, which was less than the overall average variance $\bar{\sigma}=2.63 \pm 0.15$. In addition, the variance on the evaluation landmarks was even less with  $\bar{\sigma}=2.30 \pm 0.19$ than the average variance for bone landmarks (see tab. \ref{tab:landmarks-model-performance}). This suggests that we have selected robust landmarks for evaluation. 
Since the LMSE is based on comparing predicted landmark scores to a mean anatomical shape, the interpretability is limited in the case of degenerated anatomies like scoliosis. For example, in figure \ref{fig:scolioses-validation-set} the CT volume with the highest LMSE in the validation set is visible. It can be observed that the patient has scoliosis. 
In our evaluation, a high LMSE resulted from deformed or degenerated spinal anatomy does not necessarily imply a poor model performance. 
A limitation that comes with the LMSE metric is the manual annotation effort. Moreover, some landmarks are overrepresented because some body regions appear more frequently than others.
It can be seen that for the head and neck region, far fewer landmarks are annotated than for the abdomen region (see tab. \ref{tab:count-annotated-landmarks}). Therefore, the model's precision in the abdomen region has a bigger influence on the LMSE metric than in the head and neck region. 

\textbf{Aggregation Technique: } The overestimation of some landmark can be counteracted through different strategies. For example, we could annotate the same amount of volumes for each landmark. The disadvantage of this approach would be, that we could not calculate the LMSE per volume any more because only a few and not all available landmarks in the volume are labeled. Another approach would be to redefine the LMSE to 
\begin{equation}
    \bar{\phi} = \frac{1}{L}\sum_{k=1}^{L}\frac{1}{N_k}\sum_{i=1}^{N_{k}}
    \left(\frac{\bar{s}_{l_{k}} - s_{il_{k}}}{d}\right)^{2}, 
\end{equation} 
where $N_i$ is the number of annoated volumes for landmark $k$ and $L$ the total amount of annotated landmarks. 
The difference to the original definition is that we now aggregate first over the volumes for one landmark and then average the LMSEs per landmark to one value. 
In this metric, all landmarks are weighted equally, while the contribution of each volume depends on the number of annotated landmarks. The aggregation technique matters in our case since we have not for every volume all landmarks annotated because of different fields of view. We decided for the definition in section \ref{sec:lmse} to average the LMSE per volume to the final LMSE $\bar{\phi}$ because we wanted to estimate how good the model generalizes to new volumes and not how well the model generalizes to new landmarks. 

\textbf{Accuracy: }
The accuracy was already used in previous literature \cite{yan2018unsupervised}. The main advantage of the accuracy is its easy interpretability. However, the accuracy metric only considers a subset of landmarks, which are the boundaries of defined classes. Moreover, it needs annotated data as well. 
The accuracy converts the problem to a classification task. Discontinuous behavior within a class can not be detected. Additionally, the performance on landmarks between two classes has higher importance than landmarks on the border.
Therefore, the accuracy is not as well-suited and precise as the LMSE metric. 

\textbf{BUSN $R^{2}$-Metric: } The $R^{2}$-metric was proposed by Tang et al. to evaluate the BUSN model \cite{tang2021body}. The advantage of the proposed method is that no additional annotated data is needed, because the method only measures how well the predicted scores of a volume fit to a straight line. 
The main disadvantage of the metric is that it can not represent inter-patient anatomical variation. 
If a model predicts different slice scores for the same anatomy, the metric would not recognize it as long as the slice scores increase linearly within a volume. Looking at the $R^{2}$-metric the BUSN model has a better performance than the $\text{SSBR}^{\star}$ model (see tab. \ref{tab:busn-comparison}), although we can visually inspect in figure \ref{fig:evaluation-landmarks-examples-busn} and figure \ref{fig:evaluation-slice-scores-busn}, that the $\text{SSBR}^{\star}$ model has a better quantitative performance because no outliers are visible. 

\section{Application and Deployment}
The body part regression model was integrated into Kaapana and published on GitHub (see sec. \ref{sec:results-deployment}). It can be easily downloaded and used to automatically analyze the scanned body range in CT volumes, crop CT volumes to the right size for deep learning algorithms, and apply basic data sanity checks.  

\textbf{Body Part Examined: }From the qualitative analysis of the  \textit{body part examined dictionary} $d$ and the  \textit{body part examined tag} $t$ we have seen, what the literature already reported. The \textit{BodyPartExamined} tag from the DICOM header is often missing, imprecise or wrong
 \cite{fedorov2016dicom}. Based on the experiments in section \ref{sec:results-bpe}, we have seen that the developed \textit{body part examined tag} $t$ and \textit{body part examined dictionary} $d$ work pretty well and give a precise and reasonable overview of CT datasets. Additionally, the developed \textit{body part examined dictionary} gives us an even more fine-grained and richer view of the scanned body range compared to a single tag. 
Based on the results of section \ref{sec:results-bpe}, it can be recommended to use the deployed python package to label CT images with the examined body part instead of using the information in the DICOM header. This can significantly reduce the manual expenditure of cleaning and sorting CT datasets.

\textbf{Known Region Cropping: } Many medical deep learning algorithms are dependent on specific body parts. Because of this, a too large field of view can lead to false-positive predictions in unknown regions. Therefore, we have analyzed the benefits of the proposed \textit{known region cropping} processing step based on various medical segmentation algorithms. 
The proposed known region cropping processing step can detect false-positive predictions in invalid regions pretty well. In the experiments, over 80\% of the predictions in invalid regions were found, and all other false-positive predictions lied in regions, which the model has seen during training (see tab. \ref{tab:data-driven-slice-score-cutting}). 
For 13 volumes, the actual prediction was truncated (see tab. \ref{tab:data-driven-slice-score-cutting}). This problem can be solved by customizing the minimum quantile $q_{\text{min}}$ and maximum quantile $q_{\text{max}}$, which are used to define the known region. 
By extending the known region, fewer segmentation predictions are incorrectly truncated.
Customizing the hyperparameters $q_{\text{min}}$ and $q_{\text{max}}$ for each model might be beneficial and can avoid truncating correct segmentations. 

\begin{figure}
    \centering
    \includegraphics[width=\textwidth]{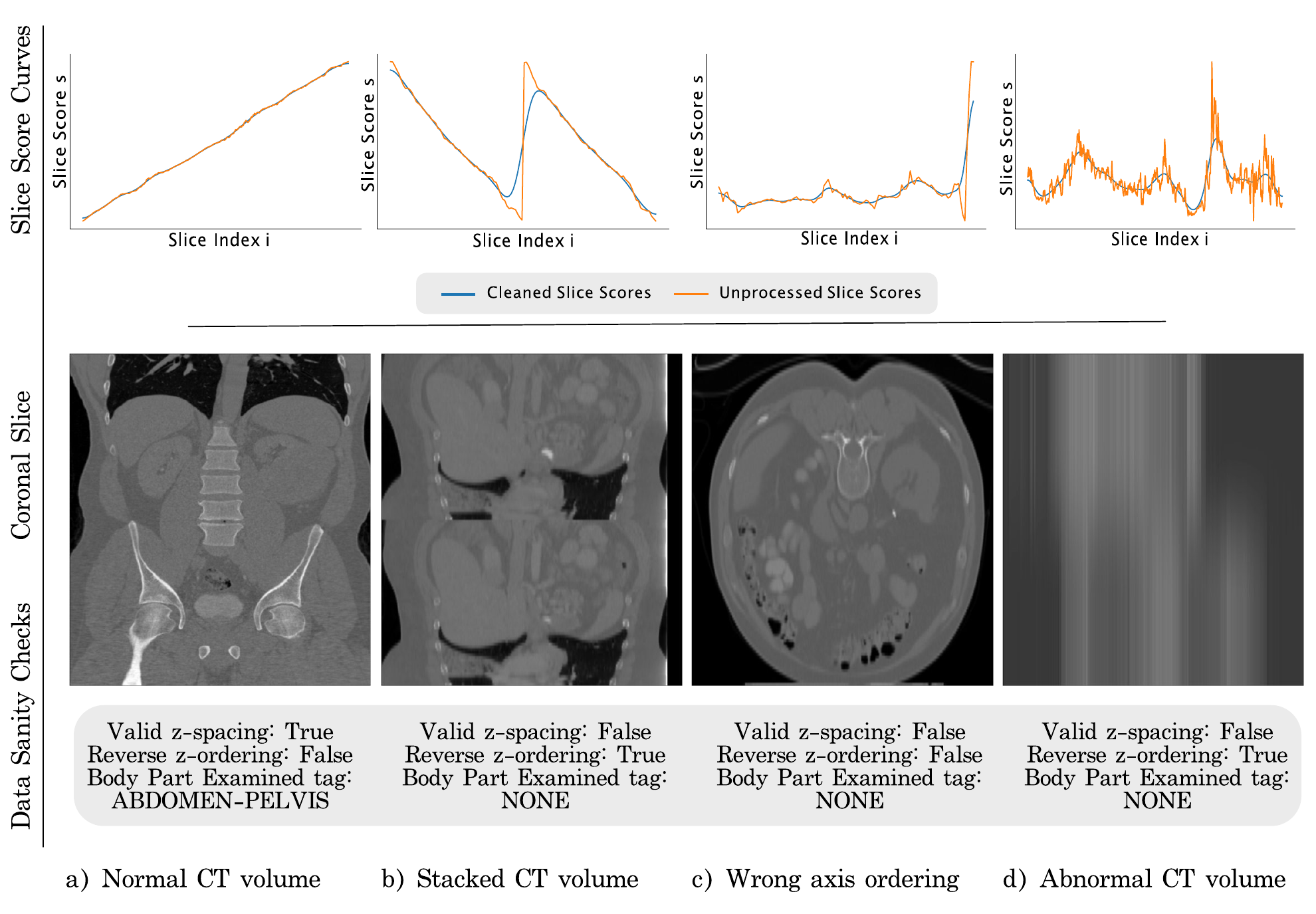}
    \caption{Identified corrupted CT scans based on predicted slice score curves and data sanity checks. In figure a), a normal CT scan and the corresponding slice score curve are visible. In figure b) a stacked CT scan and the slice score curve can be seen. This leads to a discontinuity in the slice score curve. In figure c) the axes were not arranged properly, and in figure d), no body part is visible at all. The visible CT scans are from the TCGA-KIRC dataset \cite{clark2013cancer,akin2016radiology}.}
    \label{fig:corrupted-ct-volumes}
\end{figure}

\textbf{Data Sanity Checks: }
Based on the predicted slice scores and the proposed data sanity checks, we can find corrupted CT volumes. For example, CT volumes, with unexpected axis ordering or stacked CT volumes from different sources.
Figure \ref{fig:corrupted-ct-volumes} shows examples of corrupted CT scans together with the predicted slice scores and the output from the data sanity checks. 
Although it is possible to find corrupted CT scans with the proposed data sanity checks, it is not possible to detect minor relative deviations in the z-spacing (see fig. \ref{fig:detected-invalid-zspacings}). Only for relative errors of more than $+40\%$ or less than $-25\%$ of the z-spacing, more than  $50\%$ of the invalid z-spacings can be detected. This emphasizes that the validation of the z-spacing is conservative and has its strength in detecting substantial outliers.

\textbf{Run Time Comparison: }Finding the scanned body region can be done by several approaches besides body part regression. Further, approaches can be the classification of the analyzed body parts, registration of the volume to an atlas, segmentation of the main bones like the spine's vertebrae. Regarding classification, we have already discussed that a regression approach gives us a more fine-grained view of the scanned volume. Moreover, it fits better to the problem because the human body is a continuous object and not a concatenation of distinct classes. The disadvantage of registration and segmentation is that these algorithms have a long inference time. For example, the nnU-Net segmentation of the vertebrae took about 20 min per volume on an Nvidia V100 SXM2  GPU with a GPU memory of 32 GB. The method deployed in this thesis needs on a GeForce RTX 2080 GBU with 11 GB GPU memory, about 0.14 s inference time. The end-to-end process needs 2s, and if the inference runs on a CPU, the process takes about 5s for an average CT volume (see tab. \ref{tab:ram-time}).
A further possible approach to obtain the scanned body part range inside a CT volume is to register the volume to a reference CT volume. For registration, run times of over a minute were reported by Criminisi et al.  \cite{criminisi2010regression}.
The computation time of the body part regression model is with a few seconds by far less than for the segmentation task and the registration task. Therefore, the body part regression approach is more practical and easier usable in a clinical environment. 

\chapter{Conclusion}
In the introduction, three aims were stated. In section \ref{sec:summary} the three main contributions of this thesis are summarized. Finally, in section \ref{sec:future-work} promising future application fields and further open research questions are presented. 

\section{Summary}\label{sec:summary}
\begin{itemize}

    \item [\textbf{C1: }] \textit{Obtaining insights into the underlying optimization problem of robust body part regression to train a better-performing and generalizable body part regression model for CT images.}
    
    We have identified that the monotony and independence condition are crucial for the training of a well-performing body part regression model
    (see sec. \ref{sec:architecture}). Moreover, it was discovered that a linear constraint is stabilizing the learning procedure. To ensure robust deep learning, a set of different data augmentation techniques and a large diverse dataset was used for training, validation, and testing. Different experiments were performed to highlight the importance of using extensive data augmentations. Additionally, the loss function used by Yan et al. \cite{yan2018unsupervised} was derived theoretically, and a new loss function was proposed, which naturally already contains the linear constraint. We showed that the body part regression model trained on a variety of CT studies in combination with extensive data augmentation strategies generalizes well to new CT datasets and shows excellent qualitative and quantitative results (see sec. \ref{sec:model-evaluation}). Furthermore, the trained body part regression model performs superior compared to the state-of-the-art 2-stage BUSN model proposed by Tang et al. \cite{tang2021body} and the original SSBR model proposed by Yan et al. \cite{yan2018unsupervised, yan2018deep}. 

    \item [\textbf{C2: }] \textit{Proposing a more fine-grained, thorough and robust evaluation strategy for cross-model performance comparison for body part regression models.}
    
    The Landmark Mean Square Error (LMSE) was proposed and used for comparing body part regression models. 
    We demonstrated that this evaluation method is well-suited for measuring whether a model fulfills the independence condition. Compared to other evaluation methods, such as the accuracy (used by Yan et al. \cite{yan2018unsupervised}) or the $R^{2}$-metric (proposed by Tang et al. \cite{tang2021body}), the LMSE is superior in taking inter-patient anatomical variations into account and delivering a fine-grained body region accuracy measure. 
    The experimental results of the quantitative LMSE metric were consistend with the qualitative observations (see sec. \ref{sec:model-evaluation}). 

    \item [\textbf{C3: }] \textit{Proposing three use cases of body part regression and simplify their application for the research community and the clinical environment.}
    
    It was explained how a \emph{body part examined tag} $t$ can be derived through the body part regression model. Additionally, we were able to show that the \emph{body part examined tag} is reliable and leads to more robust information compared to the DICOM tag \textit{BodyPartExamiend}. Additionally, a method was proposed to gain a \emph{body part examined dictionary}, including a more fine-grained view of the examined body parts. 
    Furthermore, the \textit{known region cropping} procedure was introduced.
    It was shown that this cropping method can help deep learning algorithms to be robustly applied across different fields of view. By cutting the appropriate body region for an algorithm, wrong predictions outside the training scope of the algorithm can be avoided. 
    Moreover, two basic data sanity checks were introduced, which can detect corrupted CT images within a dataset. Finally, the body part regression model was integrated in Kaapana and published as python package on GitHub. This enables access to an easy-to-use image analysis method for the research and the clinical community. 

\end{itemize}

\section{Future Work}\label{sec:future-work}
The body part regression model shows excellent generalizability on heterogeneous CT images. Future work could focus on the transfer of the algorithm to other imaging modalities like \ac{MRI} and the extension of the scope of the algorithm to the legs so that full-body CT scans can be precisely processed as well. For including the legs to the algorithm's scope, more publicly available whole-body CT images would be precious. 

Extending the algorithm for MRI data will encounter a few challenges. On the one hand, MRI data has no physical uniform intensity scale as CT scans. On the other hand, 
MRI images can be recorded with many different sequences, which leads to fundamentally different appearances of anatomical structures in the images. A solution for these problems could include a broader range of data augmentation techniques along with a well-considered normalization strategy of the MRI images. 

Regarding the integration into Kaapana and the proposed use cases, the next step would be to apply and evaluate the benefit of the use cases across different medical imaging tasks in an actual clinical setup.

\chapter{Appendix}
\setcounter{figure}{0}   
\renewcommand\thefigure{A.\arabic{figure}} 
\setcounter{table}{0}
\renewcommand{\thetable}{A.\arabic{table}}

\begin{figure}[H]
    \centering
    \includegraphics[width=0.6\textwidth]{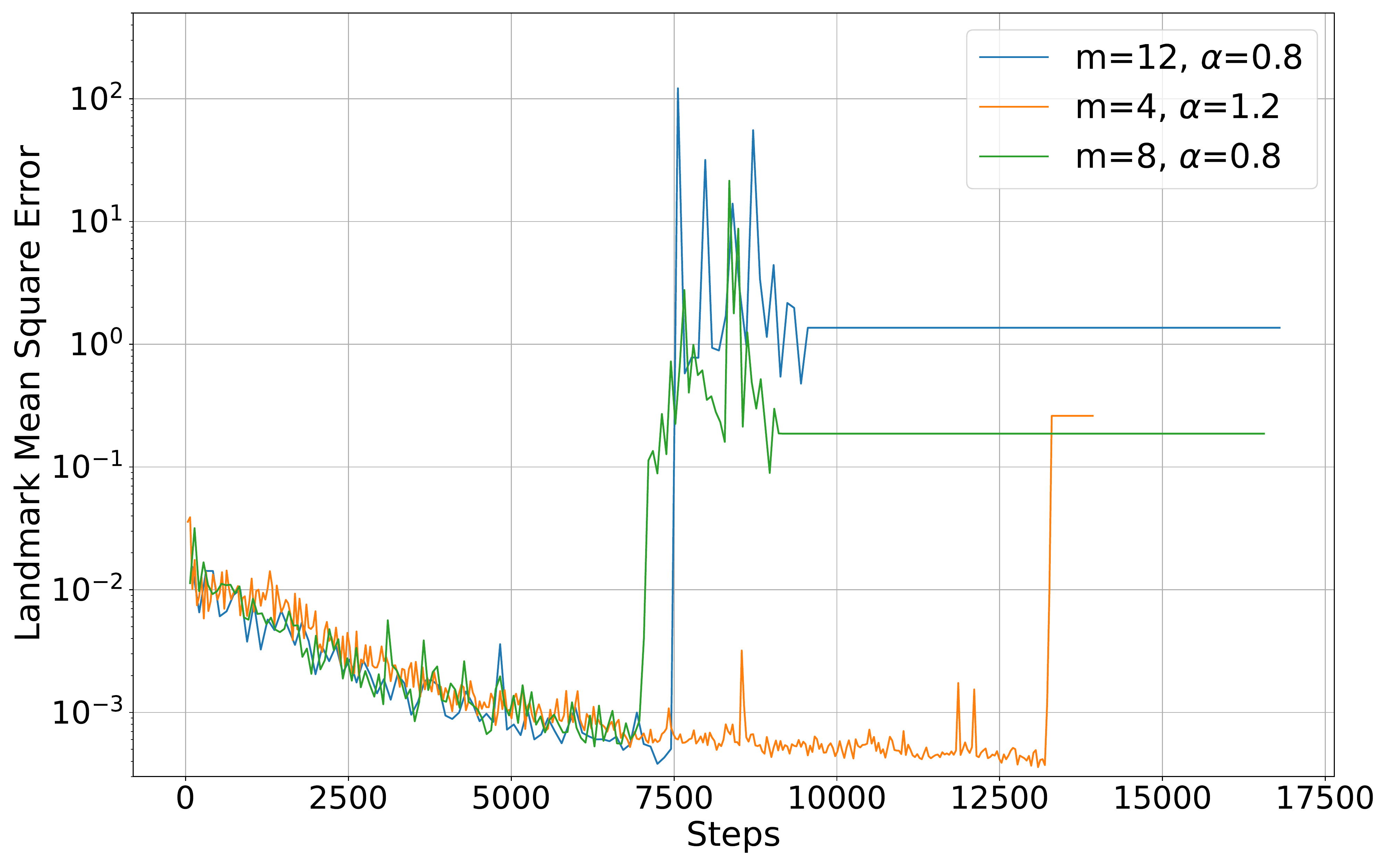}
    \caption{Illustration of diverging Landmark Mean Square Error during training with uncorrected order loss evaluated on the validation set.}
    \label{fig:illustration-uncorrected-order-loss}
\end{figure} 

\begin{figure}[H]
    \centering
    \includegraphics[width=\textwidth]{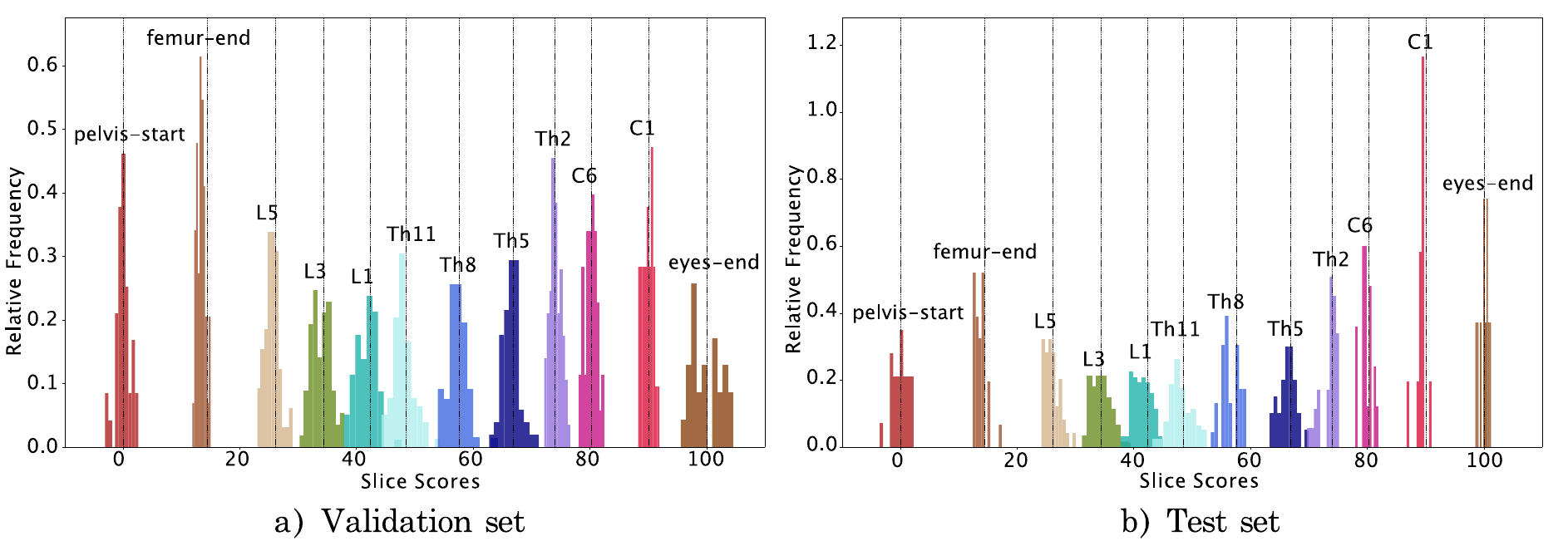}
    \caption{Relative frequency of predicted landmark scores on the test and the validation set for the public model. Each histogram represents the distributions of the predicted slice scores at a specific landmark. The black lines represent the pseudo-labels for the landmarks from the slice score reference table.}
    \label{fig:public-evaluate-landmarks}
\end{figure}

\begin{figure}
    \centering
    \includegraphics[width=0.9\textwidth]{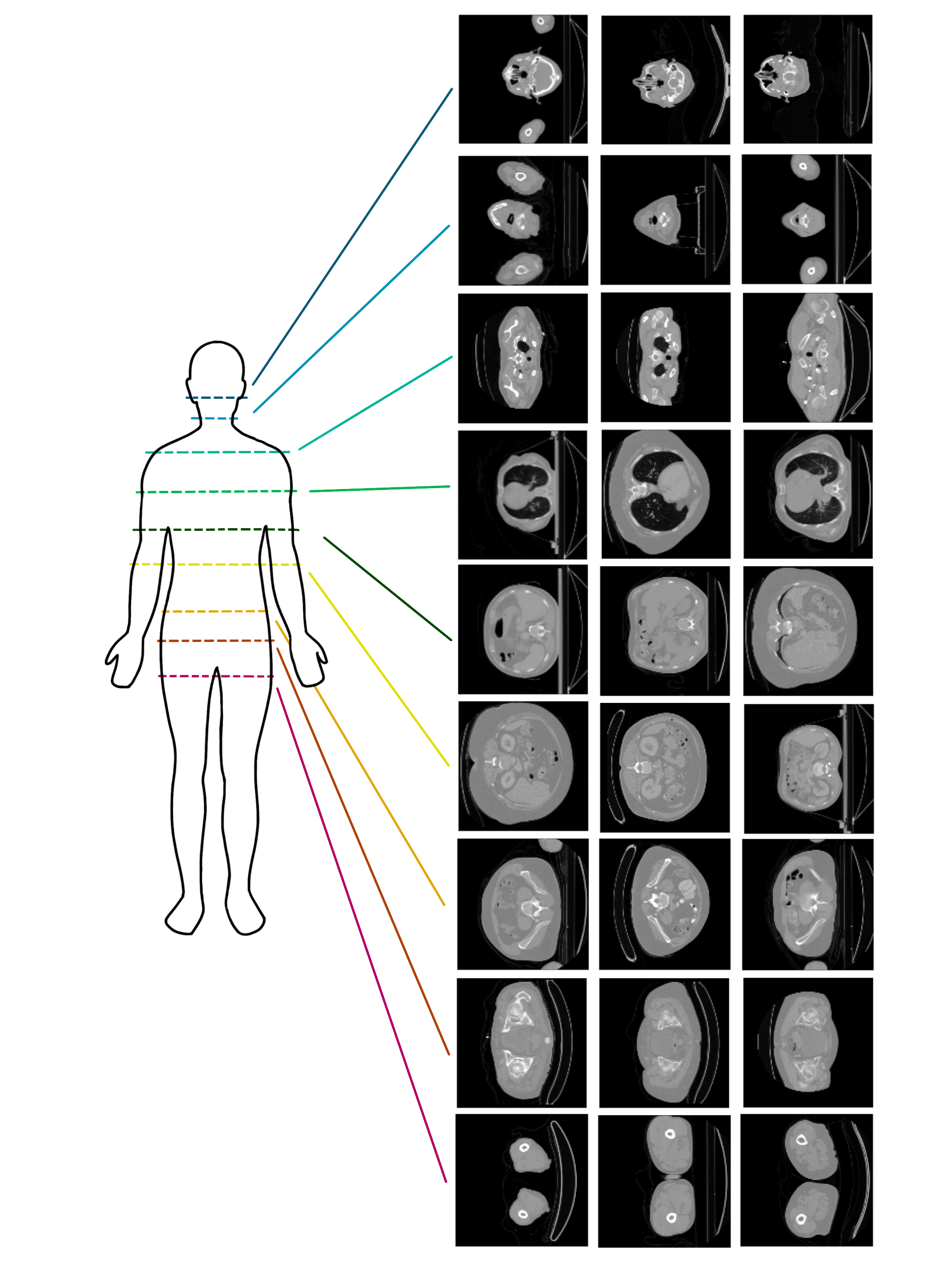}
    \caption{Appearance of different body regions in CT images. In each line three CT slices of one body region from random CT images and patients can be seen. }
    \label{fig:anatomy}
\end{figure}

\begin{figure}
    \centering
    \includegraphics[width=0.9\textwidth]{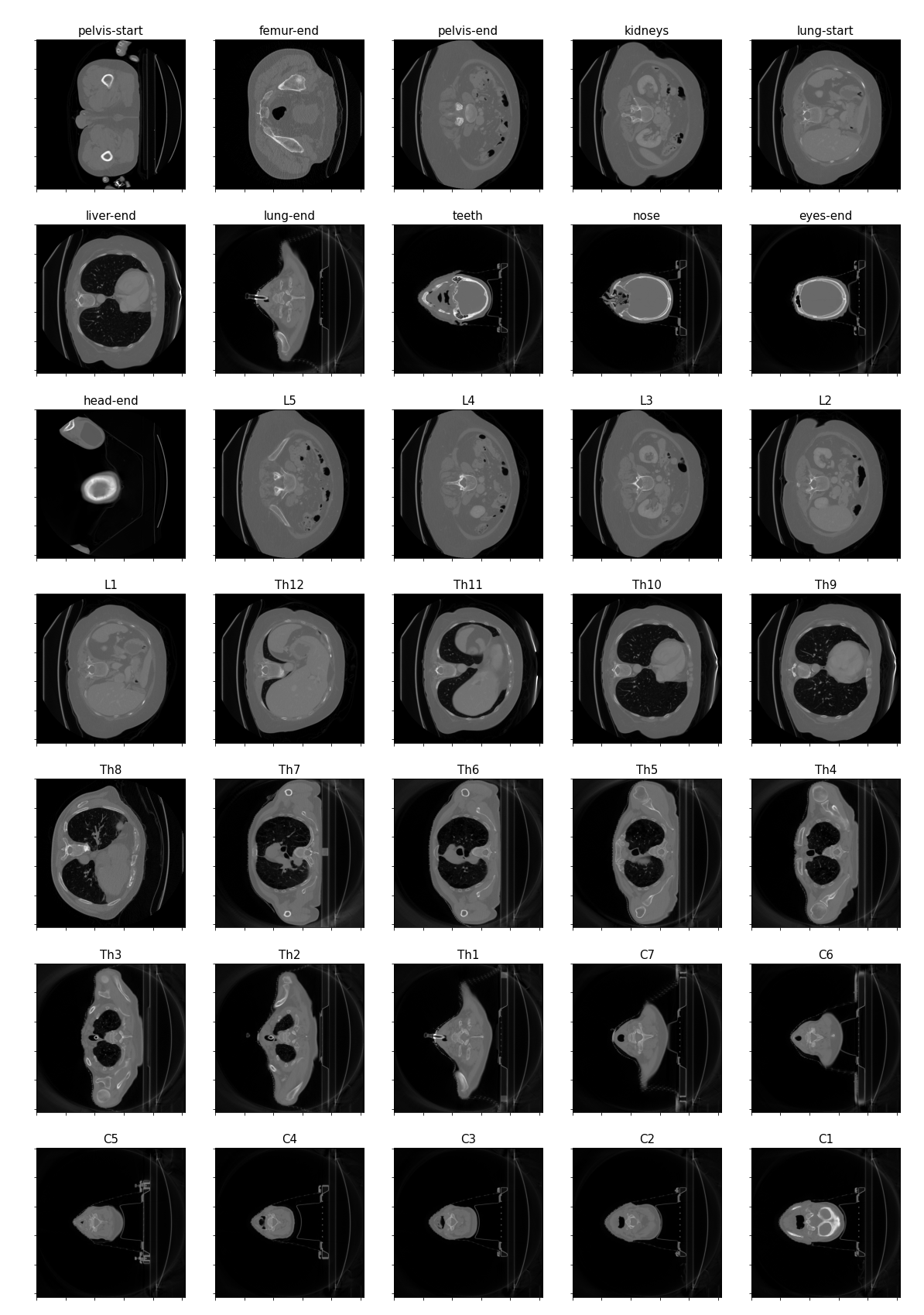}
    \caption{Example axial CT slices for each defined landmark of table \ref{tab:landmark-definitions}.}
    \label{fig:example-slices}
\end{figure}

\begin{table}
    \centering
    \caption{Counts of annotated landmarks in the training, validation and test set for the dataset described in section \ref{sec:dataset}.}
    \begin{tabular}{llll||l}
    \hline
    Dataset &  Training set &  Validation set &  Test set &    Sum \\
    \hline
    pelvis-start &     19 &          42 &    25 &     86 \\
    femur-end    &     24 &          47 &    31 &    102 \\
    L5           &     30 &          54 &    42 &    126 \\
    pelvis-end   &     33 &          62 &    60 &    155 \\
    L4           &     32 &          64 &    63 &    159 \\
    L3           &     37 &          75 &    74 &    186 \\
    kidney       &     39 &          80 &    78 &    197 \\
    L2           &     40 &          81 &    84 &    205 \\
    L1           &     42 &          83 &    87 &    212 \\
    lung-start   &     43 &          89 &    81 &    213 \\
    Th12         &     44 &          93 &    88 &    225 \\
    Th11         &     44 &          79 &    86 &    209 \\
    Th10         &     43 &          85 &    75 &    203 \\
    Th9          &     39 &          78 &    51 &    168 \\
    liver-end    &     41 &          87 &    66 &    194 \\
    Th8          &     33 &          67 &    38 &    138 \\
    Th7          &     31 &          62 &    29 &    122 \\
    Th6          &     28 &          62 &    27 &    117 \\
    Th5          &     29 &          61 &    30 &    120 \\
    Th4          &     26 &          65 &    32 &    123 \\
    Th3          &     24 &          66 &    31 &    121 \\
    Th2          &     29 &          66 &    33 &    128 \\
    lung-end     &     30 &          69 &    32 &    131 \\
    Th1          &     29 &          66 &    32 &    127 \\
    C7           &     25 &          52 &    30 &    107 \\
    C6           &     16 &          42 &    21 &     79 \\
    C5           &     13 &          31 &    18 &     62 \\
    C4           &     11 &          29 &    17 &     57 \\
    C3           &     13 &          29 &    15 &     57 \\
    teeth        &     20 &          36 &    19 &     75 \\
    C2           &     16 &          32 &    13 &     61 \\
    C1           &     13 &          30 &    12 &     55 \\
    nose         &     16 &          31 &    11 &     58 \\
    eyes-end     &     14 &          27 &    10 &     51 \\
    head-end     &      4 &           7 &    10 &     21 \\
    \hline
    \hline
    Sum          &   3880 &        8116 &  5804 &  17800 \\
    \hline
    \end{tabular}
    \label{tab:count-annotated-landmarks}
\end{table}

\begin{table}
    \centering
    \caption{Slice score reference table computed on all landmarks defined on the training and validation dataset. For computing the slice scores the private model from section \ref{sec:model-evaluation} was used. The mean slice score $\bar{s}$ and the standard deviation $\sigma_{s}$ are given for each landmark. With a checkmark  $\checkmark$ the evaluation landmarks and the organ landmarks are marked.}
    \begin{tabular}{llll} 
        \hline
        Landmark & $\bar{s} \pm  \sigma_{s}$ & Evaluation landmark & Organ landmark\\
        \hline 
        pelvis-start &    0.0 $\pm$ 2.1 &               \checkmark &                  \\
        femur-end    &   13.4 $\pm$ 1.1 &               \checkmark &                  \\
        L5           &   25.8 $\pm$ 2.4 &                     &                  \\
        pelvis-end   &   30.4 $\pm$ 2.9 &                     &                  \\
        L4           &   31.1 $\pm$ 3.0 &                     &                  \\
        L3           &   35.4 $\pm$ 2.5 &               \checkmark &                  \\
        kidney       &   39.0 $\pm$ 4.2 &                     &            \checkmark \\
        L2           &   39.1 $\pm$ 2.1 &                     &                  \\
        L1           &   42.6 $\pm$ 2.3 &               \checkmark &                  \\
        lung-start   &   45.1 $\pm$ 4.0 &                     &            \checkmark \\
        Th12         &   45.7 $\pm$ 2.6 &                     &                  \\
        Th11         &   48.5 $\pm$ 2.4 &               \checkmark &                  \\
        Th10         &   51.0 $\pm$ 2.2 &                     &                  \\
        Th9          &   53.3 $\pm$ 2.5 &                     &                  \\
        liver-end    &   54.0 $\pm$ 3.7 &                     &            \checkmark \\
        Th8          &   56.1 $\pm$ 2.6 &               \checkmark &                  \\
        Th7          &   59.0 $\pm$ 3.1 &                     &                  \\
        Th6          &   62.2 $\pm$ 2.8 &                     &                  \\
        Th5          &   64.7 $\pm$ 2.5 &               \checkmark &                  \\
        Th4          &   67.1 $\pm$ 2.0 &                     &                  \\
        Th3          &   69.2 $\pm$ 1.7 &                     &                  \\
        Th2          &   71.5 $\pm$ 2.2 &               \checkmark &                  \\
        lung-end     &   74.0 $\pm$ 2.4 &                     &            \checkmark \\
        Th1          &   74.5 $\pm$ 2.3 &                     &                  \\
        C7           &   76.5 $\pm$ 3.0 &                     &                  \\
        C6           &   78.4 $\pm$ 1.8 &               \checkmark &                  \\
        C5           &   80.5 $\pm$ 1.9 &                     &                  \\
        C4           &   83.0 $\pm$ 2.2 &                     &                  \\
        C3           &   85.2 $\pm$ 2.4 &                     &                  \\
        teeth        &   87.0 $\pm$ 4.6 &                     &                  \\
        C2           &   87.8 $\pm$ 2.2 &                     &                  \\
        C1           &   89.7 $\pm$ 1.7 &                     &                  \\
        nose         &   91.9 $\pm$ 5.7 &                     &            \checkmark \\
        eyes-end     &  100.0 $\pm$ 3.5 &               \checkmark &                  \\
        head-end     &  108.7 $\pm$ 2.1 &                     &                  \\
        \hline
    \end{tabular}
    \label{tab:slice-score-reference-table-train-val}
\end{table}

\begin{figure}
    \centering
    \includegraphics[width=\textwidth]{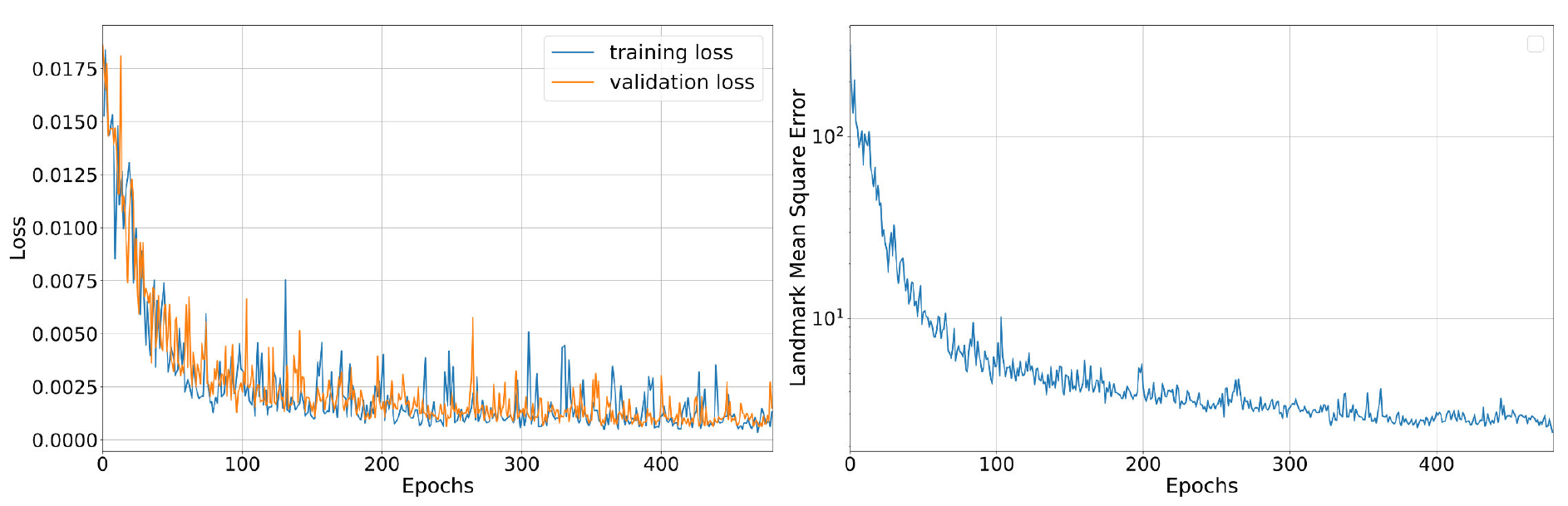}
    \caption{Insights of training for the model evaluated in section \ref{sec:model-evaluation}. In the left figure the validation loss and the training loss during training is visible, and in the right figure the LMSE on the validation set during training is shown.}
    \label{fig:best-model-loss-curves}
\end{figure}
\begin{figure}
    \centering
    \includegraphics[width=\textwidth]{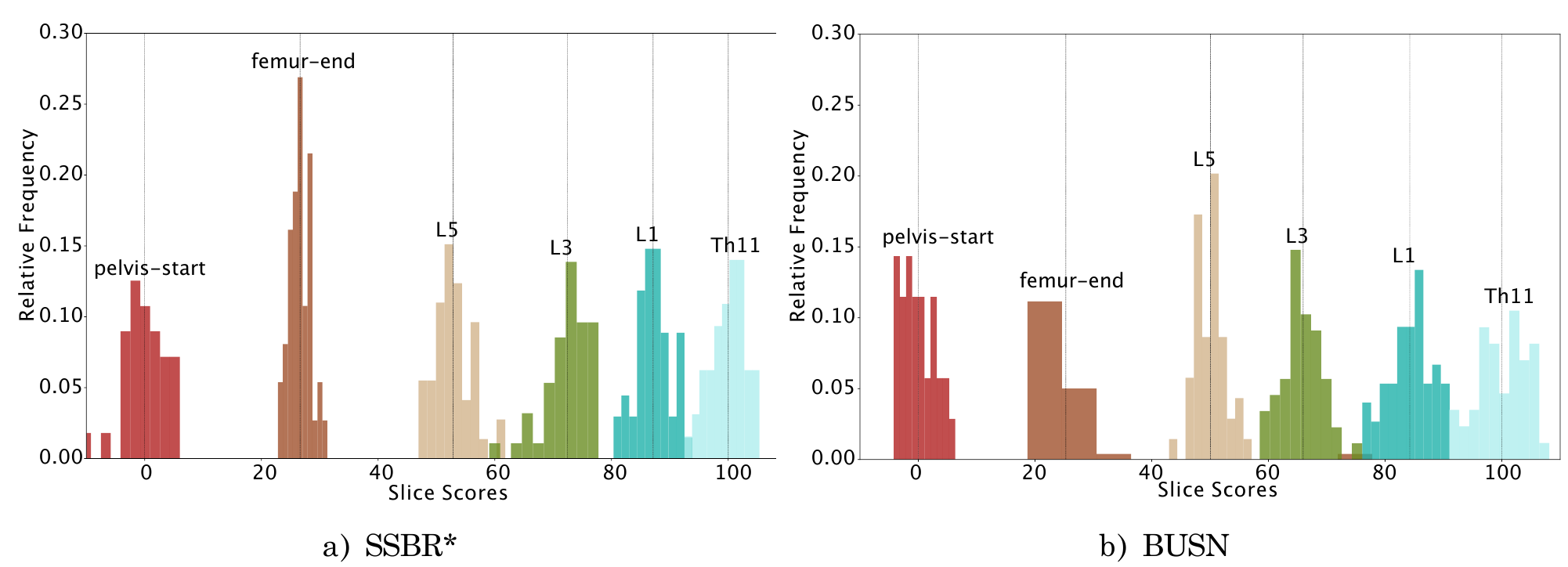}
    \caption{Relative frequency of landmark slice scores on the BTCV-dataset for the $\text{SSBR}^{\star}$ and the BUSN model. The black lines represent the pseudo-labels of the landmarks, which were calculated based on the mean landmark slice score values of the BCTV-dataset. The slice scores were linearly transformed, so that the mean landmark score for pelvis-start is mapped to 0 and the mean landmark score for Th11 is mapped to 100. }
    \label{fig:evaluation-landmarks-examples-busn}
\end{figure}

\begin{figure}
    \centering
    \includegraphics[width=0.5\textwidth]{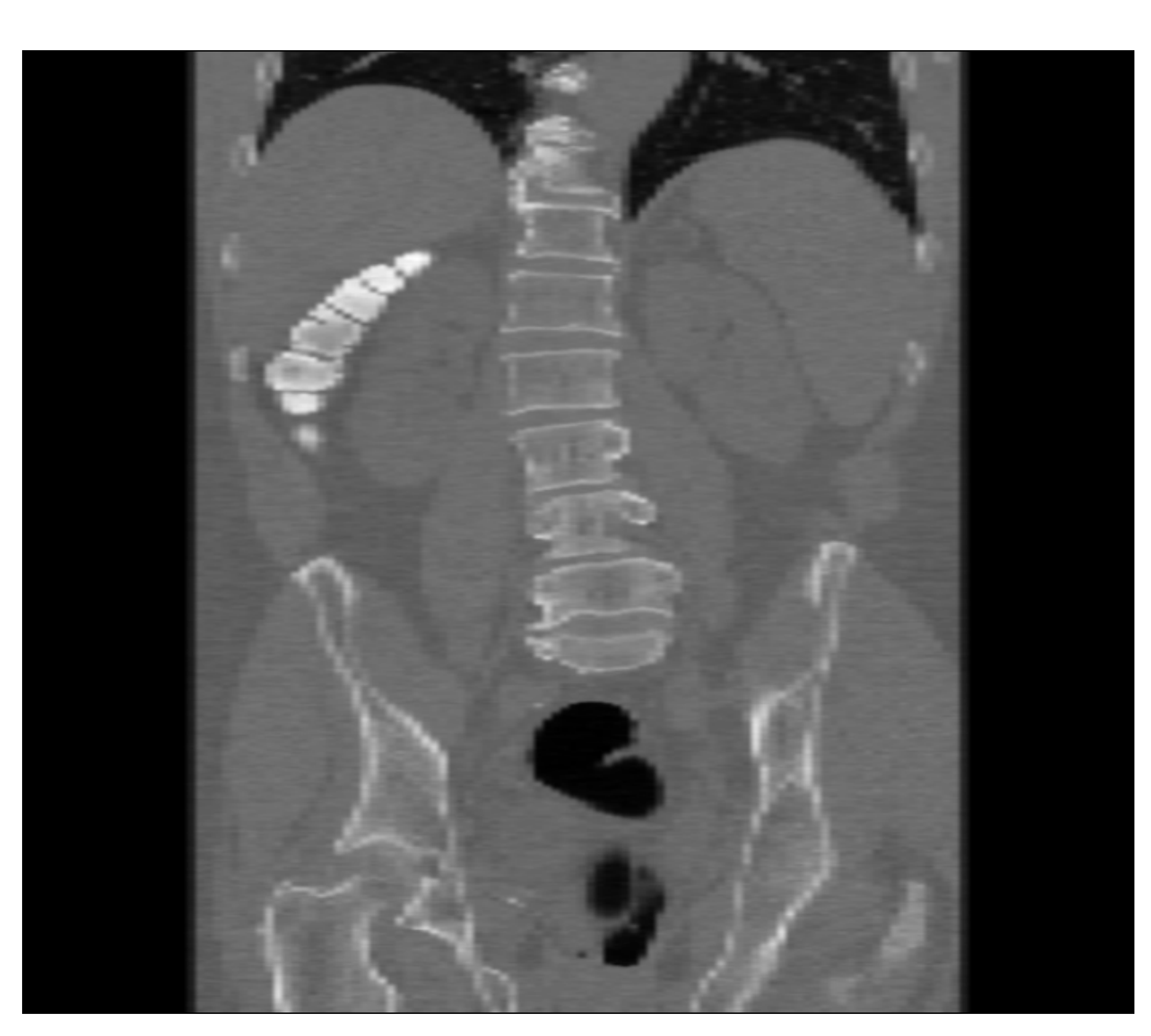}
    \caption{CT image from the validation set with the highest LMSE with $\phi=15.86$. It can be observed that the patient has scoliosis based on the curved spine. The CT volume comes from the CT-COLONOGRAPHY dataset \cite{clark2013cancer,johnson2008accuracy, dataCOLONOGRAPHY}. }
    \label{fig:scolioses-validation-set}
\end{figure}

{
    \chapter*{Acronyms}
    \renewcommand{\leftmark}{ACRONYMS}
    \renewcommand{\rightmark}{ACRONYMS}
    \addcontentsline{toc}{chapter}{Acronyms}
    
    \markboth{Acronyms}{} 
    
    \begin{acronym}[MMMMMMMM]
        \acro{BUSN}{Blind Unsupervised-Supervision Network}
        \acro{CT}{Computed Tomography}
        \acro{CNN}{Convolutional Neural Network}
        \acro{DAG}{Directed Acyclic Graph}
        \acro{DICOM}{Digital Imaging and Communication in Medicine}
        \acro{HU}{Hounsfield unit}
        \acro{JIP}{Joint Imaging Platform}
        \acro{JSON}{JavaScript Object Notation}
        \acro{LMSE}{Landmark Mean Square Error}
        \acro{MITK}{the Medical Imaging Interaction Toolkit}
        \acro{MRI}{Magnetic Resonance Imaging}
        \acro{MSE}{Mean Square Error}
        \acro{NIfTI}{Neuroimaging Informatics Technology Initiative}
        \acro{nnU-Net}{no new U-Net}
        \acro{PACS}{Picture Archiving And Communication Systems}
        \acro{ReLU}{Rectified Linear Unit}
        \acro{SSBR}{Self-Supervised Body Part Regression}
        \acro{TCIA}{The Cancer Imaging Archive}
        \acro{UC}{use case}

    \end{acronym}
}
\listoffigures
\addcontentsline{toc}{chapter}{List of Figures}
\listoftables
\addcontentsline{toc}{chapter}{List of Tables}

\addcontentsline{toc}{chapter}{Bibliography}
\printbibliography
\chapter*{Acknowledgement}
\addcontentsline{toc}{chapter}{Acknowledgement}
I want to express my special gratitude to my internal supervisors from the Medical Image Computing Division in the German Cancer Research Center, Lisa Kausch, and Dr. Fabian Isensee, which guided me throughout this project. Furthermore, I wish to show my appreciation to Prof. Dr. Klaus H. Maier-Hein, the head of the Medical Image Computing Division, for giving feedback on the project and making it possible to write my master thesis in the group. Moreover, I want to thank Dr. Peter Neher, Jonas Scherer, and Jan Sellner, who suggested the research project and accompanied me on my journey. Additionally, I would like to thank the whole Medical Image Computing group for the great scientific discussion and exchange. I wish to extend my thanks to Prof. Dr. Carsten Rother for making it possible for me to write my thesis at the German Cancer Research Center. Finally, I want to thank Jakob Henrichs, Maximilian Zenk, Malte Prinzler, and André Klein, who gave me valuable feedback on the thesis and helped me to finalize it.

\newpage\null\thispagestyle{empty}\newpage
\thispagestyle{empty}
\setlength{\parindent}{0em}

Erklärung:\par
\vspace{3\baselineskip}
Ich versichere, dass ich diese Arbeit selbstständig verfasst habe und keine
anderen als die angegebenen Quellen und Hilfsmittel benutzt habe.\par
\vspace{5\baselineskip}
Heidelberg, den \hspace{4cm}\dotfill

\end{document}